%% file: main.tex
\documentclass[pdfa, a4paper,USenglish,cleveref,autoref,thm-restate]{lipics-v2021}

\pdfoutput=1

\hideLIPIcs
\nolinenumbers

\input{macros}

\usepackage{amsfonts}
\usepackage{amsmath}
\usepackage{stmaryrd}

\theoremstyle{definition}
\newtheorem*{problem*}{Problem}

\usepackage[ruled,linesnumbered,noend]{algorithm2e}
\SetKwBlock{Repeat}{repeat}{}

\usepackage{pgfplots}
\pgfplotsset{compat=1.17}

\usepackage{tikz}
\usetikzlibrary{shapes}

\tikzset{
	iso/.style={kite, kite vertex angles=120,  minimum size=0.865cm, outer sep=0pt}
}

\newcommand{\isogrid}[2]{
	\foreach \i [count=\row from 0, remember=\row as \lastrow (initially 0)] in {0,...,#1}{
		\foreach \j [count=\col from 0, remember=\col as \lastcol (initially 0)] in {0,...,#2}{
			\ifnum\row=0
			\ifnum\col=0
			\node[iso] (\row-\col) {};
			\else
			\node[iso, anchor=left vertex] (\row-\col) at (\row-\lastcol.right vertex) {};
			\fi
			\else
			\ifnum\col=0
			\node[iso, anchor=upper vertex] (\row-\col) at (\lastrow-\col.lower vertex) {};
			\else
			\node[iso, anchor=left vertex] (\row-\col) at (\row-\lastcol.right vertex) {};
			\fi
			\fi
		}
	}
}

\title{{P}areto-Rational Verification}

\author{Véronique Bruyère}{Université de Mons (UMONS), Mons, Belgium}{}{}{}

\author{Jean-François Raskin}{Université libre de Bruxelles (ULB), Brussels, Belgium}{}{}{}

\author{Clément Tamines}{Université de Mons (UMONS), Mons, Belgium}{}{}{}

\authorrunning{V. Bruyère, J.-F. Raskin, and C. Tamines}

\Copyright{Véronique Bruyère, Jean-François Raskin, and Clément Tamines}

\ccsdesc[100]{Software and its engineering~Formal methods}
\ccsdesc[100]{Theory of computation~Logic and verification}
\ccsdesc[100]{Theory of computation~Solution concepts in game theory}

\keywords{Rational verification, Model-checking, Pareto-optimality, $\omega$-regular objectives}

\funding{This work is partially supported by the two PDR projects \emph{Subgame perfection in graph games} and \emph{Rational} (F.R.S.-FNRS), the ARC project \emph{Non-Zero Sum Game Graphs: Applications to Reactive Synthesis and Beyond} (Fédération Wallonie-Bruxelles), the EOS project \emph{Verifying Learning Artificial Intelligence Systems} (F.R.S.-FNRS and FWO), and the COST Action 16228 \emph{GAMENET} (European Cooperation in Science and Technology).}

\begin{document}

\maketitle

\begin{abstract}
We study the rational verification problem which consists in verifying the correctness of a system executing in an environment that is assumed to behave rationally. We consider the model of rationality in which the environment only executes behaviors that are Pareto-optimal with regard to its set of objectives, given the behavior of the system (which is committed in advance of any interaction). We examine two ways of specifying this behavior, first by means of a deterministic Moore machine, and then by lifting its determinism. In the latter case the machine may embed several different behaviors for the system, and the universal rational verification problem aims at verifying that all of them are correct when the environment is rational. For parity objectives, we prove that the Pareto-rational verification problem is \conpComplete{} and that its universal version is in \pspace{} and both \npHard{} and \conpHard{}. For Boolean B\"uchi objectives, the former problem is \piComplete{} and the latter is \pspaceComplete{}. We also study the case where the objectives are expressed using LTL formulas and show that the first problem is \pspaceComplete{}, and that the second is \twoExptime{}-complete. Both problems are also shown to be fixed-parameter tractable (\FPT{}) for parity and Boolean B\"uchi objectives. Finally, we evaluate two variations of the \FPT{} algorithm proposed to solve the  Pareto-rational verification problem on a parametric toy example as well as on randomly generated instances.
\end{abstract}

\section{Introduction}

Formal verification is essential to ensure the correctness of systems responsible for critical tasks. 
Many advancements have been made in the field of formal verification both in terms of theoretical foundations and tool development, and computer-aided verification techniques, such as \emph{model-checking}~\cite{DBLP:books/daglib/0020348,DBLP:reference/mc/BloemCJ18}, are now widely used in industry.
In the classical approach to verification, it is assumed that the system designer provides \emph{(i)} a model of the system to verify, together with \emph{(ii)} a model of the environment in which the system will be executed, and \emph{(iii)} a specification $\varphi$ (e.g. an $\omega$-regular property) that must be enforced by the system.
Those models are usually nondeterministic automata that cover all possible behaviors of both the system and the environment. The model-checking algorithm is then used to decide if all executions of the system in the environment are correct with regard to $\varphi$. 
Unfortunately, in some settings, providing a faithful and sufficiently precise model of the environment may be \emph{difficult or even impossible}. This is particularly true in heterogeneous systems composed of software entities interacting with human users, e.g. self-driving cars interacting with human drivers. Alternative approaches are thus needed in order to verify such complex multi-agent systems. 
One possible solution to this problem is to consider more \emph{declarative} ways of modeling the environment. Instead of considering an operational model of each agent composing the environment, in this paper, we propose instead to identify the \emph{objectives} of those agents. 
We then consider only the behaviors of the environment that concur to those objectives, instead of all behaviors described by some model of the system. 
We study the problem of \emph{rational verification}: the system needs to be proven correct with regard to property $\varphi$, not in all the executions of the environment, but only in those executions that are \emph{rational} with regard to the objectives of the environment. 

There are several ways to model rationality. For instance, a famous model of rational behavior for the agents is the concept of \emph{Nash equilibrium} (NE)~\cite{Nas50}. Some promising exploratory works, based on the concept of NE, exist in the literature, like in verification of non-repudiation and fair exchange protocols~\cite{DBLP:journals/jcs/KremerR03,DBLP:conf/vmcai/ChatterjeeR12}, planning of self-driving cars interacting with human drivers~\cite{DBLP:conf/rss/SadighSSD16}, or the automatic verification of an LTL specification in multi-agent systems that behave according to an NE~\cite{DBLP:journals/ai/GutierrezNPW20}. Another classical approach is to model the environment as a single agent with multiple objectives. In that setting, trade-offs between (partially) conflicting objectives need to be made, and a rational agent will behave in a way to satisfy a \emph{Pareto-optimal} set of its objectives. 
Pareto-optimality and multi-objective formalisms have been considered in computer science, see for instance~\cite{DBLP:conf/focs/PapadimitriouY00} and references therein, and in formal methods, see e.g.~\cite{DBLP:conf/fossacs/AlurDMW09,DBLP:conf/cav/BrenguierR15}.

Nevertheless, we have only scratched the surface and there is a lack of a general theoretical background for marrying concepts from game theory and formal verification. This is the motivation of our work. 
We consider the setting in which a designer specifies the behavior of a system and identifies its objective $\Obj_0$ as well as the multiple objectives $\{ \Omega_1, \dots, \Omega_t\}$ of the environment in a underlying game arena $G$. The behavior of the system is usually modeled by the designer using a \emph{deterministic Moore machine} that describes the strategy of the system opposite the environment. The designer can also use the model of \emph{nondeterministic} Moore machine in order to describe a set of multiple possible strategies for the system instead of some single specific strategy. Given a strategy $\sigma_0$ for the system, the environment being rational only executes behaviors induced by $\sigma_0$ which result in a Pareto-optimal payoff with regard to its set of objectives $\{ \Omega_1, \dots, \Omega_t\}$. When the Moore machine $\mathcal{M}$ is deterministic, the \emph{Pareto-rational verification} (PRV) problem asks whether all behaviors that are induced by the machine $\mathcal{M}$ in $G$ and that are Pareto-optimal for the environment all satisfy the objective $\Obj_0$ of the system. When the Moore machine is nondeterministic, the \emph{universal} \problemVerifAb{} asks whether for all strategies $\sigma_0$ of the system described by $\mathcal{M}$, all behaviors induced by $\sigma_0$ that are Pareto-optimal for the environment satisfy $\Obj_0$. The latter problem is a clear generalization of the former and is conceptually more challenging, as it asks to verify the correctness of the possibly infinite set of strategies described by $\mathcal{M}$. The universal \problemVerifAb{} is also a well motivated problem, as typically, in the early stages of a development cycle, not all implementation details are fixed, and the use of nondeterminism is prevailing. In this last setting, we want to guarantee that a positive verification result is transferred to all possible implementations of the nondeterministic model of the system.

\begin{example}
\label{ex:intersection_general}

\input{intersection-figure}

Consider Figure \ref{fig:intersection_example} where three cars aim to cross straight ahead at an intersection managed by three traffic lights. Car $c_1$ is a self-driving car seen as a system that interacts with an environment composed of two other cars ($c_2$ and $c_3$, which can be self-driving or controlled by a human driver) and a traffic management system which controls the lights. 
The objective of car $c_1$ is to eventually cross the intersection without accident. The environment has several objectives, such as eventually turning each light green, allowing cars to cross without accidents or making several cars cross at the same time.

It is sound to ask that car $c_1$ fulfils its objective whatever the rational behavior of the environment which satisfies a Pareto-optimal set of its objectives. This example is further developed later in \autoref{sec:implementation}.

\end{example}

\subparagraph{Technical Contributions.} We introduce the Pareto-rational verification (PRV) problem and its universal variant. The objective $\Obj_0$ of the system and the set of objectives $\{ \Omega_1, \dots, \Omega_t\}$ of the environment are $\omega$-regular objectives. We consider several ways of specifying these objectives: either by using parity conditions (a canonical way to specify $\omega$-regular objectives), Boolean B\"uchi conditions (a generic way to specify B\"uchi, co-B\"uchi, Streett, Rabin, and other objectives), or using LTL formulas. Our technical results, some of which are summarized in \autoref{table:summary-comp}, are as follows. 
\begin{table}[ht]
    \centering
    \caption{Summary of complexity results for the \problemVerifAb{} and \problemUVerifAb{}.}
    \label{table:summary-comp}
    \resizebox{\textwidth}{!}{%
    \begin{tabular}{ l|l|l }
    Objective           & \problemVerifAb{} complexity                            & \problemUVerifAb{} complexity               \\ 
    \hline
    Parity              & \conpComplete{} (\autoref{thm:PRV-parity-complexity})      & \pspace{}, \npHard{}, \conpHard{} (\autoref{thm:PSPACE})  \\ 
    Boolean B\"uchi     & \piComplete{} (\autoref{thm:PRV-BB-complexity})        & \pspaceComplete{} (\autoref{thm:PSPACE})                   \\ 
    LTL                 & \pspaceComplete{} (\autoref{thm:ltl_verif_pspace})      & \twoExptime{}-complete (\autoref{thm:ltl_universal_verif_pspace})      
    \end{tabular}
    }%
\end{table} 

First, we study the complexity class of the \problemVerifAb{}. We prove that it is \conpComplete{} for parity objectives, \piComplete{} for Boolean B\"uchi objectives, and \pspaceComplete{} for LTL objectives.

Second, we consider the universal variant of the \problemVerifAb{}. We prove that it is in \pspace{} and both \npHard{} and \conpHard{} for parity objectives, \pspaceComplete{} for Boolean B\"uchi objectives, and $\mathsf{2EXPTIME}$-complete for LTL objectives.

Third, we establish the fixed-parameter tractability (\FPT{}) of the universal \problemVerifAb{} where the parameters are the number $t$ of objectives of the environment as well as the highest priorities used in the parity objectives or the size of the formulas used in the Boolean B\"uchi objectives. For the particular case of the \problemVerifAb{} with parity conditions, the parameters reduce to $t$ only. Since this number is expected to be limited in practice, our result is of practical relevance. We further propose two approaches to efficiently implement this \FPT{} result for the case of the \problemVerifAb{}. The first optimizes the construction of the set of Pareto-optimal payoffs for the environment, while the second is a counterexample-based algorithm that builds an under-approximation of the set of Pareto-optimal payoffs on demand.

Finally, we have implemented the two variations of our \FPT{} algorithm for the \problemVerifAb{} in the case of parity objectives, and compared their performances on a series of parametric instances generalizing \autoref{ex:intersection_general} as well as on a family of randomly generated instances.

\subparagraph{Related Work.} The concept of nondeterminism for strategies has been studied in the particular context of two-player zero-sum games where one player is opposed to the other one, under the name of permissive strategy, multi-strategy, or nondeterministic strategy in~\cite{BernetJW02,BouyerDMR09,BouyerFM15,DBLP:journals/corr/abs-0806-2923,pinchinat}. Those works concern synthesis and not verification.

Several fundamental results have been obtained on multi-player games played on graphs where the objectives of the players are Boolean or quantitative (see e.g. the book chapter~\cite{GU08} or the surveys~\cite{DBLP:conf/lata/BrenguierCHPRRS16,Bruyere17,DBLP:journals/siglog/Bruyere21}). Several notions of rational behavior of the players have been studied such as NEs, subgame perfect equilibria (SPEs)~\cite{selten}, secure equilibria~\cite{DBLP:journals/tcs/ChatterjeeHJ06}, or profiles of admissible strategies~\cite{DBLP:conf/stacs/Berwanger07}. The existing results in the literature are mainly focused on the existence of equilibria or the synthesis of such equilibria when they exist.
Multidimensional energy and mean-payoff objectives for two-player games played on graphs have been studied in~\cite{DBLP:conf/fsttcs/ChatterjeeDHR10,DBLP:journals/iandc/VelnerC0HRR15,DBLP:conf/fossacs/VelnerR11} and the Pareto curve of multidimensional mean-payoff games has been studied in~\cite{DBLP:conf/cav/BrenguierR15}. Two-player games with heterogeneous multidimensional quantitative objectives have been investigated in~\cite{DBLP:conf/concur/BruyereHR16}.

Recent results concern the synthesis of strategies for a system in a way to satisfy its objective when facing an environment that is assumed to behave rationally with respect to the objectives of all his components. In~\cite{FismanKL10,KupfermanPV16,KupfermanS22}, the objectives are expressed as LTL formulas and the considered models of rationality are NEs or SPEs. Algorithmic questions about this approach are studied in~\cite{ConduracheFGR16} for different types of $\omega$-regular objectives. In~\cite{DBLP:conf/concur/BruyereRT21}, the objectives are $\omega$-regular and the environment is assumed to behave rationally by playing in a way to obtain Pareto-optimal payoffs with respect to its objectives. We consider the concepts of~\cite{DBLP:conf/concur/BruyereRT21} as a foundation for Pareto-rational verification.

The previously mentioned results all deal with the existence or the synthesis of solutions. Rational verification (instead of synthesis) is studied in~\cite{DBLP:journals/ai/GutierrezNPW20} (see also the survey~\cite{DBLP:journals/apin/AbateGHHKNPSW21}), where the authors study how to verify a given specification for a multi-agent system with agents that behave rationally according to an NE when all objectives are specified by LTL formulas. They prove that this problem is $\mathsf{2EXPTIME}$-complete and design an algorithm that reduces this problem to solving a collection of parity games. This approach is implemented in the Equilibrium Verification Environment tool. In this paper, we study Pareto-optimality as a model of rationality instead of the concepts of NE or SPE. Our framework is more tractable as the \problemVerifAb{} is \pspaceComplete{} for LTL specifications. 

\subparagraph{Structure of the Paper.} \autoref{sec:prelim} recalls some classical notions about games played on graphs, describes the class of games considered in this paper and introduces the (universal) \problemVerifAb{}. \autoref{sec:complexity_class_PRV} discusses the complexity of the \problemVerifAb{} both for games with parity and Boolean B\"uchi objectives and \autoref{sec:complexity_class_UPRV} does so for the universal version of the problem. In \autoref{sec:fpt}, we prove that both problems are in \FPT{} for games with parity or Boolean B\"uchi objectives. We also provide two variations of the proposed \FPT{} algorithm for the \problemVerifAb{} in order to efficiently implement it. It is shown in \autoref{sec:LTL} that the \problemVerifAb{} is \pspaceComplete{} when the objectives are defined by LTL formulas and that the \problemUVerifAb{} is \twoExptime{}-complete in that setting. We evaluate in \autoref{sec:implementation} the two variations of the \FPT{} algorithm for the \problemVerifAb{} in the case of parity objectives on parametric instances generalizing \autoref{ex:intersection_general} as well as on randomly generated instances. The last section contains a conclusion.

\section{Definitions and the Pareto-Rational Verification Problem}
\label{sec:prelim}

We start by recalling several classical concepts of game theory, and in particular the model of (nondeterministic) Moore machines. We then present the verification problem studied in this paper and illustrate it on an example. We end the section by discussing the complexity of useful checks performed in several algorithms throughout this paper.

\subsection{Definitions} 
\label{subsec:preliminaries}

\subparagraph{Game Arena and Plays.}
A \emph{game arena} is a tuple \sloppy $G = (V, V_0, V_1, E, v_0)$ where $(V,E)$ is a finite directed graph such that: \emph{(i)} $V$ is the set of vertices and $(V_0, V_1)$ forms a partition of $V$ where $V_0$ (resp.\!\ $V_1$) is the set of vertices controlled by Player~$0$ (resp.\!\ Player~$1$), \emph{(ii)} $E \subseteq V \times V$ is the set of edges such that each vertex $v$ has at least one successor $v'$, i.e., $(v,v') \in E$, and \emph{(iii)} $v_0 \in V$ is the initial vertex. We denote by $|G|$ the size of $G$. A \emph{sub-arena} $G'$ with a set $V' \subseteq V$ of vertices and initial vertex $v'_0 \in V'$ is a game arena defined from $G$ as expected. A \emph{single-player} game arena is a game arena where $V_0 = \emptyset$ and $V_1 = V$ (all vertices are thus controlled by Player~$1$).

\subparagraph{Plays.}
A \emph{play} in a game arena $G$ is an infinite sequence of vertices $\rho = v_0 v_1 \dots \in V^{\omega}$ such that it starts with the initial vertex $v_0$ and $(v_j,v_{j+1}) \in E$ for all $j \in \N$. \emph{Histories} in $G$ are finite non-empty sequences $h = v_0 \dots v_j \in V^+$ defined similarly. The set of plays in $G$ is denoted by $\Plays_G$ and the set of histories (resp.\!\ histories ending with a vertex in $V_i$) is denoted by $\Hist_G$ (resp.\!\ $\Hist_{G,i}$). Notations $\Plays$, $\Hist$, and $\Hist_i$ are used when $G$ is clear from the context. The set of vertices occurring (resp.\! occurring infinitely often) in a play $\rho$ is written $\occ{\rho}$  (resp.\! $\infOcc{\rho}$).

\subparagraph{Strategies and Moore Machines.}
A \emph{strategy} $\sigma_i$ for Player~$i$ is a function $\sigma_i\colon \Hist_i \rightarrow V$ assigning to each history $hv \in \Hist_i$ a vertex $v' = \sigma_i(hv)$ such that $(v,v') \in E$. A play $\rho = v_0 v_1 \dots$ is \emph{consistent} with $\sigma_i$ if $v_{j+1} = \sigma_i(v_0 \dots v_j)$ for all $j \in \N$ such that $v_j \in V_i$. Consistency is naturally extended to histories. The set of plays (resp.\!\ histories) consistent with strategy $\sigma_i$ is written $\Playsigma{\sigma_i}$ (resp.\!\ $\Histsigma{\sigma_i}$).

A strategy $\sigma_i$ for Player~$i$ is \emph{finite-memory}~\cite{2001automata} if it can be encoded by a \emph{deterministic Moore machine} ${\mathcal{M}} = (M, m_0, \alpha_U, \alpha_N)$ where $M$ is the finite set of states (the memory of the strategy), $m_0 \in M$ is the initial memory state, $\alpha_U : M \times V \rightarrow M$ is the update function, and $\alpha_N : M \times V_i \rightarrow V$ is the next-move function. Such a machine defines the strategy $\sigma_i$ such that $\sigma_i(hv) = \alpha_N(\widehat{\alpha}_U(m_0,h),v)$ for all histories $hv \in \Hist_i$, where $\widehat{\alpha}_U$ extends $\alpha_U$ to histories as expected. In this paper, we consider the broader notion of \emph{nondeterministic} Moore machine ${\mathcal{M}}$ (see e.g.~\cite{BernetJW02}) with a next-move function $\alpha_N : M \times V_i \rightarrow 2^V$. Such a machine embeds a (possibly infinite) set of strategies $\sigma_i$ for Player~$i$ such that $\sigma_i(hv) \in \alpha_N(\widehat{\alpha}_U(m_0,h),v)$ for all histories $hv \in \Hist_i$\footnote{Notice that this definition is different from simply making the machine deterministic by fixing a single next vertex $v' \in \alpha_N(m, v)$ for each $m \in M$ and $v \in V_i$.}. We denote by $\llbracket \mathcal{M} \rrbracket$ the set of all strategies defined by $\mathcal{M}$. The \emph{size} of $\mathcal{M}$ is equal to the number $|M|$ of its memory states. \autoref{ex:instance} illustrates these concepts.

When $\mathcal{M}$ is a deterministic Moore machine with $|M| = 1$, then it defines a \emph{memoryless} strategy $\sigma_i$ where $\sigma_i(hv) = \sigma_i(h'v)$ for all $hv, h'v$ ending with the same vertex $v \in V_i$. When $\mathcal{M}$ is a nondeterministic Moore machine with $|M| = 1$ and such that $\alpha_N(m_0,v) = \{v' \mid (v,v') \in E \}$, then $\llbracket \mathcal{M} \rrbracket$ is exactly the set of \emph{all possible strategies} for Player~$i$.

\subparagraph{Objectives.}
An \emph{objective} for Player~$i$ is a set of plays $\Obj \subseteq \Plays$. A play $\rho$ \emph{satisfies} the objective $\Obj$ if $\rho \in \Obj$. The \emph{opposite objective} of $\Omega$ is written $\overline{\Omega} = \Plays \setminus \Omega$. We consider the following objectives in this paper:
\begin{itemize}
    \item Let $c : V \rightarrow \{0, \dots, d\}$ be a function called a \emph{priority function} which assigns an integer to each vertex in the arena (we assume that $d$ is even). The set of priorities occurring infinitely often in a play $\rho$ is $\infOcc{c(\rho)} = \{c(v) \mid v \in \infOcc{\rho}\}$. The \emph{parity} objective $\parity{c} = \{\rho \in \Plays \mid \min(\infOcc{c(\rho)}) \text{ is even}\}$ asks that the minimum priority visited infinitely often be even.
    The opposite objective $\overline{\Omega}$ of a parity objective $\Omega$ is again a parity objective (the priority function $c'$ of $\overline{\Omega}$ is such that $c'(v) = c(v) + 1$ for all $v \in V$).
    
    \item Given $m$ sets $\target_1, \dots, \target_m$ such that $\target_i \subseteq V$, $i \in \{1, \dots, m\}$, and $\phi$ a Boolean formula over the set of variables $X = \{x_1, \dots, x_m\}$, the \emph{Boolean B\"uchi}\footnote{This objective is also called \emph{Emerson-Lei} objective.}~\cite{DBLP:journals/scp/EmersonL87,BruyereHR18} objective $\BooleanBuchi{\phi, \target_1, \dots, \target_ m} = \{\rho \in \Plays \mid \rho \text{ satisfies } (\phi, \target_1, \dots, \target_ m) \}$ is the set of plays whose valuation of the variables in $X$ satisfy formula $\phi$. Given a play $\rho$, its valuation is such that $x_i = 1$ if and only if $\infOcc{\rho} \cap \target_i \neq  \emptyset$ and $x_i = 0$ otherwise. That is, a play satisfies the objective if the Boolean formula describing the sets to be visited (in)finitely often by a play is satisfied. It is assumed that negations only appear in literals of $\phi$ and we denote by $|\phi|$ the size of $\phi$ equal to the number of symbols in $\{\land, \lor, \lnot\} \cup X $ in $\phi$.

    The opposite objective $\overline{\Omega}$ of a Boolean B\"uchi objective $\Omega$ is again a Boolean B\"uchi objective (the formula $\neg \phi$ of $\overline{\Omega}$ is obtained from $\phi$ by replacing each symbol $\lor$ (resp.\! $\land$) by $\land$ (resp.\! $\lor)$ and each literal by its negation).
\end{itemize}
We recall that parity and Boolean B\"uchi objectives $\Obj$ are \emph{prefix-independent}, i.e., whenever $\rho \in \Obj$, then all suffixes of $\rho$ also satisfy $\Obj$.

\subparagraph{Zero-Sum Games.} A two-player \emph{zero-sum game} $\mathcal{G} = (G, \Obj)$ is a game on a game arena $G$ where Player~$0$ has objective $\Obj$ and Player~$1$ has the opposite objective $\overline{\Obj}$. Given an initial vertex $v_0$, we say that a player is \emph{winning from $v_0$} if he has a strategy such that all plays starting with $v_0$ and consistent with this strategy satisfy his objective. We assume that the reader is familiar with this concept, see e.g.~\cite{2001automata}.

\subparagraph{Lattices and Antichains.} 
A \emph{complete lattice} is a partially ordered set $(S, \leq)$ where $S$ is a set, ${\leq} \subseteq S \times S$ is a partial order on $S$, and for every pair of elements $s, s' \in S$, their greatest lower bound and their least upper bound both exist. A subset $A \subseteq S$ is an \emph{antichain} if all of its elements are pairwise incomparable with respect to $\leq$. Given $T \subseteq S$ and an antichain $A \subseteq S$, we denote by $\lceil T \rceil$ the set of maximal elements of $T$ (which is thus an antichain) and by $\downarrow^{<} \! A$ the set of all elements $s \in S$ for which there exists some $s' \in A$ such that $s < s'$. Given two antichains $A, A' \subseteq S$, we write $A \sqsubseteq A'$ when for all $s \in A$, there exists $s'\in A'$ such that $s \leq s'$, and we write $A \sqsubset A'$ when $A \sqsubseteq A'$ and $A \neq A'$.

\subsection{Pareto-Rational Verification Problem}
\label{subsec:PRVP}

We start by recalling the class of two-player games considered in this paper and the notion of payoffs in those games.

\subparagraph{Stackelberg-Pareto Games.} 
A \emph{\game{}} (\gameAb{}) $\mathcal{G} = (G, \ObjPlayer{0},\ObjPlayer{1}, \dots, \ObjPlayer{\nbrObjectives})$ is composed of a game arena $G$, an objective $\ObjPlayer{0}$ for Player~$0$, and $\nbrObjectives \geq 1$ objectives $\ObjPlayer{1}, \dots, \ObjPlayer{\nbrObjectives}$ for Player~$1$ \cite{DBLP:conf/concur/BruyereRT21}. An \gameAb{} where all objectives are parity (resp.\! Boolean B\"uchi) objectives is called a \emph{parity} (resp.\! \emph{Boolean B\"uchi}) \emph{\gameAb{}}.

\subparagraph{Payoffs.}
The \emph{payoff} of a play $\rho \in \Plays$ corresponds to the vector of Booleans $\payoff{\rho} \in \{0,1\}^{\nbrObjectives}$ such that for all $i \in \{1, \dots, \nbrObjectives\}$, $\payoffObj{i}{\rho} = 1$ if $\rho \in \ObjPlayer{i}$, and $\payoffObj{i}{\rho} = 0$ otherwise. Notice that we omit to include the objective of Player~$0$ when discussing the payoff of a play. Instead we say that a play $\rho$ is \emph{won} by Player~$0$ if $\rho \in \ObjPlayer{0}$ and we write $\won{\rho} = 1$, otherwise it is \emph{lost} by Player~$0$ and we write $\won{\rho} = 0$. We write $(\won{\rho},\payoff{\rho})$ for the \emph{extended payoff} of $\rho$. A payoff $p$ (resp.\! extended payoff $(w,p)$) is \emph{realizable} if there exists a play $\rho \in \Plays $ such that $\payoff{\rho} = p$ (resp.\! $(\won{\rho},\payoff{\rho}) = (w,p)$); we say that $\rho$ \emph{realizes} $p$ (resp.\! $(w,p)$).

We consider the following partial order on payoffs. Given two payoffs $p = (p_1, \dots, p_\nbrObjectives)$ and $p' = (p'_1, \dots, p'_\nbrObjectives)$ such that $p, p' \in \{0,1\}^{\nbrObjectives}$, we say that $p'$ is \emph{larger} than $p$ and write $p  \leq p'$ if $p_i \leq p'_i$ for all $i \in \{1, \dots, \nbrObjectives\}$. Moreover, when it also  holds that $p_i < p'_i$ for some $i$, we say that $p'$ is \emph{strictly larger} than $p$ and we write $p < p'$. Notice that the pair $(\{0,1\}^{\nbrObjectives}, \leq)$ is a complete lattice with size $2^\nbrObjectives$ and that the size of any antichain on $(\{0,1\}^{\nbrObjectives}, \leq)$ is thus upper bounded by $2^\nbrObjectives$. \autoref{fig:lattice} depicts the lattice of payoffs for $\nbrObjectives = 3$ objectives. The sets $A = \{(0,1,1),(1,1,0)\}$ and $A' = \{(0,1,0)\}$ are two antichains such that $A' \sqsubset A$ and $\downarrow^< \! A$ is composed of the payoffs highlighted in bold. 

\begin{figure}[t]
	\centering
	\resizebox{0.5\textwidth}{!}{%
    	\begin{tikzpicture}
    		
    		      \node (max) at (0,3.5) { $(1,1,1)$};
                  \node (a) at (-2.5,2) {$(0,1,1)$};
                  \node (b) at (0,2) {$(1,0,1)$};
                  \node (c) at (2.5,2) {$(1,1,0)$};
                  \node (d) at (-2.5,0.5) {$\mathbf{(0,0,1)}$};
                  \node (e) at (0,0.5) {$\mathbf{(0,1,0)}$};
                  \node (f) at (2.5,0.5) {$\mathbf{(1,0,0)}$};
                  \node (min) at (0,-1) {$\mathbf{(0,0,0)}$};
                  \draw (min) -- (d) -- (a) -- (max) -- (b) -- (f)
                  (e) -- (min) -- (f) -- (c) -- (max)
                  (d) -- (b);
                  \draw[preaction={draw=white, -,line width=6pt}] (a) -- (e) -- (c);
    	\end{tikzpicture}
	}%
    \caption{The lattice of payoffs for $\nbrObjectives = 3$ objectives.}
    \label{fig:lattice}
\end{figure}

\subparagraph{} Let $\mathcal{G} = (G, \ObjPlayer{0},\ObjPlayer{1}, \dots, \ObjPlayer{\nbrObjectives})$ be an \gameAb{} and let $\sigma_0$ be a strategy of Player~$0$. We can consider the set of payoffs of plays consistent with $\sigma_0$ which are \emph{Pareto-optimal}, i.e., maximal with respect to $\leq$. We write this set $\paretoSet{\sigma_0} =  \max \{\payoff{\rho} \mid \rho \in \Playsigmazero \}$. Notice that this set is an antichain. In this paper, we study the following \emph{verification} problem.

\begin{problem*}
	Let $\mathcal{G}$ be an \gameAb{} and let $\mathcal{M}$ be a nondeterministic Moore machine for Player~$0$. The \emph{\problemUVerif{}} (\problemUVerifAb{}) is to decide whether for all $\sigma_0 \in \llbracket \mathcal{M} \rrbracket$, it holds that every play $\rho \in \Playsigmazero$ such that $\payoff{\rho} \in \paretoSet{\sigma_0}$ satisfies the objective of Player~$0$. When $\mathcal{M}$ is deterministic, we consider the single strategy $\sigma_0 \in \llbracket \mathcal{M} \rrbracket$ and speak about the \emph{\problemVerif{}} (\problemVerifAb{}).
\end{problem*}

The \problemUVerifAb{} models the situation where the system may employ one of several possible strategies in a nondeterministic manner and we therefore want to verify that all of them are correct. We do so in the context where the environment is rational and only executes behaviors which result in a Pareto-optimal payoff with regard to its set of objectives. In the following sections, we study the complexity of the (U)PRV problem in terms of $|G|$ the size of the game arena, $|M|$ the size of the Moore machine, $\nbrObjectives$ the number of objectives of Player~$1$, $\max d_i$ the maximum of all maximum priorities $d_i$ according to each parity objective $\Omega_i$ in case of parity \gamesAb{}, and $\max |\phi_i|$ the maximum of all sizes $|\phi_i|$ such that $\phi_i$ is the formula for objective $\Omega_i$ in case of Boolean B\"uchi \gamesAb{}.

\begin{figure}[t]
	\centering
		\resizebox{\textwidth}{!}{%
		\begin{tikzpicture}
		
		\node[draw, diamond, minimum size=0.7cm] (m0) at (10,0.75){$m_0$};
		\node[draw, diamond, minimum size=0.7cm] (m1) at (12.75,0.75){$m_1$};
		
		\draw[-stealth, auto] (m0) edge [in=30,out=60,loop] node[xshift=3pt, yshift=-12pt] { $V\setminus \{v_3\}$} (m0);
		\draw[-stealth, auto] (m0) edge [below] node { $v_3/v_5$} (m1);
		\draw[-stealth, auto] (m1) edge [in=30,out=60,loop] node[xshift=3pt, yshift=-12pt] { $V\setminus \{v_3\}$} (m1);
		\draw[-stealth, auto] (m1) edge [in=300,out=330,loop] node[xshift=3pt, yshift=12pt] { $v_3/v_7$} (m1);
		
		\node[draw, diamond] (n0) at (10,-1.25){$m_0$};
		\node[draw, diamond] (n1) at (12.75,-1.25){$m_1$};
		
		\draw[-stealth, auto] (n0) edge [in=30,out=60,loop] node[xshift=3pt, yshift=-12pt] { $V\setminus \{v_3\}$} (n0);
		\draw[-stealth, auto] (n0) edge [below] node { $v_3/\{v_5\}$} (n1);
		\draw[-stealth, auto] (n1) edge [in=30,out=60,loop] node[xshift=3pt, yshift=-12pt] { $V\setminus \{v_3\}$} (n1);
		\draw[-stealth, auto] (n1) edge [in=300,out=330,loop] node[xshift=3pt, yshift=12pt] { $v_3/\{v_5,v_7\}$} (n1);
		
		\node[draw, diamond, minimum size=0.8cm] (o0) at (8.625,-3.25){$m_0$};
		\node[draw, diamond, minimum size=0.8cm] (o1) at (11.375,-3.25){$m_1$};
		\node[draw, diamond, minimum size=0.8cm] (o2) at (14.125,-3.25){$m_2$};

		\draw[-stealth, auto] (o0) edge [in=30,out=60,loop] node[xshift=3pt, yshift=-12pt] { $V\setminus \{v_3\}$} (o0);
		\draw[-stealth, auto] (o0) edge [below] node { $v_3/\{v_5\}$} (o1);
		\draw[-stealth, auto] (o1) edge [in=30,out=60,loop] node[xshift=3pt, yshift=-12pt] { $V\setminus \{v_3\}$} (o1);
		\draw[-stealth, auto] (o1) edge [below] node { $v_3/\{v_5, v_7\}$} (o2);
		\draw[-stealth, auto] (o2) edge [in=30,out=60,loop] node[xshift=3pt, yshift=-12pt] { $V\setminus \{v_3\}$} (o2);
		\draw[-stealth, auto] (o2) edge [in=300,out=330,loop] node[xshift=3pt, yshift=12pt] { $v_3/\{v_7\}$} (o2);
		
		\node[draw, rectangle, minimum size=0.8cm, inner sep = 0.5pt] (v0) at (-0.75,-0.58){$v_0$};
		\node[draw, rectangle, minimum size=0.8cm, inner sep = 0.5pt] (v1) at (1,0.75){$v_1$};
		\node[draw, rectangle, minimum size=0.8cm, inner sep = 0.5pt] (x) at (1,-1.91){$v_2$};
		\node[draw, circle, minimum size=0.8cm, inner sep = 0.5pt] (v2) at (2.75,-0.58){$v_3$};
		\node[draw, rectangle, minimum size=0.8cm, inner sep = 0.5pt] (v3) at (2.75,-3.25){$v_4$};
		\node[draw, rectangle, minimum size=0.8cm, inner sep = 0.5pt] (v4) at (4.25,0.75){$v_5$};			
		\node[draw, rectangle, minimum size=0.8cm, inner sep = 0.5pt] (v5) at (4.25,-1.91){$v_7$};
		\node[draw, rectangle, minimum size=0.8cm, inner sep = 0.5pt] (v6) at (5.75,0.75){$v_6$};
				
		\draw[-stealth, shorten >=1pt,auto] (v0) to [] (v1);
		\draw[-stealth, shorten >=1pt,auto] (v0) edge [] node {} (x);
		\draw[-stealth, shorten >=1pt,auto] (x) edge [] node {} (v2);
		\draw[-stealth, shorten >=1pt,auto] (x) edge [] node {} (v3);

	    \draw[-stealth, shorten >=1pt,auto] (v1) edge [loop right] node {$(0, (0,0,1))$} (v1);

	    \draw[-stealth, shorten >=1pt,auto] (v3) edge [loop right] node {$( 0, (1,0,0))$} (v3);

		\draw[-stealth, shorten >=1pt,auto] (v2) edge [] node {} (v4);
		\draw[-stealth, shorten >=1pt,auto] (v4) edge [bend left] node {} (v2);

		\draw[-stealth, shorten >=1pt,auto] (v2) edge [] node {} (v5);
		
		\draw[-stealth, shorten >=1pt,auto] (v5) edge [loop right] node  {$(1, (1,1,0))$} (v5);

		\draw[-stealth, shorten >=1pt,auto] (v4) edge [] node {} (v6);
		
		\draw[-stealth, shorten >=1pt,auto] (v6) edge [loop right] node {$(1, (0,1,1))$} (v6);

		\end{tikzpicture}
		}%
	
	\caption{A parity \gameAb{} (left), one deterministic Moore machine $\mathcal{M}_t$ and two nondeterministic Moore machines $\mathcal{M}_c$ and $\mathcal{M}_b$ (respectively top, center, and bottom right).}
	\label{example_memory}
\end{figure}

\begin{example} \label{ex:instance}
Consider the parity \gameAb{} $\mathcal{G}$ with arena $G$ depicted in \autoref{example_memory} (left) in which Player~$1$ has $\nbrObjectives = 3$ objectives~\cite{DBLP:conf/concur/BruyereRT21}. The vertices of Player~$0$ (resp.\!\ Player~$1$) are depicted as circles (resp.\!\ squares)\footnote{This convention is used throughout this paper.}.
We do not explicitly define the parity objective $\Omega_0$ of Player~$0$ nor the three parity objectives of Player~$1$. Instead, the extended payoff of plays reaching vertices from which they can only loop is displayed in the arena next to those vertices, and we set the extended payoff of play $v_0 v_2 (v_3v_5)^\omega$ to $(0, (0,1,0))$.

Consider the memoryless strategy $\sigma_0$ of Player~$0$ such that he chooses to always move to $v_5$ from $v_3$. The set of payoffs of plays consistent with $\sigma_0$ is $\{(0,0,1), (0,1,0), (1, 0, 0), (0, 1, 1)\}$ and the set of those that are Pareto-optimal is $\paretoSet{\sigma_0} = \{(1, 0, 0), (0, 1, 1)\}$. Notice that play $\rho = v_0 v_2 v_4^\omega$ is consistent with $\sigma_0$, has payoff $(1, 0, 0) \in \paretoSet{\sigma_0}$ and is lost by Player~$0$. Together with $\mathcal{G}$, strategy $\sigma_0$ is therefore a negative instance of the \problemVerifAb{}. 

Let us now consider the finite-memory strategy $\sigma'_0$ such that $\sigma'_0(v_0 v_2 v_3) = v_5$ and $\sigma'_0(v_0 v_2 v_3 v_5 v_3) = v_7$. Contrarily to the previous strategy, $\mathcal{G}$ and $\sigma'_0$ constitute a positive instance of the  \problemVerifAb{}. Indeed, the set of Pareto-optimal payoffs is $\paretoSet{\sigma'_0} = \{(0, 1, 1),(1, 1, 0)\}$ and Player~$0$ wins every play consistent with $\sigma'_0$ whose payoff is in this set. A deterministic Moore machine $\mathcal{M}_t$ for $\sigma'_0$ is depicted in \autoref{example_memory} (top right). It has two memory states with state $m_1$ indicating that $v_3$ has been visited. Each edge from $m$ to $m'$ is labeled by $v/v'$ with an optional $v'$ such that $\alpha_U(m,v) = m'$ and $\alpha_N(m,v) = v'$ if $v \in V_0$.

Finally, we provide two nondeterministic Moore machines in \autoref{example_memory} (center right and bottom right). Each edge from $m$ to $m'$ is now labeled by $v/T$ such that $\alpha_N(m,v) = T \subseteq V$ when $v \in V_0$. Let us show that the \gameAb{} $\mathcal{G}$ with machine $\mathcal{M}_c$ (resp.\! machine $\mathcal{M}_b$) is a negative (resp.\! positive) instance of the \problemUVerifAb{}.

One can check that the memoryless strategy $\sigma_0$ mentioned above (always move to $v_5$ from $v_3$) belongs to the set $\llbracket \mathcal{M}_c \rrbracket$. It follows that $\mathcal{G}$ and $\mathcal{M}_c$ are a negative instance of the \problemUVerifAb{}. Notice that all the other strategies $\sigma_0^k$, $k \geq 1$, of $\llbracket \mathcal{M}_c \rrbracket$ are such that $\sigma_0^k(hv_3) = v_5$ except when $h = v_0v_2(v_3v_5)^k$ in which case $\sigma_0^k(hv_3) = v_7$ (the strategy allows to cycle between $v_3$ and $v_5$ $k$ times before dictating that $v_7$ be visited).

The machine $\mathcal{M}_b$ has three memory states such that $m_1$ (resp.\! $m_2$) records one visit (resp.\! at least two visits) to $v_3$. The set $\llbracket \mathcal{M}_b \rrbracket$ contains exactly two strategies: one is the finite-memory strategy $\sigma'_0$ given before and the other one is the strategy $\sigma''_0$ such that $\sigma''_0(v_0 v_2 v_3) = \sigma''_0(v_0 v_2 v_3 v_5 v_3) = v_5$ and $\sigma''_0(v_0 v_2 v_3 (v_5 v_3)^2) = v_7$. One can verify that $\mathcal{G}$ and $\mathcal{M}_b$ are a positive instance of the \problemUVerifAb{}. 
\lipicsEnd
\end{example}

\begin{remark} \label{rem:product}
In the sequel, we often consider the Cartesian product $G \times \mathcal{M}$ with initial vertex $(v_0, m_0)$ of the arena $G$ of $\mathcal{G}$ with the (nondeterministic) Moore machine $\mathcal{M}$ for Player~$0$. When $\mathcal{M}$ is nondeterministic, this finite graph $G \times \mathcal{M}$ is a two-player game arena (as the vertices of Player~$0$ may have several successors). The strategies $\sigma'_0$ for Player~$0$ in this product correspond exactly to the strategies $\sigma_0 \in \llbracket \mathcal{M}\rrbracket$. With this in mind, we can reformulate the \problemUVerifAb{} to take a game arena as input. Given $G' = G \times \mathcal{M}$, the \problemUVerifAb{} is to decide whether for all strategies $\sigma'_0$ of Player~$0$ in $G'$, every play $\rho \in \Plays_{\sigma'_0}$ such that $\payoff{\rho} \in \paretoSet{\sigma'_0}$ satisfies the objective of Player~$0$. When $\mathcal{M}$ is deterministic, this product is a finite graph whose infinite paths, starting from the initial vertex, are exactly the plays consistent with the single strategy $\sigma_0 \in \llbracket \mathcal{M}\rrbracket$. This graph can be seen as a single-player game arena $G'$ (as every vertex of Player~$0$ only has a single successor). In that setting, given a single-player arena $G' = G \times \mathcal{M}$, the \problemVerifAb{} is to decide whether every play $\rho \in \Plays_{G'}$ such that $\payoff{\rho} \in \max \{\payoff{\rho} \mid \rho \in \Plays_{G'} \}$ satisfies the objective of Player~$0$.
\end{remark}

\subparagraph{Payoff Realizability and Lassoes.}
In order to study the (U)PRV problem, we need to perform specific checks on payoffs as described in the next proposition.

\begin{proposition}
\label{prop:parity_payoff_existence}
    Let $\mathcal G = (G, \Obj_1, \ldots, \Obj_\nbrObjectives)$ be an \gameAb{} and let $p$ (resp.\! $(w,p)$) be a payoff (resp.\! extended payoff). The existence of a play $\rho$ realizing payoff $p$ (resp.\! extended payoff $(w,p)$) can be decided with the following complexities.
    \begin{itemize}
        \item For parity objectives: in time polynomial in $|G|$, $\nbrObjectives$, and $\max d_i$.
        \item For Boolean B\"uchi objectives: in time polynomial in $|G|$, and exponential in $\nbrObjectives$ and $\max |\phi_i|$.
    \end{itemize}
    Checking whether a realizable payoff $p$ is Pareto-optimal is decided with the same complexities.
\end{proposition}
\begin{proof}
We start with the case of parity objectives, and then explain how to modify the approach for Boolean B\"uchi objectives. Let $\mathcal{G}$ be a parity \gameAb{} and $p$ be a payoff in $\{0,1\}^\nbrObjectives$. Deciding the existence of a play $\rho \in \Plays$ with $\payoff{\rho} = p$ can be performed as follows. We want to decide the existence of a play in $G$ which satisfies the intersection of parity objectives 
\begin{equation} \label{eq:realizable}
    \Obj' = \bigcap_{i = 1}^\nbrObjectives \Omega'_i \mbox{~~where $\Omega'_i = \Omega_i$ if $p_i = 1$ and  $\Omega'_i = \overline{\Omega}_i$ otherwise.}
\end{equation}
 That is, to decide the existence of a play that satisfies the objectives which are satisfied in the payoff $p$ and that does not satisfy those which are not. To do so, we use the following results. We recall that given $q$ pairs of sets $(E_1, F_1), \dots, (E_q, F_q)$ such that $E_i, F_i \subseteq V$ with $i \in \{1, \dots, q\}$, the Streett objective asks that for every pair $(E_i, F_i)$ if $F_i$ is visited infinitely often then $E_i$ is also visited infinitely often.
\begin{itemize}
    \item Checking for the existence of a play which satisfies a Streett objective in an arena $G$ has the same complexity as the \emph{emptiness check problem for a Streett automaton}\footnote{The arena is considered as an automaton as in this case the partition of the vertices between the two players does not matter.}. The latter check can be solved in polynomial time $\mathcal{O}((|G|^2+ b) \cdot \min (|G|, q))$ with $q$ the number of Streett pairs $(E_i,F_i)$ and $b = \sum^q_{i = 1} |E_i| + |F_i|$~\cite{DBLP:conf/swat/HenzingerT96,DBLP:journals/fuin/LatvalaH00}. Notice that $b \leq 2q \cdot |G|$.
    \item The conjunction of $\nbrObjectives$ parity objectives can be expressed as a Streett objective with $q = \sum_{i = 1}^\nbrObjectives d_i/2$ pairs, with $d_i$ the maximal priority for objective $\Omega_i$~\cite{DBLP:conf/fossacs/ChatterjeeHP07}. 
\end{itemize}
Therefore, it follows that we can encode the intersection of parity objectives $\Obj'$ corresponding to payoff $p$ into a Streett objective and decide the existence of a play which satisfies this objective $\Obj'$. This is done in $\mathcal{O}((|G|^2 + 2q \cdot |G|) \cdot \min (|G|, q))$ with $q = \sum_{i = 1}^\nbrObjectives d_i/2$. Parameter $q$ is polynomial in the number $\nbrObjectives$ of objectives of Player~$1$ and the maximum priority $d = \max d_i$.

\smallskip Suppose now that $\mathcal{G}$ is a Boolean B\"uchi \gameAb{}, meaning that the objective $\Obj'$ in (\ref{eq:realizable}) is now an intersection of Boolean B\"uchi objectives. This objective $\Obj'$ is defined by a formula $\phi'$ whose size is in $\mathcal{O}(t \cdot (\max |\phi_i|))$. It is proved in~\cite{DBLP:conf/atva/BaierBD00S19} that the \emph{emptiness check problem for a Boolean B\"uchi automaton} can be solved in time polynomial in the size of the automaton and exponential in the size of the formula defining the Boolean B\"uchi condition. This leads to the complexity announced in \autoref{prop:parity_payoff_existence}. 

\smallskip Finally, deciding the existence of an extended payoff $(w,p)$ is done similarly by adding objective $\Omega_0$ to the conjunction in (\ref{eq:realizable}) if $w = 1$ and $\overline{\Omega}_0$ if $w = 0$. We conclude by discussing below how the Pareto-optimality of a realizable payoff can be checked.
\begin{romanenumerate}
    \item \label{rem:improve-item1} Deciding the existence of a play $\rho$ with $\payoff{\rho} \geq p$ can be performed as described above. The only difference is that we now consider the following intersection of parity objectives instead of (\ref{eq:realizable}): $\Obj' = \bigcap_{p_i = 1} \Omega_i$. We therefore have the same complexity as
    announced in \autoref{prop:parity_payoff_existence} for both parity \gamesAb{} and Boolean B\"uchi \gamesAb{}.

    \item \label{rem:improve-item2} When a payoff $p$ is realizable, we also need to test whether it is Pareto-optimal. This can be done as follows. We consider the set of payoffs $\{(p_1, \dots, p_{i-1}, 1, p_{i+1}, \dots, p_\nbrObjectives) \mid i \in \{1, \dots, \nbrObjectives\} \textnormal{ such that }p_i = 0\}$, that is all possible payoffs which satisfy exactly one more objective than $p$. The maximal number of such payoffs $p'$ is $\nbrObjectives$, and for each $p'$ we check the existence of a play with a payoff larger than or equal to $p'$ as described in (\ref{rem:improve-item1}). Overall we obtain a complexity for checking whether a payoff is Pareto-optimal that is polynomial in $|G|$, $\nbrObjectives$, and $\max d_i$ for parity \gamesAb{}, and polynomial in $|G|$ and exponential in $\nbrObjectives$ and $\max |\phi_i|$ for Boolean B\"uchi \gamesAb{}. 
\end{romanenumerate}
\end{proof}

We also need the next property which shows that when a play satisfies a parity or a Boolean B\"uchi objective, there exists another such play that is a lasso of polynomial size.
\begin{lemma}{\cite{BouyerBMU15}} \label{lem:patricia}
For any play~$\rho \in \Plays$, there exists a lasso $\rho' = gh^\omega$ such that $\rho$ and $\rho'$ start with the same vertex, $\occ{\rho} = \occ{\rho'}$, $\infOcc{\rho} = \infOcc{\rho'}$, and $|gh|$ is quadratic in $|G|$.
\end{lemma}

\subparagraph{Related Synthesis Problem.} Our verification problem is related to the \emph{\problemSynt{}} introduced in~\cite{DBLP:conf/concur/BruyereRT21}. This synthesis problem asks, given a two-player \gameAb{}, whether there exists a strategy $\sigma_0$ for Player~$0$ such that every play in $\Playsigmazero$ with a Pareto-optimal payoff satisfies the objective of Player~$0$. This problem is solved in~\cite{DBLP:conf/concur/BruyereRT21} for parity and reachability objectives. It is shown that the problem is \nexptime{}-complete, and that finite-memory strategies are sufficient for Player~$0$ to have a solution $\sigma_0$ to the problem.

\section{Complexity Class of the \problemVerifAb{}}
\label{sec:complexity_class_PRV}

In this section, we provide the complexity class of the \problemVerifAb{} for both parity \gamesAb{} and Boolean B\"uchi \gamesAb{}. The complexity class of the \problemUVerifAb{} is studied in \autoref{sec:complexity_class_UPRV}. In this whole section, we assume that an instance of the \problemVerifAb{} is an \gameAb{} with a \emph{single-player} game arena (see \autoref{rem:product}). This is not problematic with respect to the algorithmic complexities since the size of the single-player game arena is $|G|\cdot|M|$. 

\subsection{Parity Objectives}

We begin by studying the complexity class of the \problemVerifAb{} for parity \gamesAb{}.

\begin{theorem} \label{thm:PRV-parity-complexity}
    The \problemVerifAb{} is \conpComplete{} for parity \gamesAb{}.
\end{theorem}

\subparagraph{Membership to \conp{}.}
The \conp{}-membership stated in \autoref{thm:PRV-parity-complexity} is easily proved by showing that the complement of the \problemVerifAb{} is in \np{}. Given a single-player \gameAb{} $\mathcal G$, we guess a payoff $p \in \{0,1\}^\nbrObjectives$, and we check \emph{(i)} whether $p$ is realizable and Pareto-optimal, and \emph{(ii)} whether there exists a play $\rho$ with payoff $p$ which is lost by Player~$0$. In the case of parity objectives, those two checks can be performed in polynomial time by \autoref{prop:parity_payoff_existence}.

\subparagraph{} The proof of \conp{}-hardness is more involved. In order to show this result, we provide a reduction from the \coSAT{} problem to the \problemVerifAb{}.

\subparagraph{The \coSAT{} Problem.} 
We consider a formula $\psi = D_1 \land \dots \land D_r$ in 3-Conjunctive Normal Form (3CNF) consisting of $r$ clauses, each containing exactly 3 literals over the set of variables $X = \{x_1, \dots, x_m\}$. We assume that each variable $x$ occurs as a literal $\ell \in \{x, \neg x\}$ in at least one clause of $\psi$. The satisfiability problem, called \SAT{}, is to decide whether there exists a valuation of the variables in $X$ such that the formula $\psi$ evaluates to true. This problem is well-known to be \npComplete{} \cite{Cook71,Karp72}. We can consider the complement of this problem, which is to decide for such a formula $\psi$ whether \emph{all valuations} of the variables in $X$ falsify the formula i.e., make at least one of the clauses evaluate to false. This problem, called \coSAT{}, being the complement of an \npComplete{} problem, is \conpComplete{} \cite{DBLP:books/daglib/0072413}.

\subparagraph{Intuition of the Reduction.} 
Given an instance of \coSAT{}, we create a parity \gameAb{} $\mathcal{G}$ with a single-player game arena $G$ consisting of two sub-arenas $G_1$ and $G_2$ reachable from the initial vertex $v_0$ as depicted in \autoref{fig:reduction_parity}.
The intuition behind this construction is the following. A play in the arena starts in $v_0$ and will either enter $G_1$ through $v_1$ and stay in that sub-arena forever or enter $G_2$ through $v_2$, visit some vertex $s_i$ with $i \in \{1,\ldots,r\}$, and stay forever in the corresponding sub-arena $S_i$. The objectives are devised such that a payoff contains one objective per literal of $X$ and one objective per literal, per clause of $\psi$. A play in $G_1$ has a payoff corresponding to a valuation of $X$ and the literals in the clauses of $\psi$ satisfied by that valuation. In addition, the objective of Player~$0$ is not satisfied in those plays. Therefore, it must be the case that the payoffs of plays in $G_1$ are not Pareto-optimal in order for the instance of the \problemVerifAb{} to be positive. This is only the case when the instance of \coSAT{} is also positive due to the fact that plays in $G_2$, which all satisfy the objective of Player~$0$, then have payoffs strictly larger than that of plays in $G_1$. This is not the case if some play in $G_1$ corresponds to a valuation of $X$ which satisfies $\psi$.

\begin{figure}[t]
    \centering
	\resizebox{0.85\textwidth}{!}{%
    \begin{tikzpicture}
    
		\draw[loosely dashed,  rounded corners] (0.25,10.75) rectangle (3.5,3.25) {};
        \node[] at (3.5 - 0.4,10.75 - 0.4) {$G_1$};
		
		\node[draw, rectangle, minimum size=0.75cm, inner sep = 0.5pt] (v1) at (2,10){$v_1$};
		\node[draw, rectangle, minimum size=0.75cm, inner sep = 0.5pt] (x1) at (1,9){$x_1$};
		\node[draw, rectangle, minimum size=0.75cm, inner sep = 0.5pt] (nx1) at (3,9){$\neg x_1$};
		\node[draw, rectangle, minimum size=0.75cm, inner sep = 0.5pt] (mid1) at (2,8){};
		\node[minimum size=0.75cm, inner sep = 0.5pt] (dots1l) at (1,7){$\dots$};
		\node[minimum size=0.75cm, inner sep = 0.5pt] (dots1r) at (3,7){$\dots$};
		\node[draw, rectangle, minimum size=0.75cm, inner sep = 0.5pt] (midm) at (2,6){};
		\node[draw, rectangle, minimum size=0.75cm, inner sep = 0.5pt] (xm) at (1,5){$x_m$};
		\node[draw, rectangle, minimum size=0.75cm, inner sep = 0.5pt] (nxm) at (3,5){$\neg x_m$};
		\node[draw, rectangle, minimum size=0.75cm, inner sep = 0.5pt] (end1) at (2,4){};
		
	    \draw[-stealth, shorten >=1pt, auto] (v1) to [] node []{} (x1);
	    \draw[-stealth, shorten >=1pt, auto] (v1) to [] node []{} (nx1);
	    \draw[-stealth, shorten >=1pt, auto] (x1) to [] node []{} (mid1);
	    \draw[-stealth, shorten >=1pt, auto] (nx1) to [] node []{} (mid1);
	    \draw[-stealth, shorten >=1pt, auto] (mid1) to [] node []{} (dots1l);
	    \draw[-stealth, shorten >=1pt, auto] (mid1) to [] node []{} (dots1r);
	    \draw[-stealth, shorten >=1pt, auto] (dots1r) to [] node []{} (midm);
	    \draw[-stealth, shorten >=1pt, auto] (dots1l) to [] node []{} (midm);
	    \draw[-stealth, shorten >=1pt, auto] (midm) to [] node []{} (xm);
	    \draw[-stealth, shorten >=1pt, auto] (midm) to [] node []{} (nxm);
	    \draw[-stealth, shorten >=1pt, auto] (xm) to [] node []{} (end1);
	    \draw[-stealth, shorten >=1pt, auto] (nxm) to [] node []{} (end1);

		\draw[] (end1) to [] (0.45, 4);
		\draw[] (0.45, 4) to [] (0.45, 10);
	    \draw[-stealth, shorten >=1pt, auto] (0.45, 10) to [] node []{} (v1);
	    
		\draw[rounded corners, fill=gray, opacity=0.15] (4.25,10.75) rectangle (7.5,3.5) {};
        \node[] at (7.5 - 0.4,10.75 - 0.4) {$S_1$};
        
        \node[draw, rectangle, minimum size=0.75cm, inner sep = 0.5pt] (v1s1) at (6,10){$s_1$};
		\node[draw, rectangle, minimum size=0.75cm, inner sep = 0.5pt] (x1s1) at (5,9){$x_1$};
		\node[draw, rectangle, minimum size=0.75cm, inner sep = 0.5pt] (nx1s1) at (7,9){$\neg x_1$};
		\node[draw, rectangle, minimum size=0.75cm, inner sep = 0.5pt] (mid1s1) at (6,8){};
		\node[minimum size=0.75cm, inner sep = 0.5pt] (dots1ls1) at (5,7){$\dots$};
		\node[minimum size=0.75cm, inner sep = 0.5pt] (dots1rs1) at (7,7){$\dots$};
		\node[draw, rectangle, minimum size=0.75cm, inner sep = 0.5pt] (midms1) at (6,6){};
		\node[draw, rectangle, minimum size=0.75cm, inner sep = 0.5pt] (xms1) at (5,5){$x_m$};
		\node[draw, rectangle, minimum size=0.75cm, inner sep = 0.5pt] (nxms1) at (7,5){$\neg x_m$};
		\node[draw, rectangle, minimum size=0.75cm, inner sep = 0.5pt] (end1s1) at (6,4){};
		
	    \draw[-stealth, shorten >=1pt, auto] (v1s1) to [] node []{} (x1s1);
	    \draw[-stealth, shorten >=1pt, auto] (v1s1) to [] node []{} (nx1s1);
	    \draw[-stealth, shorten >=1pt, auto] (x1s1) to [] node []{} (mid1s1);
	    \draw[-stealth, shorten >=1pt, auto] (nx1s1) to [] node []{} (mid1s1);
	    \draw[-stealth, shorten >=1pt, auto] (mid1s1) to [] node []{} (dots1ls1);
	    \draw[-stealth, shorten >=1pt, auto] (mid1s1) to [] node []{} (dots1rs1);
	    \draw[-stealth, shorten >=1pt, auto] (dots1rs1) to [] node []{} (midms1);
	    \draw[-stealth, shorten >=1pt, auto] (dots1ls1) to [] node []{} (midms1);
	    \draw[-stealth, shorten >=1pt, auto] (midms1) to [] node []{} (xms1);
	    \draw[-stealth, shorten >=1pt, auto] (midms1) to [] node []{} (nxms1);
	    \draw[-stealth, shorten >=1pt, auto] (xms1) to [] node []{} (end1s1);
	    \draw[-stealth, shorten >=1pt, auto] (nxms1) to [] node []{} (end1s1);

		\draw[] (end1s1) to [] (4.45, 4);
		\draw[] (4.45, 4) to [] (4.45, 10);
	    \draw[-stealth, shorten >=1pt, auto] (4.45, 10) to [] node []{} (v1s1);
	    
		\draw[rounded corners, fill=gray, opacity=0.15] (9.5,10.75) rectangle (12.75,3.5) {};
        \node[] at (9.5 + 0.4,10.75 - 0.4) {$S_r$};
        
        \node[draw, rectangle, minimum size=0.75cm, inner sep = 0.5pt] (v1sp) at (11,10){$s_r$};
		\node[draw, rectangle, minimum size=0.75cm, inner sep = 0.5pt] (x1sp) at (10,9){$x_1$};
		\node[draw, rectangle, minimum size=0.75cm, inner sep = 0.5pt] (nx1sp) at (12,9){$\neg x_1$};
		\node[draw, rectangle, minimum size=0.75cm, inner sep = 0.5pt] (mid1sp) at (11,8){};
		\node[minimum size=0.75cm, inner sep = 0.5pt] (dots1lsp) at (10,7){$\dots$};
		\node[minimum size=0.75cm, inner sep = 0.5pt] (dots1rsp) at (12,7){$\dots$};
		\node[draw, rectangle, minimum size=0.75cm, inner sep = 0.5pt] (midmsp) at (11,6){};
		\node[draw, rectangle, minimum size=0.75cm, inner sep = 0.5pt] (xmsp) at (10,5){$x_m$};
		\node[draw, rectangle, minimum size=0.75cm, inner sep = 0.5pt] (nxmsp) at (12,5){$\neg x_m$};
		\node[draw, rectangle, minimum size=0.75cm, inner sep = 0.5pt] (end1sp) at (11,4){};
		
	    \draw[-stealth, shorten >=1pt, auto] (v1sp) to [] node []{} (x1sp);
	    \draw[-stealth, shorten >=1pt, auto] (v1sp) to [] node []{} (nx1sp);
	    \draw[-stealth, shorten >=1pt, auto] (x1sp) to [] node []{} (mid1sp);
	    \draw[-stealth, shorten >=1pt, auto] (nx1sp) to [] node []{} (mid1sp);
	    \draw[-stealth, shorten >=1pt, auto] (mid1sp) to [] node []{} (dots1lsp);
	    \draw[-stealth, shorten >=1pt, auto] (mid1sp) to [] node []{} (dots1rsp);
	    \draw[-stealth, shorten >=1pt, auto] (dots1rsp) to [] node []{} (midmsp);
	    \draw[-stealth, shorten >=1pt, auto] (dots1lsp) to [] node []{} (midmsp);
	    \draw[-stealth, shorten >=1pt, auto] (midmsp) to [] node []{} (xmsp);
	    \draw[-stealth, shorten >=1pt, auto] (midmsp) to [] node []{} (nxmsp);
	    \draw[-stealth, shorten >=1pt, auto] (xmsp) to [] node []{} (end1sp);
	    \draw[-stealth, shorten >=1pt, auto] (nxmsp) to [] node []{} (end1sp);

		\draw[] (end1sp) to [] (12.55, 4);
		\draw[] (12.55, 4) to [] (12.55, 10);
	    \draw[-stealth, shorten >=1pt, auto] (12.55, 10) to [] node []{} (v1sp);
	    
	    \draw[loosely dashed, rounded corners] (4,12.35) rectangle (13,3.25) {};
        \node[] at (13 - 0.4,12.35 - 0.4) {$G_2$};
        
	    \node[draw, rectangle, minimum size=0.75cm, inner sep = 0.5pt] (v2) at (8.5,11.75){$v_2$};
	    \node[minimum size=0.75cm, inner sep = 0.5pt] (dots) at (8.5, 10){$\vdots$};
	    \node[draw, rectangle, minimum size=0.75cm, inner sep = 0.5pt] (v0) at (2,11.75){$v_0$};
	    \draw[-stealth, shorten >=1pt, auto] (v0) to [] (v1);
	    \draw[-stealth, shorten >=1pt, auto] (v0) to [] (v2);
		\draw[-stealth, shorten >=1pt, auto] (v2) to [] (dots);
		\draw[-stealth, shorten >=1pt, auto] (v2.225) to [] (v1s1.45);
		\draw[-stealth, shorten >=1pt, auto] (v2.315) to [] (v1sp.135);

    \end{tikzpicture}
    }%
	\caption{The single-player arena $G$ used in the reduction from \coSAT{} for parity objectives.}
	\label{fig:reduction_parity}
\end{figure}

\subparagraph{Structure of a Payoff.} We now detail the objectives used in the reduction and the corresponding structure of a payoff in $G$. Player~$0$ has a single parity objective $\Omega_0$. Player~$1$ has $1 + 2 \cdot m + 3 \cdot r$ parity objectives (assuming each clause is composed of exactly 3 literals). The payoff of a play in $G$ therefore consists in a vector of $1 + 2 \cdot m + 3 \cdot r$ Booleans for the following objectives:
\[(\Omega_{1}, \Omega_{x_1}, \Omega_{\neg x_1}, \dots, \Omega_{x_m}, \Omega_{\neg x_m}, \Omega_{\ell^{1,1}}, \Omega_{\ell^{1,2}}, \Omega_{\ell^{1,3}}, \dots,\Omega_{\ell^{r,1}}, \Omega_{\ell^{r,2}},\Omega_{\ell^{r,3}}).\]
The objective $\Omega_0$ is equal to objective $\Omega_1 = \parity{c}$ with $c(v) = 2$ if $v\in G_2$ and $c(v) = 1$ otherwise. It is direct to see that these objectives are only satisfied for plays in $G_2$. We define the objective $\Omega_{x} = \parity{c}$ (resp.\! $\Omega_{\neg x} = \parity{c'}$) with $c(x) = 2$ and $c(\neg x) = 1$ (resp.\!  $c'(\neg x) = 2$ and $c'(x) = 1$) for the vertices labelled $x$ and $\neg x$ in $G_1$ and $G_2$, and such that every other vertex has priority 2 according to $c$ (resp.\! $c'$). Objective $\Omega_{x}$ (resp.\! $\neg \Omega_{x}$) is satisfied if and only if vertex $x$ (resp.\! $\neg x$) is visited infinitely often and $\neg x$ (resp.\! $x$) is not. If both $x$ and $\neg x$ are visited infinitely often, neither $\Omega_{x}$ not $\Omega_{\neg x}$ are satisfied. These objectives are used to encode valuations of $X$ into payoffs. The objective $\Omega_{\ell^{i,j}}$ corresponds to the objective for the $j\textsuperscript{th}$ literal of the $i\textsuperscript{th}$ clause of $\psi$, written $\ell^{i,j} \in \{x_k, \neg x_k\}$ for some $k \in \{1, \dots, m \}$, we define the priority function for this objective later for each sub-arena.

\subparagraph{Payoff of Plays Entering Sub-Arena $G_1$.}
We define the priority function $c$ of objective $\Omega_{\ell^{i,j}}$ in $G_1$ such that $c(\ell^{i,j}) = 2$ and $c(\neg \ell^{i,j}) = 1$ for vertices labeled $\ell^{i,j}$ and $\neg \ell^{i,j}$ in $G_1$. Notice that a play in $G_1$ corresponds to repeatedly making the choice of visiting $x_i$ or $\neg x_i$ for $i \in \{1, \dots, m\}$. We call plays which visit both $x_i$ and $\neg x_i$ infinitely often for some $i$ \emph{unstable} plays and those which visit infinitely often either $x_i$ or $\neg x_i$ for each $i$ \emph{stable} plays. We introduce the following lemma on the stability of plays in $G_1$. 
\begin{lemma}
    \label{lem:stability}
	Unstable plays in $G_1$ do not have a Pareto-optimal payoff.
\end{lemma}
\begin{proof}
Let $\rho = v_0 v_1 \dots$ be an unstable play in $G_1$. Let $I \subseteq \{1, \dots, m\}$ be the set of indexes such that for all $i \in I$, both $x_i$ and $\neg x_i$ are visited infinitely often in $\rho$. Let us consider the stable play $\rho'$ in $G_1$ which visits vertex $x_k$ (resp.\! $\neg x_k$) if and only if $x_k$ (resp.\! $\neg x_k$) is visited infinitely often in $\rho$ for $k \not \in I$ and which only visits vertex $x_i$ infinitely often for each $i \in I$. Let us show that $\payoff{\rho} < \payoff{\rho'}$, which implies that $\rho$ does not have a Pareto-optimal payoff. To do so, we proceed per group of objectives. First, $\Omega_1$ is not satisfied in $\rho$ nor in $\rho'$. Second, the payoff of $\rho$ for the objectives $\Omega_{x_1}, \Omega_{\neg x_1}, \dots, \Omega_{x_m},\Omega_{\neg x_m}$ is strictly smaller than that of $\rho'$ as $\Omega_{x_k}$ (resp.\! $\Omega_{\neg x_k}$) is satisfied in $\rho'$ if and only if it is satisfied in $\rho$ for $k \not \in I$ and $\Omega_{x_i}$ is satisfied in $\rho'$ while neither $\Omega_{x_i}$ nor $\Omega_{\neg x_i}$ are satisfied in $\rho$ for $i \in I$. Finally, the payoff of $\rho$ for the objectives $\Omega_{\ell^{1,1}}, \Omega_{\ell^{1,2}}, \Omega_{\ell^{1,3}}, \dots,\Omega_{\ell^{r,1}}, \Omega_{\ell^{r,2}},\Omega_{\ell^{r,3}}$ is smaller than or equal to that of $\rho'$. This is because $\rho'$ satisfies the objectives for the same literals of the variables $x_k$ with $k \not \in I$ as $\rho$ but it may additionally satisfy some objectives for the literals of $x_i$ with $i \in I$ while $\rho$ does not satisfy any. Overall, it follows that $\payoff{\rho} < \payoff{\rho'}$.
\end{proof}

In the sequel, we therefore only consider stable plays $\rho$ in $G_1$. The objective $\Omega_0$ of Player~$0$ and $\Omega_1$ of Player~$1$ are not satisfied in $\rho$ and such a play satisfies either the objective $\ObjPlayer{x_i}$ or $\ObjPlayer{\neg x_i}$ for each $x_i \in X$. The part of the payoff of $\rho$ for these objectives can be seen as a valuation of the variables in $X$, expressed as a vector of $2 \cdot m$ Booleans. The objective $\Omega_{\ell^{i,j}}$ is satisfied in the payoff of $\rho$ if and only if the literal $\ell^{i,j}$ is satisfied by that valuation. That is if either $\ell^{i,j} = x_k$ and $\ObjPlayer{x_k}$ is satisfied or $\ell^{i,j} = \neg x_k$ and $\ObjPlayer{\neg x_k}$ is satisfied, for $x_k \in X$. Given a positive instance of the \coSAT{} problem, it holds that none of the valuations of $X$ satisfy the formula $\psi$. Therefore, since stable plays in $G_1$ encode valuations of $X$ and the corresponding satisfied literals of the clauses of $\psi$, the following lemma holds.
\begin{lemma}
    \label{lem:clause_not_satisfied}
    Given a positive instance of the \coSAT{} problem and any stable play $\rho$ in $G_1$, there exists a clause $D_i$ for $i \in \{1, \dots, r\}$ such that $\Omega_{\ell^{i,j}}$ is not satisfied in $\rho$ for $j\in \{1,2,3\}$.
\end{lemma}
\begin{proof}
Let $\rho$ be a stable play in $G_1$ in the arena corresponding to the instance of the \coSAT{} problem. By definition, since the payoff of $\rho$ corresponds to a valuation of the variables in $X$ and the literals of the clauses of $\psi$ satisfied by that valuation, it follows that there exists a clause $D_i$ for $i \in \{1, \dots, r\}$ such that $\Omega_{\ell^{i,j}}$ is not satisfied in $\payoff{\rho}$ for $j\in \{1,2,3\}$.
\end{proof}

In order for the instance of the \problemVerifAb{} to be positive in case of a positive instance of \coSAT{}, since plays in $G_1$ do not satisfy the objective of Player~$0$, it must be the case that the payoff of these plays are not Pareto-optimal when considering the whole arena $G$. Therefore, given any play in $G_1$, there must exists a play with a strictly larger payoff in $G_2$ which also satisfies the objective of Player~$0$.

\subparagraph{Payoff of Plays Entering Sub-Arena $G_2$.}
We define the priority function $c$ of objective $\Omega_{\ell^{i,j}}$ in $G_2$ such that $c(s_i) = 1$ and $c(v) = 2$ for $v \neq s_i$ in $G_2$. Therefore, any play entering $S_i$ satisfies every objective for the literals of the clauses of $\psi$, except for objectives $\Omega_{\ell^{i,j}}$, $j \in \{1,2,3\}$. After entering a sub-arena $S_j$, plays in $G_2$ can visit infinitely often either or both $x_i$ and $\neg x_i$ for $i \in \{1, \dots, m\}$ and we therefore introduce the following lemma on the stability of plays in $G_2$.
\begin{lemma}
    \label{lem:stability2}
	Unstable plays in $G_2$ do not have a Pareto-optimal payoff.
\end{lemma}
\begin{proof}
Let $\rho = v_0 v_2 s_j \dots$ be an unstable play in $G_2$ entering sub-arena $S_j$. Let $I \subseteq \{1, \dots, m\}$ be the set of indexes such that for all $i \in I$, both $x_i$ and $\neg x_i$ are visited infinitely often in $\rho$. Let us consider the stable play $\rho'$ in $G_2$ which also enters $S_j$, visits vertex $x_k$ (resp.\! $\neg x_k$) if and only if $x_k$ (resp.\! $\neg x_k$) is visited infinitely often in $\rho$ for $k \not \in I$ and only visits vertex $x_i$ infinitely often for $i \in I$. Let us show that $\payoff{\rho} < \payoff{\rho'}$, which implies that $\rho$ does not have a Pareto-optimal payoff. To do so, we proceed per group of objectives. First, $\Omega_1$ is satisfied in both $\rho$ and $\rho'$. Second, the payoff of $\rho$ for the objectives $\Omega_{x_1}, \Omega_{\neg x_1}, \dots, \Omega_{x_m},\Omega_{\neg x_m}$ is strictly smaller than that of $\rho'$ as $\Omega_{x_k}$ (resp.\! $\Omega_{\neg x_k}$) is satisfied in $\rho'$ if and only if it is satisfied in $\rho$ for $k \not \in I$ and $\Omega_{x_i}$ is satisfied in $\rho'$ while neither $\Omega_{x_i}$ nor $\Omega_{\neg x_i}$ are satisfied in $\rho$ for $i \in I$. Finally, the payoff of $\rho$ for the objectives $\Omega_{\ell^{1,1}}, \Omega_{\ell^{1,2}}, \Omega_{\ell^{1,3}}, \dots,\Omega_{\ell^{r,1}}, \Omega_{\ell^{r,2}},\Omega_{\ell^{r,3}}$ is the same as in $\rho'$ as the payoff for these objectives for a play in $G_2$ only depends on the sub-arena $S_j$ entered by that play. Overall, it follows that $\payoff{\rho} < \payoff{\rho'}$.
\end{proof}
We therefore only consider stable plays in $G_2$. Such a play $\rho$ satisfies either the objective $\ObjPlayer{x_i}$ or $\ObjPlayer{\neg x_i}$ for each $x_i \in X$. The objectives corresponding to the literals in the clauses of $\psi$ which are satisfied in $\rho$ only depend on the sub-arena $S_j$ entered by $\rho$. It can easily be shown that every such objective is satisfied by $\rho$ except for $\Omega_{\ell^{j, 1}}, \Omega_{\ell^{j, 2}}$ and $\Omega_{\ell^{j, 3}}$ for clause $D_j$.

\begin{proposition}
    \label{prop:reduction_correctness_parity}
    The instance of the \problemVerifAb{} in $G$ is positive if and only if the corresponding instance of \coSAT{} is positive.
\end{proposition}

\begin{proof}
We start by showing the contrapositive of the first implication in the equivalence. Let us assume that the instance of \coSAT{} is negative and show that the instance of the \problemVerifAb{} is also negative in the corresponding arena $G$. Since the instance of \coSAT{} is negative, there exists a valuation $val: X \rightarrow \{0, 1\}$ of the variables in $X$ such that this valuation satisfies $\psi$. By construction, there exists a stable play $\rho$ in $G_1$ corresponding to this valuation $val$, that is visiting infinitely often $x_i$ if and only if $val(x_i) = 1$ and visiting infinitely often $\neg x_i$ otherwise. It follows that, in the payoff of $\rho$ for the objectives $\Omega_{\ell^{1,1}}, \Omega_{\ell^{1,2}}, \Omega_{\ell^{1,3}}, \dots,\Omega_{\ell^{r,1}}, \Omega_{\ell^{r,2}},\Omega_{\ell^{r,3}}$, at least one objective for some literal of each clause is satisfied (as this valuation satisfies $\psi$ and as the payoff for these objectives correspond to the literals satisfied by that valuation). This play does not satisfy the objective of Player~$0$. Let us show that its payoff is Pareto-optimal and therefore that the \problemVerifAb{} is not satisfied. First, we show that the payoff of $\rho$ is incomparable to that of every play in $G_2$. Let $\rho'$ be a play in $G_2$. By construction, $\rho'$ enters some sub-arena $S_i$ with $i \in \{1, \dots, r\}$ and it follows that none of the objectives $\Omega_{\ell^{i,1}}, \Omega_{\ell^{i,2}}, \Omega_{\ell^{i,3}}$ are satisfied in the payoff of $\rho'$ while some are satisfied in the payoff of $\rho$. It also holds that $\Omega_1$ is not satisfied in the payoff of $\rho$ while it is satisfied in that of $\rho'$. It follows that the payoff of $\rho$ and $\rho'$ are incomparable. In addition, the payoff of $\rho$ is also incomparable to the payoff of any other stable play in $G_1$ because of its valuation of the variables in $X$ and is therefore Pareto-optimal.

Let us now assume that the instance of \coSAT{} is positive and show that there is a solution to the \problemVerifAb{} in the corresponding arena $G$. It suffices to show that for any stable play in $G_1$, there exists a stable play with a strictly larger payoff in $G_2$. Since all plays in $G_2$ satisfy the objective of Player~$0$, it follows that the \problemVerifAb{} is satisfied in $G$ as all plays with a Pareto-optimal payoff will satisfy the objective of Player~$0$. Let $\rho = v_0 \: (v_1 \: z_1 \boxempty \dots \boxempty z_m \boxempty)^\omega$ be a stable play in $G_1$ where $z_i$ is either $x_i$ or $\neg x_i$. Since the instance of \coSAT{} is positive and given \autoref{lem:clause_not_satisfied}, there exists a clause $D_i$ with $i \in \{1, \dots, r\}$ of $\psi$ such that $\Omega_{\ell^{i,j}}$ is not satisfied in $\rho$ for $j\in \{1,2,3\}$. Let us consider the stable play $\rho' = v_0 \: v_2 \: (s_i \: z_1 \boxempty \dots \boxempty z_m \boxempty)^\omega$ in $G_2$. This play satisfies the same objectives in $\Omega_{x_1}, \Omega_{\neg x_1}, \dots, \Omega_{x_m}, \Omega_{\neg x_m}$ as $\rho_1$ as it visits exactly the same valuation of $X$. It also holds that the objectives $\Omega_{\ell^{i,j}}$ for $j\in \{1,2,3\}$ are not satisfied in the payoff of $\rho$ nor in that of $\rho'$. Since the objectives $\Omega_{\ell^{k,j}}$ for $k \neq i$ are all satisfied in $\rho'$, it holds that this part of the payoff of $\rho'$ is equal or larger to that of $\rho$. However, since $\Omega_1$ is satisfied in $\rho'$ and not in $\rho$, it follows that $\payoff{\rho} < \payoff{\rho'}$.
\end{proof}

\subsection{Boolean B\"uchi Objectives}

We now study the complexity class of the \problemVerifAb{} for Boolean B\"uchi \gamesAb{}. We recall that the class \sigmaClass{} in the second level of the polynomial hierarchy is the class \npNp, also equal to the class \npConp{}~\cite{DBLP:books/daglib/0072413}, and that its complement is the class \piClass{}. 

\begin{theorem}
\label{thm:PRV-BB-complexity}
 The \problemVerifAb{} is \piComplete{} for Boolean B\"uchi \gamesAb{.}
\end{theorem}

In order to show \autoref{thm:PRV-BB-complexity}, we consider the complement of the \problemVerifAb{} and show that it is \sigmaComplete{}. Given a single-player game arena $G$ and the Boolean B\"uchi objectives $\Omega_0, \Omega_1, \dots, \Omega_\nbrObjectives$, the complement of the \problemVerifAb{} is to decide whether there exists a play $\rho \in \Plays$ such that $\payoff{\rho}$ is Pareto-optimal and $\won{\rho} = 0$.

\subparagraph{Membership to \sigmaClass{}.} We start by guessing a play $\rho \in \Plays$, which can be done in polynomial time as by \autoref{lem:patricia}, $\rho$ can be guessed in a lasso form $gh^\omega$. Then, we retrieve $\payoff{\rho}$ and $\won{\rho}$ in polynomial time by evaluating which Boolean B\"uchi objectives are satisfied by $\rho$ using $h$ to retrieve their variable valuation. It remains to verify that $p = \payoff{\rho}$ is a Pareto-optimal payoff in $G$. We can devise an \np{} algorithm which checks that $p$ is \emph{not} Pareto-optimal by guessing a play $\rho'$ such that $\payoff{\rho'} > p$, the arguments used are similar to those detailed above. Checking that $p$ is Pareto-optimal can therefore be done using a call to a \conp{} algorithm. Overall, the algorithm works in \np{} with a call to a \conp{} oracle and is therefore in \sigmaClass{}.

\begin{figure}[t]
	\centering
	\resizebox{\textwidth}{!}{%
		\begin{tikzpicture}
		
		\node[draw, rectangle, minimum size=0.8cm] (o) at (-2.5,-0.25){$v_0$};
		
		
		\draw[dashed, rounded corners] (2.9,-2) rectangle (10.1, 1.25)  {};
        \node[] at (10.1 - 0.4, 1.25 - 0.3) {$G_1$};
		 
		\node[draw, rectangle, minimum size=0.8cm] (b1b) at (3.5,-0.25){$g_1$};
		\node[draw, rectangle, minimum size=0.8cm] (x11b) at (4.5,-1.25){$\neg y^1_1$};
		\node[draw, rectangle, minimum size=0.8cm] (nx11b) at (4.5,0.75){$y^1_1$};
		\node[draw, rectangle, minimum size=0.8cm] (b2b) at (5.5,-0.25){};
		\node[minimum size=0.8cm] (b2i1b) at (6.5,-1.25){$\dots$};
		\node[minimum size=0.8cm] (b2i2b) at (6.5,0.75){$\dots$};
		\node[draw, rectangle, minimum size=0.8cm] (bmb) at (7.5,-0.25){};
		\node[draw, rectangle, minimum size=0.72cm, inner sep=0.05pt] (x1mb) at (8.5,-1.25){$\neg y^1_n$};
		\node[draw, rectangle, minimum size=0.8cm] (nx1mb) at (8.5,0.75){$y^1_n$};
		\node[draw, rectangle, minimum size=0.8cm] (end1b) at (9.5,-0.25){};
		
		\draw[-stealth, shorten >=1pt,auto] (b1b) to [] node []{} (x11b);
		\draw[-stealth, shorten >=1pt,auto] (b1b) to [] node []{} (nx11b);
		\draw[-stealth, shorten >=1pt,auto] (x11b) to [] node []{} (b2b);
		\draw[-stealth, shorten >=1pt,auto] (nx11b) to [] node []{} (b2b);
		\draw[-stealth, shorten >=1pt,auto] (b2b) to [] node []{} (b2i1b);
		\draw[-stealth, shorten >=1pt,auto] (b2b) to [] node []{} (b2i2b);
		\draw[-stealth, shorten >=1pt,auto] (b2i1b) to [] node []{} (bmb);
		\draw[-stealth, shorten >=1pt,auto] (b2i2b) to [] node []{} (bmb);
		\draw[-stealth, shorten >=1pt,auto] (bmb) to [] node []{} (x1mb);
		\draw[-stealth, shorten >=1pt,auto] (bmb) to [] node []{} (nx1mb);		
		\draw[-stealth, shorten >=1pt,auto] (x1mb) to [] node []{} (end1b);
		\draw[-stealth, shorten >=1pt,auto] (nx1mb) to [] node []{} (end1b);
		
		\draw[] (end1b) to [] (9.5, -1.8125);
		\draw[] (9.5, -1.8125) to [] (3.5, -1.8125);
		\draw[-stealth,shorten >=1pt,auto] (3.5, -1.8125) to [] (b1b);

		
		\draw[dashed, rounded corners] (-3.1,-5.5) rectangle (10.1, -2.25) {};
		\node[] at (10.1 - 0.4,-2.25 - 0.3) {$G_2$};

	    \node[draw, rectangle, minimum size=0.8cm] (s1) at (-2.5,-3.75){$g_2$};
		\node[draw, rectangle, minimum size=0.8cm] (s2) at (-1.5,-4.75){$\neg y^2_1$};
		\node[draw, rectangle, minimum size=0.8cm] (s3) at (-1.5,-2.75){$y^2_1$};
		\node[draw, rectangle, minimum size=0.8cm] (s4) at (-0.5,-3.75){};
		\node[minimum size=0.8cm] (s5) at (0.5,-4.75){$\dots$};
		\node[minimum size=0.8cm] (s6) at (0.5,-2.75){$\dots$};
		\node[draw, rectangle, minimum size=0.8cm] (s7) at (1.5,-3.75){};
		\node[draw, rectangle, minimum size=0.8cm] (s8) at (2.5,-4.75){$\neg y^2_n$};
		\node[draw, rectangle, minimum size=0.8cm] (s9) at (2.5,-2.75){$y^2_n$};

		\node[draw, rectangle, minimum size=0.8cm] (d1) at (3.5,-3.75){};
		\node[draw, rectangle, minimum size=0.8cm] (x31) at (4.5,-4.75){$\neg x_1$};
		\node[draw, rectangle, minimum size=0.8cm] (nx31) at (4.5,-2.75){$x_1$};
		\node[draw, rectangle, minimum size=0.8cm] (d2) at (5.5,-3.75){};
		\node[minimum size=0.8cm] (d2i1) at (6.5,-4.75){$\dots$};
		\node[minimum size=0.8cm] (d2i2) at (6.5,-2.75){$\dots$};
		\node[draw, rectangle, minimum size=0.8cm] (dm) at (7.5,-3.75){};
		\node[draw, rectangle, minimum size=0.8cm] (x3m) at (8.5,-4.75){$\neg x_m$};
		\node[draw, rectangle, minimum size=0.8cm] (nx3m) at (8.5,-2.75){$x_m$};
		\node[draw, rectangle, minimum size=0.8cm] (eng3) at (9.5,-3.75){};
		
		\draw[-stealth, shorten >=1pt,auto] (s1) to [] node []{} (s2);
		\draw[-stealth, shorten >=1pt,auto] (s1) to [] node []{} (s3);
		\draw[-stealth, shorten >=1pt,auto] (s2) to [] node []{} (s4);
		\draw[-stealth, shorten >=1pt,auto] (s3) to [] node []{} (s4);
		\draw[-stealth, shorten >=1pt,auto] (s4) to [] node []{} (s5);
		\draw[-stealth, shorten >=1pt,auto] (s4) to [] node []{} (s6);
		\draw[-stealth, shorten >=1pt,auto] (s5) to [] node []{} (s7);
		\draw[-stealth, shorten >=1pt,auto] (s6) to [] node []{} (s7);
		\draw[-stealth, shorten >=1pt,auto] (s7) to [] node []{} (s8);
		\draw[-stealth, shorten >=1pt,auto] (s7) to [] node []{} (s9);
		\draw[-stealth, shorten >=1pt,auto] (s8) to [] node []{} (d1);
		\draw[-stealth, shorten >=1pt,auto] (s9) to [] node []{} (d1);
		
		\draw[-stealth, shorten >=1pt,auto] (d1) to [] node []{} (x31);
		\draw[-stealth, shorten >=1pt,auto] (d1) to [] node []{} (nx31);
		\draw[-stealth, shorten >=1pt,auto] (x31) to [] node []{} (d2);
		\draw[-stealth, shorten >=1pt,auto] (nx31) to [] node []{} (d2);
		\draw[-stealth, shorten >=1pt,auto] (d2) to [] node []{} (d2i1);
		\draw[-stealth, shorten >=1pt,auto] (d2) to [] node []{} (d2i2);
		\draw[-stealth, shorten >=1pt,auto] (d2i1) to [] node []{} (dm);
		\draw[-stealth, shorten >=1pt,auto] (d2i2) to [] node []{} (dm);
		\draw[-stealth, shorten >=1pt,auto] (dm) to [] node []{} (x3m);
		\draw[-stealth, shorten >=1pt,auto] (dm) to [] node []{} (nx3m);
		\draw[-stealth, shorten >=1pt,auto] (x3m) to [] node []{} (eng3);
		\draw[-stealth, shorten >=1pt,auto] (nx3m) to [] node []{} (eng3);

		\draw[] (eng3) to [] (9.5,-5.3125);
		\draw[] (9.5,-5.3125) to [] (-2.5,-5.3125);
		\draw[-stealth,shorten >=1pt,auto] (-2.5,-5.3125) to [] (s1);
		
		\draw[-stealth, shorten >=1pt,auto] (o) to [] node []{} (s1);		
		\draw[-stealth, shorten >=1pt,auto] (o) to [] node []{} (b1b);
		
		\end{tikzpicture}
		}%
	\caption{The single-player arena $G$ used in the reduction from \sigmaQBF{} for Boolean B\"uchi objectives.}
	\label{PRV-fig-QBF}
\end{figure}

\medskip
The lower bound is established by reduction from the following \sigmaComplete{} variant of the Quantified Boolean Formula (QBF) problem.

\subparagraph{The \sigmaQBF{} problem.} Let $\exists y_1 \dots \exists y_n \forall x_1 \dots \forall x_m \psi$ be a fully quantified Boolean formula over the set of variables $X \cup Y$ with $X = \{x_1, \dots, x_m\}$ and $Y = \{y_1,\dots, y_n\}$ such that it contains two blocks of alternating quantifiers beginning with $\exists$. The \sigmaQBF{} problem, which is to decide whether such a formula is true, is \sigmaComplete{}~\cite{DBLP:books/daglib/0072413}.

\subparagraph*{Intuition of the Reduction} Given an instance $\exists y_1 \dots \exists y_n \forall x_1 \dots \forall x_m \psi$ of the \sigmaQBF{} problem, we devise an instance of the \problemVerifAb{} for Boolean B\"uchi objectives consisting of an arena composed of two sub-arenas $G_1$ and $G_2$ as depicted in \autoref{PRV-fig-QBF}. The intuition behind this construction is as follows. A payoff contains one objective per literal of $Y$ and one objective for formula $\psi$. The payoff of plays in $G_1$ correspond to every possible valuation $val_Y$ of the existentially quantified variables in $Y$. In addition, the objective of Player 0 is not satisfied in those plays. The payoffs of plays in $G_2$ again correspond to every possible valuation of $Y$, but these plays also visit every possible valuation of the variables in $X$. Plays in $G_2$ all satisfy the objective of Player $0$. If in $G_2$, together a valuation $val_Y$ of $Y$ and $val_X$ of $X$ falsify formula $\psi$, the payoff of the resulting play is strictly larger than that of the play in $G_1$ for $val_Y$. If the instance of the \problemVerifAb{} is positive, it means that whatever the valuation $val_Y$ of $Y$, there exists a valuation $val_X$ of $X$ such that together they falsify $\psi$, as every payoff of $G_1$ is strictly smaller than some payoff of $G_2$. Conversely, if the instance is negative (that is, the complement of the problem is positive), there exists a valuation of $Y$ such that whatever the valuation of $X$, $\psi$ is true and there is therefore a Pareto-optimal payoff in $G_1$ lost by Player~$0$.

\subparagraph{Objectives.} Player~$0$ has a single Boolean B\"uchi objective $\Omega_0$ and Player~$1$ has $2 + 2 \cdot n$ Boolean B\"uchi objectives. The payoff of a play in $G$ therefore consists in a vector of $2 + 2 \cdot n$ Booleans for the following objectives:
$(\Omega_1, \Omega_{y_1}, \Omega_{\neg y_1}, \dots, \Omega_{y_n}, \Omega_{\neg y_n}, \Omega_{\neg \psi})$.
As all the sets $\target$ used in the Boolean B\"uchi objectives for this reduction only contain a single vertex $v$, in the formulas defining the objectives we use $v$ to mean \emph{true if $v$ is visited infinitely often}, and $\overline{v}$ to mean \emph{true if $v$ is visited finitely often}. Let us define the Boolean B\"uchi objectives:
\begin{itemize}
    \item $\Omega_0 = \Omega_1 = g_2 \land \phi_{stable}$,
    \item $\Omega_{y_i} = (y^1_i \lor y^2_i) \land \phi_{stable}$,
    \item $\Omega_{\neg y_i} = (\neg y^1_i \lor \neg y^2_i) \land \phi_{stable}$,
    \item $\Omega_{\neg \psi} = (g_1 \lor B\!B(\neg \psi)) \land \phi_{stable}$.
\end{itemize}
All these objectives contains a conjunction with formula 
\[\phi_{stable} = \Bigg{(}g_1 ~\land \bigwedge_{i \in \{1, \dots, n\}}\big{(}\overline{y}^1_i \lor \overline{\neg y}^1_i\big{)}\Bigg{)} \lor \Bigg{(}g_2 ~\land \bigwedge_{i \in \{1, \dots, n\}}\big{(}\overline{y}^2_i \lor \overline{\neg y}^2_i\big{)} ~\land \bigwedge_{j \in \{1, \dots, m\}}\big{(}\overline{x}_j \lor \overline{\neg x}_j\big{)}\Bigg{)}.\] 
Moreover the objective $\Omega_{\neg \psi}$ contains the formula $B\!B(\neg \psi)$ equal to $\neg \psi$ expressed as a Boolean B\"uchi objective by replacing each variable with the corresponding vertex in $G_2$.

\subparagraph{Stability.} We call \emph{stable} those plays in $G_1$ (resp.\! $G_2$) which visit finitely often either $y^1_i$ or $\neg y^1_i$ for each $i \in \{1, \dots, n\}$ (resp.\! either $y^2_i$ or $\neg y^2_i$ for each $i \in \{1, \dots, n\}$ and either $x_j$ or $\neg x_j$ for each $j \in \{1, \dots, m\}$), and \emph{unstable} those plays which do not. It is easily checked that unstable plays do not have a Pareto-optimal payoff. Indeed, formula $\phi_{stable}$ must be true for any objective to be satisfied, and $\phi_{stable}$ is only true if the play is in $G_1$ and is stable with regard to $y^1_i$ and $\neg y^1_i$ for each $i \in \{1, \dots, n\}$ or the play is in $G_2$ and is stable with regard to $y^2_i$ and $\neg y^2_i$ for each $i \in \{1, \dots, n\}$ as well as to $x_j$ and $\neg x_j$ for each $j \in \{1, \dots, m\}$. Unstable plays therefore do not satisfy any objective and stable plays do (e.g., objective $\Omega_1$ in $G_2$).
\begin{lemma}
    Unstable plays do not have a Pareto-optimal payoff.
\end{lemma}
The vertices visited in a stable play in $G_1$ (resp.\! $G_2$) can be interpreted as a valuation of the variables in $Y$ (resp.\! $Y$ and $X$). For the variables in $Y$, these valuations are encoded in the payoff of those plays using the objectives for the literals of those variables.

\subparagraph{Satisfying $\Omega_{\neg \psi}$.} The objective $\Omega_{\neg \psi}$ is satisfied in every stable play in $G_1$. It is satisfied in a stable play in $G_2$ if and only if the vertices it visits infinitely often correspond to a valuation of $X$ and $Y$ which together \emph{falsify} formula $\psi$.

\begin{proposition}
The instance of the \sigmaQBF{} problem is positive if and only if the corresponding instance of the \problemVerifAb{} is negative.
\end{proposition}
\begin{proof}
Let us assume that the instance of the \sigmaQBF{} problem is positive. Therefore, there exists a valuation $val_Y$ of the variables in $Y$ such that whatever the valuation of the remaining variables in $X$, formula $\psi$ is true. Let $\rho$ be the stable play in $G_1$ which corresponds to that valuation $val_Y$. Play $\rho$ is lost by Player $0$ and its payoff is of the form $(0, val_Y, 1)$ where $val_Y$ is here expressed as a vector of $2 \cdot n$ Booleans for the objectives $\Omega_{y_1}, \Omega_{\neg y_1}, \dots, \Omega_{y_n}, \Omega_{\neg y_n}$. Let us show that no play in $G$ has a payoff strictly larger than that of $\rho$. First, only stable plays in $G_2$ which correspond to the valuation $val_Y$ could potentially have such a payoff (as other stable plays have an incomparable payoff with regard to their satisfied objectives in $\Omega_{y_1}, \Omega_{\neg y_1}, \dots, \Omega_{y_n}, \Omega_{\neg y_n}$). Then, it suffices to consider every stable play $\rho'$ in $G_2$ corresponding to valuation $val_Y$. Such a play $\rho'$ visits $val_Y$ and then some valuation of the variables in $X$. Since $val_Y$ is a solution to the \sigmaQBF{} problem, no play $\rho'$ satisfies objective $\Omega_{\neg \psi}$ (as no valuation of $X$ together with $val_Y$ falsifies $\psi$). It follows that the payoff of each play $\rho'$ is incomparable to that of $\rho$ (as $\rho$ satisfies $\Omega_{\neg \psi}$ and $\rho'$ does not, and as $\rho$ does not satisfy $\Omega_1$ but $\rho'$ does).

Let us now assume that the instance of the \problemVerifAb{} is negative. Then, it holds that there exists some play in $G_1$ with a Pareto-optimal payoff (as only plays in $G_1$ do not satisfy the objective of Player $0$). Let $\rho$ be such a play, it holds that it is stable and corresponds to a valuation $val_Y$ of the variables in $Y$. It also holds that no play in $G_2$ has a strictly larger payoff. In particular, all plays in $G_2$ corresponding to the same valuation of $Y$ therefore must not satisfy objective $\Omega_{\neg \psi}$. It follows that given $val_Y$, for all valuations $val_X$ of $X$, together $val_Y$ and $val_X$ satisfy $\psi$. The instance of the \sigmaQBF{} problem is therefore positive.
\end{proof}

\subsection{A Related Problem}
As we have established in the previous sections, the lower bound for the \problemVerifAb{} is stronger for Boolean B\"uchi objectives than for parity objectives. We can show that this difference in complexity is even more apparent if we consider the following variant of the \emph{complement} of the \problemVerifAb{}, which we call the \emph{\problemPayoffVerifAb{}}, in which we \emph{fix a payoff} for Player~1. Indeed, this variant remains computationally hard as stated in the following theorem.

\begin{theorem}
\label{thm:DP}
Given a single-player Boolean B\"uchi \gameAb{} $\mathcal G$ and a payoff $p$, the \problemPayoffVerifAb{} is to decide whether there exists a play $\rho \in \Plays$ such that $\payoff{\rho} = p$, $\won{\rho} =0$ and $p$ is Pareto-optimal in $G$. This problem is \DPComplete{}.
\end{theorem}

\begin{remark}
This problem is in \p{} for parity \gamesAb{} as $p$ does not need to be guessed anymore in the \conp{} algorithm.
\end{remark}

Let us recall that the \DP{} class\footnote{This class is also called $\mathsf{DP}$ and must not be mistaken with the class \np{} $\cap$ \conp{}.} is the class of problems of the form $L \cap L'$ where $L$ is \np{}-easy and $L'$ is \conp{}-easy, and both have the same set of instances~\cite{DBLP:books/daglib/0072413}.

\subparagraph{Membership to \DP{}.}
Let us show that the \problemPayoffVerifAb{} is in \DP{}. Let $\mathcal G$ be a single-player Boolean B\"uchi \gameAb{} and $p$ be a payoff. We start by guessing a play $\rho$ in lasso form $\rho = hg^\omega$ where $hg$ has a polynomial size (by \autoref{lem:patricia}). We then check in polynomial time that its extended payoff is equal to $(0,p)$. This algorithm executes in nondeterministic polynomial time. Separately, we use another nondeterministic polynomial algorithm to check that $p$ is \emph{not} Pareto-optimal. This algorithm guesses a play $\rho' = h'g'^\omega$ and verifies that $\payoff{\rho'} > p$. In this way we proved that the \problemPayoffVerifAb{} is of the form $L \cap L'$ with $L \in$ \np{} and $L' \in$ \conp{}.

\medskip
To prove the \DP{}-hardness of the \problemPayoffVerifAb{}, we provide a reduction from the \SATcoSAT{} problem.

\subparagraph{\SATcoSAT{} Problem.} Given a pair of 3CNF formulas $(\psi_1,\psi_2)$, the \SATcoSAT{} problem is to decide whether $\psi_1$ is satisfiable and $\psi_2$ is not. We assume that $\psi_1$ and $\psi_2$ use the same set of variables $X = \{x_1,\ldots,x_m\}$ (if this is not the case, we simply add trivially true clauses to pad the formula missing some variables). This problem is known to be \DP{}-complete~\cite{DBLP:books/daglib/0072413}.

\begin{figure}[t]
	\centering
	\resizebox{1\textwidth}{!}{%
		\begin{tikzpicture}
		
		\node[draw, rectangle, minimum size=0.8cm] (o) at (-5.5,-0.25){$v_0$};
		
		\draw[dashed, rounded corners] (-4.6,-2) rectangle (2.6, 1.25)  {};
        \node[] at (2.6 - 0.4, 1.25 - 0.3) {$G_1$};
		 
		\node[draw, rectangle, minimum size=0.8cm] (b1b2) at (-4,-0.25){$g_1$};
		\node[draw, rectangle, minimum size=0.8cm] (x11b2) at (-3,-1.25){$\neg x^1_1$};
		\node[draw, rectangle, minimum size=0.8cm] (nx11b2) at (-3,0.75){$x^1_1$};
		\node[draw, rectangle, minimum size=0.8cm] (b2b2) at (-2,-0.25){};
		\node[minimum size=0.8cm] (b2i1b2) at (-1,-1.25){$\dots$};
		\node[minimum size=0.8cm] (b2i2b2) at (-1,0.75){$\dots$};
		\node[draw, rectangle, minimum size=0.8cm] (bmb2) at (0,-0.25){};
		\node[draw, rectangle, minimum size=0.8cm, inner sep=0.02pt] (x1mb2) at (1,-1.25){$\neg x^1_m$};
		\node[draw, rectangle, minimum size=0.8cm] (nx1mb2) at (1,0.75){$x^1_m$};
		\node[draw, rectangle, minimum size=0.8cm] (end1b2) at (2,-0.25){};
		
		\draw[-stealth, shorten >=1pt,auto] (b1b2) to [] node []{} (x11b2);
		\draw[-stealth, shorten >=1pt,auto] (b1b2) to [] node []{} (nx11b2);
		\draw[-stealth, shorten >=1pt,auto] (x11b2) to [] node []{} (b2b2);
		\draw[-stealth, shorten >=1pt,auto] (nx11b2) to [] node []{} (b2b2);
		\draw[-stealth, shorten >=1pt,auto] (b2b2) to [] node []{} (b2i1b2);
		\draw[-stealth, shorten >=1pt,auto] (b2b2) to [] node []{} (b2i2b2);
		\draw[-stealth, shorten >=1pt,auto] (b2i1b2) to [] node []{} (bmb2);
		\draw[-stealth, shorten >=1pt,auto] (b2i2b2) to [] node []{} (bmb2);
		\draw[-stealth, shorten >=1pt,auto] (bmb2) to [] node []{} (x1mb2);
		\draw[-stealth, shorten >=1pt,auto] (bmb2) to [] node []{} (nx1mb2);		
		\draw[-stealth, shorten >=1pt,auto] (x1mb2) to [] node []{} (end1b2);
		\draw[-stealth, shorten >=1pt,auto] (nx1mb2) to [] node []{} (end1b2);
		
		\draw[] (end1b2) to [] (2, -1.8125);
		\draw[] (2, -1.8125) to [] (-4, -1.8125);
		\draw[-stealth,shorten >=1pt,auto] (-4, -1.8125) to [] (b1b2);
		
		
		\draw[dashed, rounded corners] (2.9,-2) rectangle (10.1, 1.25)  {};
        \node[] at (10.1 - 0.4, 1.25 - 0.3) {$G_2$};
		 
		\node[draw, rectangle, minimum size=0.8cm] (b1b) at (3.5,-0.25){$g_2$};
		\node[draw, rectangle, minimum size=0.8cm] (x11b) at (4.5,-1.25){$\neg x^2_1$};
		\node[draw, rectangle, minimum size=0.8cm] (nx11b) at (4.5,0.75){$x^2_1$};
		\node[draw, rectangle, minimum size=0.8cm] (b2b) at (5.5,-0.25){};
		\node[minimum size=0.8cm] (b2i1b) at (6.5,-1.25){$\dots$};
		\node[minimum size=0.8cm] (b2i2b) at (6.5,0.75){$\dots$};
		\node[draw, rectangle, minimum size=0.8cm] (bmb) at (7.5,-0.25){};
		\node[draw, rectangle,minimum size=0.8cm, inner sep=0.02pt] (x1mb) at (8.5,-1.25){$\neg x^2_m$};
		\node[draw, rectangle, minimum size=0.8cm] (nx1mb) at (8.5,0.75){$x^2_m$};
		\node[draw, rectangle, minimum size=0.8cm] (end1b) at (9.5,-0.25){};
		
		\draw[-stealth, shorten >=1pt,auto] (b1b) to [] node []{} (x11b);
		\draw[-stealth, shorten >=1pt,auto] (b1b) to [] node []{} (nx11b);
		\draw[-stealth, shorten >=1pt,auto] (x11b) to [] node []{} (b2b);
		\draw[-stealth, shorten >=1pt,auto] (nx11b) to [] node []{} (b2b);
		\draw[-stealth, shorten >=1pt,auto] (b2b) to [] node []{} (b2i1b);
		\draw[-stealth, shorten >=1pt,auto] (b2b) to [] node []{} (b2i2b);
		\draw[-stealth, shorten >=1pt,auto] (b2i1b) to [] node []{} (bmb);
		\draw[-stealth, shorten >=1pt,auto] (b2i2b) to [] node []{} (bmb);
		\draw[-stealth, shorten >=1pt,auto] (bmb) to [] node []{} (x1mb);
		\draw[-stealth, shorten >=1pt,auto] (bmb) to [] node []{} (nx1mb);		
		\draw[-stealth, shorten >=1pt,auto] (x1mb) to [] node []{} (end1b);
		\draw[-stealth, shorten >=1pt,auto] (nx1mb) to [] node []{} (end1b);
		
		\draw[] (end1b) to [] (9.5, -1.8125);
		\draw[] (9.5, -1.8125) to [] (3.5, -1.8125);
		\draw[-stealth,shorten >=1pt,auto] (3.5, -1.8125) to [] (b1b);
		\draw[-stealth,shorten >=1pt,auto] (o) to [] (b1b2.180);
		\draw[] (o) to [] (-5.5, 1.5);
		\draw[] (-5.5, 1.5) to [] (3.5, 1.5);
		\draw[-stealth,shorten >=1pt,auto] (3.5, 1.5) to [] (b1b);
		
		\end{tikzpicture}
		}%
	\caption{The single-player arena $G$ used in the reduction from \SATcoSAT{} for Boolean B\"uchi objectives.}
	\label{PRV-fig-SAT-COSAT}
\end{figure}

\subparagraph{Intuition of the Reduction.} Given an instance of the \SATcoSAT{} problem consisting of two 3CNF formulas $\psi_1$ and $\psi_2$, we construct an instance of the \problemPayoffVerifAb{} consisting of the arena depicted in \autoref{PRV-fig-SAT-COSAT} and of the payoff $p = (1, 0)$. The intuition behind this construction is as follows. The vertices visited in a play $\rho$ in $G_1$ correspond to a valuation of the variables in $X$, and $\rho$ satisfies the first objective of Player $1$ if this valuation satisfies $\psi_1$. Plays in $G_1$ are the only ones to not satisfy the objective of Player $0$ nor the second objective of Player $1$. It follows that for some play with payoff $p$ and lost by Player $0$ to be realized, there must exist some valuation of $X$ which satisfies $\psi_1$. The vertices visited in a play $\rho'$ in $G_2$ correspond to a valuation of the variables in $X$, and $\rho'$ satisfies the first objective of Player $1$ if this valuation satisfies $\psi_2$. In addition, plays in $G_2$ satisfy the objective of Player $0$ and the second objective of Player $1$. It follows that for payoff $p$ to be Pareto-optimal, no valuation of $X$ must satisfy $\psi_2$.

\subparagraph{Objectives.}  Player~$0$ has a single Boolean B\"uchi objective $\Omega_0$ and Player~$1$ has two Boolean B\"uchi objectives $\Omega_1$ and $\Omega_2$. These Boolean B\"uchi objectives are defined as follows (we use the notations described for the hardness proof of \autoref{thm:PRV-BB-complexity}):
\begin{itemize}
    \item $\Omega_0 = \Omega_2 = g_2 \land \phi_{stable}$,
    \item $\Omega_1 = ((g_1 \land B\!B(\psi_1)) \lor (g_2 \land B\!B(\psi_2))) \land \phi_{stable}$.
\end{itemize}
Objective $\Omega_{1}$ contains the Boolean B\"uchi translation of formula $\psi_1$ using vertices of $G_1$ and of $\psi_2$ using vertices of $G_2$.
All these objectives contains a conjunction with formula
\[\phi_{stable} = \Bigg{(}g_1 ~\land \bigwedge_{i \in \{1, \dots, m\}}\big{(}\overline{x}^1_i \lor \overline{\neg x}^1_i\big{)}\Bigg{)} \lor \Bigg{(}g_2 ~\land \bigwedge_{i \in \{1, \dots, m\}}\big{(}\overline{x}^2_i \lor \overline{\neg x}^2_i\big{)}\Bigg{)}.\] 
Using the same arguments developed in the previous section, we state that unstable plays do not have a Pareto-optimal payoff.

\subparagraph{Realizing Extended Payoff $(0,p)$.} In order for payoff $p = (1,0)$ to be realized by some play $\rho$ which is also lost by Player $0$, it must be the case that $\rho$ is a stable play in $G_1$ (as plays in $G_2$ satisfy $\Omega_0$). The vertices visited in a stable play in $G_1$ can be interpreted as a valuation of the variables in $X$ and objective $\Omega_1$ is satisfied if and only if this valuation satisfies $\psi_1$. It follows that for $p$ to be realized by some play lost by Player $0$, there must exist a valuation of $X$ which satisfies $\psi_1$.

\subparagraph{Ensuring $p$ is Pareto-Optimal.} All stable plays in $G_2$ satisfy objective $\Omega_2$. In order for $p$ to be Pareto-optimal, it must be the case that no play in $G_2$ satisfies in addition objective $\Omega_1$. If that were the case, some stable play in $G_2$ would have payoff $p' = (1,1)$. The vertices visited in a stable play in $G_2$ can be interpreted as a valuation of the variables in $X$ and objective $\Omega_1$ is satisfied if and only if this valuation satisfies $\psi_2$. It follows that for $p$ to be Pareto-optimal, no valuation of $X$ must satisfy $\psi_2$.

\begin{proposition}
The instance of the \SATcoSAT{} problem is positive if and only if the corresponding instance of the \problemPayoffVerifAb{} is positive.
\end{proposition}
\begin{proof}
Let us assume that the instance of the \SATcoSAT{} problem is positive. Therefore, there exists a valuation $val_X$ of $X$ such that formula $\psi_1$ is true and it holds that no valuation of $X$ satisfies formula $\psi_2$. Let $\rho$ be the stable play in $G_1$ which corresponds to valuation $val_X$. It is lost by Player $0$ (as all plays in $G_1$ are) and its payoff is $p = (1, 0)$ (which is easily shown when looking at objective $\Omega_1$ and $\Omega_2$). As no valuation of $X$ satisfies $\psi_2$, no play in $G_2$ satisfies objective $\Omega_1$. It follows that $p$ is Pareto-optimal in $G$, and that the instance of the \problemPayoffVerifAb{} is positive.

The other direction of the proof is the direct consequence of our previous remarks on the realizability of extended payoff $(0,p)$ and the Pareto-optimality of $p$.
\end{proof}

\section{Complexity Class of the \problemUVerifAb{}}
\label{sec:complexity_class_UPRV}

We study in this section the complexity class of the \problemUVerifAb{} for parity and Boolean B\"uchi \gamesAb{}. Our results are summarized in the following theorem.

\begin{theorem} \label{thm:PSPACE}
    The \problemUVerifAb{} is 
    \begin{itemize}
        \item \pspaceComplete{} for Boolean B\"uchi \gamesAb{},
        \item in \pspace{}, \np{}-hard and \conp{}-hard for parity
        \gamesAb{}.
    \end{itemize}
\end{theorem}

We show the \pspace{}-membership stated in \autoref{thm:PSPACE} in the following proposition.

\begin{proposition} \label{prop:inPspace}
The \problemUVerifAb{} is in \pspace{} for both Boolean B\"uchi \gamesAb{} and parity \gamesAb{}.
\end{proposition}

\begin{proof}
Let $\mathcal{G}$ be an \gameAb{} and $\mathcal{M}$ be a nondeterministic Moore machine for Player~$0$. By \autoref{rem:product}, the strategies of $\llbracket \mathcal{M} \rrbracket$ are exactly the strategies of the  product $G' = G \times \mathcal{M}$. In the sequel, we will shift from $G$ to $G'$ and conversely without mentioning it explicitly. 

To prove \autoref{prop:inPspace}, it is enough to show that the complement of the \problemUVerifAb{} is in \npspace{}, since \npspace{} $=$ \pspace{} and as the \pspace{} class is closed under complementation. The complement of the \problemUVerifAb{} is to decide whether there exists a strategy $\sigma_0 \in \llbracket \mathcal{M} \rrbracket$ and a play $\rho \in \Playsigmazero$ such that $\payoff{\rho} \in \paretoSet{\sigma_0}$ and $\rho$ is lost by Player~$0$.

Our algorithm works as follows in $G'$ (we detail its correctness and complexity later):
\begin{enumerate}
    \item guess a lasso $\rho' = g'h'^\omega$ in $\Plays_{G'}$ such that $g'h'$ has polynomial size,
    \item check that $\rho'$ is lost by Player~$0$,
    \item check that for each vertex $v$ of $\rho'$ controlled by Player~$1$, Player~$0$ is winning from $v$ in the two-player \emph{zero-sum} game $\mathcal{H} = (G',\Obj')$ with arena $G'$ and objective $\Omega' = \{\rho^* \in \Plays_{G'} \mid \neg(\payoff{\rho^*} > \payoff{\rho'}) \}$.
\end{enumerate}

Let us prove that this algorithm is correct. \emph{(i)} Assume first that there exists a strategy $\sigma_0 \in \llbracket \mathcal{M} \rrbracket$ and a play $\rho \in \Playsigmazero$ such that $\payoff{\rho} \in \paretoSet{\sigma_0}$ and $\rho$ is lost by Player~$0$. We see this play $\rho$ as a play in $G'$. By \autoref{lem:patricia} there exists a lasso $\rho' = g'h'^\omega$ of polynomial size in $G'$ which realises the same extended payoff and such that $\occ{\rho} = \occ{\rho'}$. This lasso is what is guessed in step 1 of the algorithm. By our assumptions on $\rho$, we know that it satisfies the check of step 2. It remains to explain why the second check also succeeds in step 3. From each vertex $v$ of $\rho'$ (and thus of $\rho$) controlled by Player~$1$, Player~$0$ is winning in $\mathcal{H}$ thanks to his strategy $\sigma_0$. Indeed, any play $\rho'_1 \in \Plays_{G'}$ consistent with $\sigma_0$ cannot have a payoff strictly larger than $\payoff{\rho'} \in \paretoSet{\sigma_0}$, and parity and Boolean B\"uchi objectives are prefix-independent. \emph{(ii)} Assume now that the two checks of our algorithm succeed for the guessed lasso $\rho'$. Let us define a strategy $\sigma_0$ for Player~$0$ in $G'$ (which is also a strategy $\sigma_0 \in \llbracket \mathcal{M} \rrbracket$) as follows: first we define $\sigma_0$ in a way to produce play $\rho'$; second after each history $hvv'$ such that $hv$ is prefix of $\rho'$ and $hvv'$ is not (meaning that $v$ belongs to Player~$1$), $\sigma_0$ acts as the winning strategy of Player~$0$ from $v$ in $\mathcal{H}$. We have thus proved that there exist a strategy $\sigma_0 \in \llbracket \mathcal{M} \rrbracket$ and a play $\rho' \in \Playsigmazero$ such that $\payoff{\rho'} \in \paretoSet{\sigma_0}$ and $\rho'$ is lost by Player~$0$.

Let us now show that our nondeterministic algorithm executes in polynomial space. Step~1 requires polynomial space to store $g'h'$. The check of step 2 requires to verify that $\rho' \not \in \Obj_0$ such that $\Obj_0$ is either a parity or a Boolean B\"uchi objective. This can be done by looking at the cycle $h'$ in polynomial space. Let us now study step 3. We are going to show that $\mathcal{H} = (G',\Obj')$ is a zero-sum game with a Boolean B\"uchi objective $\Obj'$, known to be solvable in \pspace{}~\cite{HunterD05}. Let us denote by $p = (p_1, \ldots, p_t)$ the payoff of $\rho'$. The objective $\Obj'$ is equal to
\begin{equation} \label{eq:H}
    \Big{(}\bigcap_{p_i = 0} \overline{\Obj}_i\Big{)} \cup \Big{(}\bigcup_{\substack{p_i = 1 \\ p_j = 0}} \big{(}\overline{ \Obj}_i \cap \Obj_j\big{)}\Big{)}
\end{equation}
where the the first disjunct expresses plays with payoffs less than or equal to $p$ and the second disjunct expresses plays with payoffs incomparable with $p$. Recall that any parity objective can be expressed as a Boolean B\"uchi objective using a formula of size $\mathcal{O}(d^2)$ where $d$ is the highest priority in the parity objective (see e.g. \cite{DBLP:conf/atva/BaierBD00S19}). Therefore, for both parity and Boolean B\"uchi \gamesAb{}, the objective $\Obj'$ is a Boolean B\"uchi objective defined by a formula of polynomial size. 
\end{proof}

We now turn to the hardness results stated in \autoref{thm:PSPACE}. The \conp{} hardness of the \problemUVerifAb{} for parity \gamesAb{} is easily obtained from the \conp{} hardness of the \problemVerifAb{} (\autoref{thm:PRV-parity-complexity}). We consider the other hardness results in the following proposition.

\begin{proposition}
\label{prop:hardness_generic}
    The \problemUVerifAb{} is \npHard{} for parity \gamesAb{}, and \pspaceHard{} for Boolean B\"uchi \gamesAb{}.
\end{proposition}

We begin by proving that the \problemUVerifAb{} is \npHard{} for parity \gamesAb{} and indicate later how to adapt this proof to obtain the \pspace{}-hardness for Boolean B\"uchi \gamesAb{}. For this purpose, we reduce the following \conpHard{} problem to an instance of the complement of the \problemUVerifAb{}.

\subparagraph{Generalized Parity Game.} Let us consider a two-player zero-sum generalized parity game $(G,\Obj_a \land \Obj_b)$ where the objective of Player~$0$ is a conjunction $\Obj_a \land \Obj_b$ of two parity objectives. Deciding whether Player~$0$ has a winning strategy from a vertex $v_0$ in $G$ is \conpHard{} \cite{DBLP:conf/fossacs/ChatterjeeHP07}.

\begin{figure}[t]
	\centering
	\resizebox{0.5\textwidth}{!}{%
		\begin{tikzpicture}
		
		\node[draw, rectangle, minimum size=0.8cm] (o) at (5,5){$v'_0$};
		
		\node[draw, rectangle, minimum size=0.8cm] (b1b) at (3,5){$g_1$};

		\draw[dashed, rounded corners] (6.75,4.25) rectangle (8.25, 5.75)  {};
        \node[] at (7.5, 5) {$G$};
	
		\draw[-stealth,shorten >=1pt,auto] (o) to [] (6.75, 5);
		
		\draw[-stealth, shorten >=1pt,auto] (o) to [] node []{} (b1b);		
		\draw[-stealth, shorten >=1pt,auto] (b1b) to [loop left] node []{} (b1b);
		
		\end{tikzpicture}
		}%
	\caption{The arena $G'$ used in the reduction from zero-sum games with two parity objectives.}
	\label{fig:np-arena}
\end{figure}

\subparagraph{Intuition of the Reduction.} Given a zero-sum game $(G,\Obj_a \wedge \Obj_b)$ with a conjunction $\Omega_a \land \Omega_b$ of two parity objectives for Player $0$ and a vertex $v_0$, we construct an instance of the \problemUVerifAb{} with the game arena $G'$ depicted in \autoref{fig:np-arena}. In $G'$, the dashed box labeled $G$ represents the arena of the zero-sum game and we assume that the edge from $v'_0$ goes to $v_0$ in $G$. Equivalently, the dashed box is the Cartesian product of $G$ and the nondeterministic machine $\mathcal{M}$ with one memory state embedding all possible strategies of Player~$0$ (see \autoref{rem:product}). Notice that given a play $\rho'$ of $G'$ reaching $G$, we can retrieve a corresponding play $\rho$ from $v_0$ in $G$. Any strategy $\sigma_0$ of Player~$0$ in $G'$ is a strategy in $\llbracket M \rrbracket$ and the converse also holds. We will see that the proposed construction is such that Player $0$ has a winning strategy from $v_0$ in $(G,\Omega_a \land \Omega_b)$ if and only if the corresponding instance of the \problemUVerifAb{} is \emph{negative}. 

\subparagraph{Objectives.} Player~$0$ has a single parity objective $\Omega_0$ and Player~$1$ has two parity objectives $\Omega_1$ and $\Omega_2$. We first extend the priority function $c_a$ of $\Omega_a$ (resp.\! $c_b$ of $\Omega_b$) to $G'$ such that $c_a(g_1) = c_a(v'_0) = c_b(g_1) = c_b(v'_0) = 1$ and consider the corresponding objective $\Omega'_a$ (resp.\! $\Omega'_b$) in $G'$. Notice that $\Omega'_a = \Omega_a$ (resp.\! $\Omega'_b = \Omega_b$) when considering only the plays of sub-arena $G$ in $G'$. We define the actual objectives used in the reduction as follows. Player $0$ has objective $\Omega_0 = \parity{c}$ with a priority function $c$ defined such that $\Obj_0$ is only satisfied in plays reaching $G$. The first (resp.\! second) objective of Player $1$ is such that $\Omega_1 = \overline{\Omega}'_a$ (resp.\! $\Omega_2 = \overline{\Omega}'_b$). Notice that objective $\Omega_1$ (resp.\! $\Omega_2$) is satisfied in plays reaching $G$ if and only if the objective $\Omega_a$ (resp.\! $\Omega_b$) is \emph{not} satisfied in those plays. The play $v'_0 g_1^\omega$ is consistent with any strategy of Player 0 and has extended payoff $(0, (0, 0))$. Any play reaching $G$ is of the form $\rho' = v'_0 \rho$ where $\rho$ is a play in $G$ starting from the initial vertex $v_0$. We list below the realizable extended payoffs for such a play $\rho'$: 
\begin{itemize}
    \item $(1, (0,0))$ if $\rho$ satisfies $\Omega_a$ and $\Omega_b$,
    \item $(1, (0,1))$ if $\rho$ satisfies $\Omega_a$ and not $\Omega_b$,
    \item $(1, (1,0))$ if $\rho$ satisfies $\Omega_b$ and not $\Omega_a$,
    \item $(1, (1,1))$ if $\rho$ does not satisfy $\Omega_a$ nor $\Omega_b$.
\end{itemize}

\subparagraph{Correctness.} If the instance of the \problemUVerifAb{} is negative, it holds there exists a strategy $\sigma_0 \in \llbracket \mathcal{M} \rrbracket$ such that some play in $\Plays_{\sigma_0}$ has a Pareto-optimal payoff and is lost by Player $0$. Since the play $v'_0 g_1^\omega$ with payoff $(0,0)$ is the only one in $G'$ not to satisfy $\Obj_0$, its payoff must be Pareto-optimal. It follows that all plays in $G$ that are consistent with $\sigma_0$ have payoff $(0,0)$ and therefore satisfy the conjunction $\Omega_a \land \Omega_b$. Hence, $\sigma_0$ is a winning strategy for Player $0$ from $v_0$ in the zero-sum game $(G,\Obj_a \wedge \Obj_b)$. Conversely, if Player $0$ has a winning strategy from $v_0$ in $(G,\Obj_a \wedge \Obj_b)$, it holds that this strategy is in $\llbracket \mathcal{M} \rrbracket$ and such that all consistent plays in $G$ satisfy the conjunction $\Obj_a \wedge \Obj_b$ and therefore has payoff $(0,0)$. It is easily checked that the instance of the \problemUVerifAb{} is negative.

\subparagraph{Adapting the Reduction to Boolean B\"uchi Objectives.} Let us explain how we adapt the reduction of the \np{}-hardness of the \problemUVerifAb{} for parity \gamesAb{} (see \autoref{prop:hardness_generic}) to prove that the \problemUVerifAb{} is \pspaceHard{} for Boolean B\"uchi \gamesAb{}. 
\begin{itemize}
    \item First, we consider the problem of deciding whether Player~$0$ has a winning strategy from $v_0$ in a two-player zero-sum game $(G,\Obj_a)$ where $\Obj_a = \BooleanBuchi{\phi, \target_1, \dots, \target_ m}$ is a Boolean B\"uchi objective. This problem is \pspaceComplete{}~\cite{HunterD05}.
    \item Second, given such a zero-sum game $(G,\Obj_a)$ and a vertex $v_0$, we construct an instance of the \problemUVerifAb{} on the same game arena $G'$ depicted in \autoref{fig:np-arena} which we used for the reduction of \autoref{prop:hardness_generic}. In this instance, both Player~$0$ and Player~$1$ have a single Boolean B\"uchi objective defined as follows (we again use the notations of the hardness proof of \autoref{thm:PRV-BB-complexity}):
\begin{itemize}
    \item $\Omega_0 = \overline{g_1}$,
    \item $\Omega_1 = \overline{g_1} \wedge B\!B(\neg \phi)$.
\end{itemize}
The objective $\Omega_0$ is not satisfied by the play $v'_0 g_1^\omega$ and is satisfied by all plays reaching $G$. The objective $\Omega_1$ is not satisfied by the play $v'_0 g_1^\omega$ and is satisfied by plays reaching $G$ if and only if the objective $\Omega_a$ is not satisfied in those plays. 
    \item Repeating arguments similar to the reduction of \autoref{prop:hardness_generic}, one can verify that the construction is such that Player $0$ has a winning strategy from $v_0$ in $(G,\Obj_a)$ if and only if the corresponding instance of the \problemUVerifAb{} is negative. It follows that the \problemUVerifAb{} is \pspaceHard{} for Boolean B\"uchi \gamesAb{} (as \copspace{} $=$ \pspace).
\end{itemize}

\section{Fixed-Parameter Complexity}
\label{sec:fpt}
In this section, we study the fixed-parameter complexity of the (U)PRV problem. We refer the reader to~\cite{downey2012parameterized} for the concept of fixed-parameter tractability (\FPT{}). We recall that given an \gameAb{} $\mathcal{G} = (G,\Obj_0,\ldots,\Obj_\nbrObjectives)$, $\max d_i$ is the maximum of all maximum priorities $d_i$ according to each objective $\Omega_i$ in case of parity \gamesAb{}, and that $\max |\phi_i|$ is the maximum of all sizes $|\phi_i|$ such that each $\phi_i$ defines objective $\Omega_i$ in case of Boolean B\"uchi \gamesAb{}.

\subsection{UPRV Problem}

We begin by providing an \FPT{} algorithm for the \problemUVerifAb{}.

\begin{theorem}
\label{thm:UPRV_fpt}
The \problemUVerifAb{} is in \FPT{}
\begin{itemize}
      \item with parameters $\nbrObjectives$ and $\max d_i$ for parity \gamesAb{} (with an exponential in $\nbrObjectives$ and $\max d_i$),
    \item with parameters $\nbrObjectives$ and $\max |\phi_i|$ for Boolean B\"uchi \gamesAb{} (with an exponential in $\nbrObjectives$ and $\max |\phi_i|$).
\end{itemize}
\end{theorem}
\begin{proof}
The proof uses a \emph{deterministic} variant of the algorithm given in the proof of \autoref{prop:inPspace}. Given an \gameAb{} $\mathcal{G}$ and a nondeterministic Moore machine $\mathcal{M}$, deciding whether this instance of the \problemUVerifAb{} is positive works in the following way on the Cartesian product $G' = G \times \mathcal{M}$. 
\begin{itemize}
    \item For every payoff $p \in \{0,1\}^\nbrObjectives$, consider the zero-sum game $\mathcal{H} = (G',\Obj')$ with arena $G'$ and objective $\Omega' = \{\rho' \in \Plays_{G'} \mid \neg(\payoff{\rho'} > p) \}$,
\begin{enumerate}
    \item compute the set $W_1$ of vertices $v \in V_1$ from which Player~$0$ is winning in the game $\mathcal{H}$,
    \item construct the sub-arena $G_{\upharpoonright V_0 \cup W_1}$ of $G$ restricted to $V_0 \cup W_1$,
    \item check whether there exists in $G_{\upharpoonright V_0 \cup W_1}$ a play with payoff $p$ that does not satisfy $\Obj_0$.
\end{enumerate}
\item If the test in step 3 is positive for some payoff $p$, then the given instance of the \problemVerifAb{} is negative, otherwise it is positive.
\end{itemize}

The correctness of this algorithm is proved similarly as we did in the proof of \autoref{prop:inPspace}. Let us study its complexity. The three steps are executed $2^\nbrObjectives$ times. We know from the proof of \autoref{prop:inPspace} that $\Obj'$ is equal to the Boolean B\"uchi objective (\ref{eq:H}) defined by a formula $\phi'$ of size polynomial in:
\begin{itemize}
    \item $\nbrObjectives$ and $\max |\phi_i|$ in case of Boolean B\"uchi \gameAb{} $\mathcal{G}$,
    \item $\nbrObjectives$ and $\max d_i$ in case of parity \gameAb{} $\mathcal{G}$.
\end{itemize}
Moreover, it is proved in~\cite{BruyereHR18,DBLP:journals/corr/abs-2203-01285} that computing the set $W_1$ of step 1 of our algorithm is in \FPT{} with parameter $|\phi'|$ (with an exponential in $|\phi'|$).\footnote{More precisely computing the set $W_1$ is proved in \cite{BruyereHR18} to be linear in the number of symbols $\vee, \wedge$ of $\phi'$ and double exponential in the number of variables of $\phi'$. This complexity is improved in \cite{DBLP:journals/corr/abs-2203-01285} by replacing the double exponential in $|\phi'|$ by a single exponential in $|\phi'|$.} The complexity of step 2 is polynomial. By \autoref{prop:parity_payoff_existence}, the complexity of step 3 is polynomial for parity \gamesAb{} and exponential in $\nbrObjectives$ and $\max |\phi_i|$ for Boolean B\"uchi \gamesAb{}. The overall complexity of our algorithm is therefore
\begin{itemize}
    \item exponential in $\nbrObjectives$ and $\max |\phi_i|$ for Boolean B\"uchi \gameAb{} $\mathcal{G}$,
    \item exponential in $\nbrObjectives$ and $\max d_i$ for parity \gameAb{} $\mathcal{G}$.
\end{itemize}
\end{proof}

\subsection{PRV Problem}

A corollary of \autoref{thm:UPRV_fpt} is that the \problemVerifAb{} is in \FPT{}. We now provide a simpler \FPT{} algorithm specific to this problem with improved complexity for parity \gamesAb{}.

\begin{theorem} 
\label{thm:PRV_fpt}
The \problemVerifAb{} is in \FPT{} 
\begin{itemize}
    \item with parameter $\nbrObjectives$ for parity \gamesAb{}  (with a single exponential in $\nbrObjectives$),
    \item with parameters $\nbrObjectives$ and $\max |\phi_i|$ for Boolean B\"uchi \gamesAb{}  (with a single exponential in~$\nbrObjectives$ and $\max |\phi_i|$).
\end{itemize}
\end{theorem}

\begin{proof}
Given an \gameAb{} $\mathcal{G}$ and a deterministic Moore machine $\mathcal{M}$, deciding whether this instance of the \problemVerifAb{} is positive is done in two steps on the Cartesian product $G' = G \times \mathcal{M}$ in the following way.
\begin{enumerate}
\item The algorithm considers every possible payoff $p \in \{0,1\}^\nbrObjectives$ and checks whether it is realizable in $G'$. Doing so computes the set $T$ of all realizable payoffs in $G'$ and thus the antichain $\paretoSet{\sigma_0} = \lceil T \rceil$ of Pareto-optimal payoffs in $G$ given the single strategy $\sigma_0 \in \llbracket \mathcal{M} \rrbracket$. 
\item The algorithm then checks for the existence of a play with a Pareto-optimal payoff that is lost by Player~$0$. In case of existence, the instance of the \problemVerifAb{} is negative, otherwise it is positive.
\end{enumerate}
In the first step, an existence check is performed $2^\nbrObjectives$ times, one for each payoff in the lattice of payoffs. In the second step an existence check is done for the extended payoff $(0,p)$ for each payoff $p$ of the antichain $\paretoSet{\sigma_0}$ and therefore $\mathcal{O}(2^\nbrObjectives)$ times in total. The complexity of deciding the existence of a specific (extended) payoff is described in \autoref{prop:parity_payoff_existence}. Overall, this algorithm therefore exhibits a complexity exponential in $\nbrObjectives$ for parity \gameAb{} $\mathcal{G}$, and exponential in $\nbrObjectives$ and $\max |\phi_i|$ for Boolean B\"uchi \gameAb{} $\mathcal{G}$.
\end{proof}

\subsection{Antichain Optimization Approach}
We now discuss how to modify the simpler \FPT{} algorithm for the \problemVerifAb{} presented in the previous section in order to improve its performance in practice. The resulting algorithm, described in \autoref{algo:fpt}, is the first variation that we consider.

First, notice that the algorithm presented in the proof of \autoref{thm:PRV_fpt} computes the set $\paretoSet{\sigma_0}$ by considering every possible payoff one by one. In practice, this can be avoided by going through the lattice of payoffs starting from the maximal payoff $(1, \dots, 1)$ and going down level-by-level in the lattice while testing for the existence of a play realizing a payoff. If a payoff is found to be realizable, the algorithm needs not consider the payoffs which are strictly smaller as they cannot be Pareto-optimal. In \autoref{algo:fpt}, the currently known part of $\paretoSet{\sigma_0}$ is stored in an antichain $A$ and the future potential elements of $\paretoSet{\sigma_0}$ are stored in a queue $Q$. The proposed improvement is implemented in line $13$ where a payoff $p^*$ is added to the queue only if it has not been added yet and no payoff strictly larger has been deemed realizable. Since it may be the case that such a larger payoff be found realizable after $p^*$ is added to $Q$, this check is repeated in line $5$. Notice that we only add payoffs $p^*$ strictly smaller than $p$ by one objective to the queue in order to descend level-by-level.

\begin{algorithm}[t]
    \KwIn{A single-player \gameAb{} resulting from the Cartesian product of the arena $G$ of an \gameAb{} and a deterministic Moore machine $\mathcal{M}$ for Player~$0$.}
    \KwOut{Whether the instance of the \problemVerifAb{} is positive.}
    $Q \gets \{(1, \dots, 1)\}$\\
    $A \gets \emptyset$ \\
    \While{$Q$ is not empty}{
    
        $p \gets Q$.dequeue()\\
    
        \If{$p \not\in \; \downarrow^< \! A$ }{
                 
            \eIf{$\exists \rho \in \Plays$ such that $\payoff{\rho} = p$}{
                $A \gets A \cup \{p\}$
                
                \If{$\exists \rho \in \Plays$ such that $\payoff{\rho} = p$ \textnormal{\textbf{and}} $\won{\rho} = 0$}{
                    \Return False
                }
            }{
                \For{$i \in \{1, \dots, \nbrObjectives\}$ such that $p_i = 1$}{
                    
                    $p^* \gets (p_1, \dots, p_{i-1}, 0, p_{i+1}, \dots, p_\nbrObjectives)$
                    
                    \If{$p^* \not\in \; \downarrow^< \! A$ \textnormal{\textbf{and}} $p^*$ has never been added in $Q$}{
                        $Q$.enqueue($p^*$)
                    }
                }
            }
        
        }       
    }
    \Return True
    
 \caption{Antichain optimization algorithm for the \problemVerifAb{}.}
 \label{algo:fpt}
\end{algorithm}

Second, the algorithm presented in the proof of \autoref{thm:PRV_fpt} proceeds in two steps, first computing the set $\paretoSet{\sigma_0}$ and then checking that there does not exist a play with a payoff in $\paretoSet{\sigma_0}$ that is losing for Player~$0$. In practice, merging the two steps allows the algorithm to stop early. Indeed whenever a new payoff is added to $A$, testing for the existence of a play with that payoff and losing for Player~$0$ prevents from doing unnecessary work when such a play exists. This is implemented in lines $8$-$9$.

A last improvement is made by applying the following observation about realizable (extended) payoffs.

\begin{remark}
\label{rem:better}
Let $p$ be a payoff for which we already know that there exists no play $\rho'$ in $G$ with $\payoff{\rho'} > p$. It follows that checking whether there exists a play $\rho$ such that $\payoff{\rho} = p$ amounts to checking whether there exists such a play with $\payoff{\rho} \geq p$. It is more efficient to perform this second check instead of the first as the corresponding intersection contains fewer objectives (see the proof of \autoref{prop:parity_payoff_existence}).
\end{remark}
The observation stated in the previous remark applies in the case of \autoref{algo:fpt}, as the payoffs are considered level-by-level, thanks to a descent in the lattice from $(1,\ldots,1)$. Finally, we make the following observation on \autoref{algo:fpt}.

\begin{remark}
\autoref{algo:fpt} works for any \gameAb{}. In case of parity or Boolean B\"uchi \gamesAb{}, the checks which look for the existence of a play with a specific (extended) payoff are performed as explained in the proof of \autoref{prop:parity_payoff_existence}. Given a payoff $p$, checking whether $p \not\in \; \downarrow^< \! A$ can be performed by comparing the vector of Booleans $p$ to those contained in $A$, whose size is in ${\mathcal O}(2^\nbrObjectives)$. This algorithm is thus in \FPT{} as it performs a subset of the operations used in the algorithm of the proof of \autoref{thm:PRV_fpt}.
\end{remark}

\subsection{Counterexample-Based Approach} 

We now propose an other variation of the \FPT{} algorithm for the \problemVerifAb{} provided in the proof of \autoref{thm:PRV_fpt}. Instead of computing the set $\paretoSet{\sigma_0}$ by going through the entire lattice of payoffs, we compute an \emph{under-approximation} (with respect to $\sqsubseteq$) of this antichain on demand by using counterexamples (see \autoref{algo:fpt-CE}). We first provide some intuition about this algorithm and show its correctness below. The algorithm systematically searches for plays $\rho$ losing for Player~$0$ and maintains an antichain $A$ of realizable payoffs to eliminate previous counterexamples. Initially, this antichain $A$ is empty. A potential counterexample is a play $\rho$ losing for Player~$0$ and such that for all payoffs $p$ of $A$, $\payoff{\rho}$ is not strictly smaller than $p$, that is, $\payoff{\rho} \not\in \; \downarrow^< \! A$ (line 3). When a potential counterexample $\rho$ exists, there are two possible cases. 
First, there exists a play $\rho'$ winning for Player~$0$ and such that $\payoff{\rho'} > \payoff{\rho}$ (line 4). The payoff of $\rho'$ is added to $A$ and a new approximation $A$ of $\paretoSet{\sigma_0}$ is computed (by keeping only the maximal elements, line 5). 
Second, if such a play $\rho'$ does not exist, then we have identified a counterexample (the play $\rho$), showing that the instance of the \problemVerifAb{} is negative (line 7). If there are no more potential counterexamples, then the instance is positive (line 9), otherwise we iterate. This algorithm is guaranteed to terminate as $A \sqsubset \lceil A\cup\{\payoff{\rho'}\} \rceil$ in line 5.

\begin{remark}
\autoref{algo:fpt-CE} works for any \gameAb{}. Let us explain how to perform the checks in lines~3 and~4 in case of parity or Boolean B\"uchi \gamesAb{}. The check in line~4 is similar to the checks explained in the proof of \autoref{prop:parity_payoff_existence}. Let us explain given an antichain $A$ of payoffs, how to check whether there exists a play $\rho$ such that $\payoff{\rho} \not\in \; \downarrow^< \! A$ (line~3). This is equivalent to check whether there exists a play $\rho$ such that for all $p \in A$, either $\payoff{\rho}$ is larger than or equal to $p$ or $\payoff{\rho}$ is incomparable to $p$, i.e., $\rho$ satisfies the objective $\wedge_{p \in A} ((\wedge_{p_i = 1} \Omega_i) \vee (\vee_{p_i = 0} \Omega_i))$. The latter objective can be translated into a Boolean B\"uchi objective for both parity and Boolean B\"uchi \gamesAb{} (whose size depends in particular on the size of $A$). 
\end{remark}

\begin{algorithm}[t]
    \KwIn{A single-player \gameAb{} resulting from the Cartesian product of the arena $G$ of an \gameAb{} and a deterministic Moore machine $\mathcal{M}$ for Player~$0$.}
    \KwOut{Whether the instance of the \problemVerifAb{} is positive.}
    $A \gets \emptyset$ \\
    \Repeat{
        \eIf{$\exists \rho \in \Plays$ such that $\won{\rho}=0$ and $\payoff{\rho} \not\in \; \downarrow^< \! A$}
        {
            \eIf{$\exists \rho' \in \Plays$ such that $\won{\rho'}=1$ and $\payoff{\rho'} > \payoff{\rho}$}{$A \gets \lceil A\cup\{\payoff{\rho'}\} \rceil$}{\Return False}
        
        }{\Return {\rm True}}
    }
 \caption{Counterexample-based algorithm for the \problemVerifAb{}.}
 \label{algo:fpt-CE}
\end{algorithm}

The correctness of \autoref{algo:fpt-CE} is established using the following definition and lemmas. Notice that it works for any single-player \gameAb{} (that is, not only for parity or Boolean B\"uchi objectives).

\begin{definition} \label{def:certificate}
Let $\mathcal G$ be a single-player \gameAb{}.
An antichain $A$ of payoffs is a \emph{certificate} (of correctness) if 
\begin{enumerate}
    \item each payoff of $A$ is realizable;
    \item there does not exist a play $\rho$ that is losing for Player~$0$ and such that for all $p \in A$, $\payoff{\rho}$ is not strictly smaller than $p$, i.e,
    $\neg (\exists \rho \in \Plays, \ \won{\rho} = 0 \wedge \payoff{\rho} \not\in \;  \downarrow^< \! A)$.
\end{enumerate}
\end{definition}

Notice that when presenting \autoref{algo:fpt-CE}, we have intuitively explained that this algorithm constructs a certificate if one exists. It accepts the given instance as positive in case of existence, otherwise it rejects the instance.

\begin{lemma} \label{lem:certificate}
Let $\mathcal G$ be an \gameAb{} with arena $G$ and $\mathcal{M}$ be a deterministic Moore machine defining a strategy $\sigma_0$ for Player~$0$.
This instance is a positive instance of the PRV problem if and only if there exists an antichain $A$ of payoffs that is a certificate in the single-player \gameAb{} with arena $G \times \mathcal{M}$.   
\end{lemma}

\begin{proof}
Suppose that the instance is positive. We easily check that the antichain $\paretoSet{\sigma_0}$ is a certificate in $G \times \mathcal{M}$. Suppose now that there exists an antichain $A$ that is a certificate.  Assume for the sake of contradiction that the instance is negative, that is, there exists a play $\rho$ such that $\payoff{\rho} \in \paretoSet{\sigma_0}$ and $\won{\rho} = 0$. All the payoffs of $A$ are realizable by condition~1 of \autoref{def:certificate}. Therefore, $A \sqsubseteq \paretoSet{\sigma_0}$ since $\paretoSet{\sigma_0}$ is composed of all Pareto-optimal payoffs (which are thus realizable). It follows that $\rho \not\in \; \downarrow^< \!A$. Condition~2 of \autoref{def:certificate} is therefore falsified by $\rho$ which is a contradiction of $A$ being a certificate.
\end{proof}

\begin{lemma}
\label{lem:correct}
\autoref{algo:fpt-CE} is correct.    
\end{lemma}

\begin{proof}
The following two properties are true during the execution of \autoref{algo:fpt-CE}.
\begin{enumerate}
    \item All the payoffs added to $A$ are realizable (line 4). It follows that condition 1 of \autoref{def:certificate} is satisfied by $A$, and thus $A \sqsubseteq \paretoSet{\sigma_0}$.
    \item If $A$ and $A'$ are two successive iterations of the antichain $A$ during the execution of \autoref{algo:fpt-CE}, then $A \sqsubset A'$. Indeed, when $\payoff{\rho'}$ is added to $A$ (line 5), we have that $\forall p \in A, \neg (\payoff{\rho'} \leq p)$ (lines 3-4).
\end{enumerate}
A direct consequence of the second property is that line 5 is executed only a finite number of times which is in $\mathcal{O}(2^\nbrObjectives)$. It follows that \autoref{algo:fpt-CE} terminates in either line 7 or line 9. We are now equipped to prove the correctness of \autoref{algo:fpt-CE}.  
\begin{itemize}
    \item If line 7 is reached, then the test in line 3 evaluates to True while the test in line 4 evaluates to False. As a consequence, $\rho$ is a play losing for Player~$0$ such that there is no Pareto-optimal play $\rho'$ with $\payoff{\rho'} > \payoff{\rho}$ and satisfying $\Omega_0$. Therefore the given instance of the \problemVerifAb{} is negative and the algorithm correctly returns False. 
    \item If line 9 is reached, then the test in line 3 evaluates to False and so condition~2 of \autoref{def:certificate} holds for the current antichain $A$. It follows that $A$ is a certificate, that is, the given instance of the \problemVerifAb{} is positive by \autoref{lem:certificate} and the algorithm correctly returns True.
 \end{itemize}
\end{proof}

Finally, we show that the counterexample-based algorithm is in \FPT{}.

\begin{lemma}
\label{lem:complexity}
\autoref{algo:fpt-CE} is an \FPT{} algorithm
\begin{itemize}
    \item with parameters $\nbrObjectives$ and $\max d_i$ for parity \gamesAb{} (with a double exponential in $\nbrObjectives$ and a single exponential in $\max d_i$),
    \item with parameters $\nbrObjectives$ and $\max |\phi_i|$ for Boolean B\"uchi \gamesAb{} (with a double exponential in $\nbrObjectives$ and a single exponential in $\max |\phi_i|$).
\end{itemize}
\end{lemma}

\begin{proof}
The loop is executed $\mathcal{O}(2^t)$ times (see proof of \autoref{lem:correct}).
Let us see how to perform the checks in lines~3 and~4 in case of parity or Boolean B\"uchi \gamesAb{}. The check in line~4 is similar to the checks explained in the proof of \autoref{prop:parity_payoff_existence}. It has the same complexity, that is,
 \begin{itemize}
        \item polynomial in $|G|$, $\nbrObjectives$, and $\max d_i$ for parity objectives,
        \item polynomial in $|G|$, and exponential in $\nbrObjectives$ and $\max |\phi_i|$ for Boolean B\"uchi objectives.
\end{itemize}
Given an antichain $A$ of payoffs, let us explain how to check whether there exists a play $\rho$ such that $\payoff{\rho} \not\in \; \downarrow^< \! A$ (line~3). This is equivalent to check whether there exists a play $\rho$ such that for all $p \in A$, either $\payoff{\rho}$ is larger than or equal to $p$ or $\payoff{\rho}$ is incomparable to $p$, i.e., $\rho$ satisfies the objective 
\[\Obj' = \cap_{p \in A} ((\cap_{p_i = 1} \Omega_i) \cup (\cup_{p_i = 0} \Omega_i)).\] 
Similarly to objective (\ref{eq:H}) and the explanations below it, this objective $\Obj'$ can be translated into a Boolean B\"uchi objective for both parity and Boolean B\"uchi \gamesAb{} defined by a formula $\phi'$ of size
\begin{itemize}
    \item polynomial in $|A|$, $\nbrObjectives$, and $\max d_i$ for parity objectives,
    \item polynomial in $|A|$, $\nbrObjectives$, and $\max |\phi_i|$ for Boolean B\"uchi objectives.
\end{itemize}
Checking whether there exists a play $\rho$ satisfying $\Obj'$ can be solved in time polynomial in $|G|$ and exponential in $|\phi'|$~\cite{DBLP:conf/atva/BaierBD00S19}. As the size $|A|$ of the antichain $A$ is in $\mathcal{O}(2^t)$, the check in line~4 is more costly than the check in line~3. Therefore \autoref{algo:fpt-CE} is an \FPT{} algorithm with the fixed-parameter complexity stated in \autoref{lem:complexity}.
\end{proof}

\section{LTL Pareto-Rational Verification}
\label{sec:LTL}

We have studied the complexity of the \problemVerifAb{} and the \problemUVerifAb{} for both parity and Boolean B\"uchi objectives. It is also usual to consider objectives given in Linear Temporal Logic (LTL). We show in this section that when the objectives are expressed using LTL formulas, the \problemVerifAb{}  retains the \pspaceComplete ness of the LTL model-checking problem and the \problemUVerifAb{} retains the \twoExptime{}-completeness of solving LTL games. We do not investigate the fixed-parameter complexity in this context because the completeness to \pspace{} (resp. \twoExptime{}) already holds when Player~$1$ has a single objective.

\subparagraph{LTL (Universal) Pareto-Rational Verification Problem.} 
A \emph{labeled game arena} $G_\lambda$ is a game arena where a labeling function $\lambda:V \rightarrow 2^{AP}$ maps each vertex of $G$ to a set of propositional variables in $AP$.
An \emph{LTL \gameAb{}} $\mathcal{G} = (G_\lambda, \phi_0,\phi_1, \dots, \phi_\nbrObjectives)$ is composed of a labeled game arena $G_\lambda$, an LTL formula $\phi_0$ for Player~$0$ and $\nbrObjectives \geq 1$ LTL formulas $\phi_1, \dots, \phi_\nbrObjectives$ for Player~$1$. The difference with regular \gamesAb{} is thus that the goal of the players is expressed using LTL formulas over the set of propositional variables $AP$. The payoff of plays in $G_\lambda$ is defined as expected. Given an LTL \gameAb{}, we consider the two verification problems described in \autoref{sec:prelim} and call them the \emph{LTL \problemVerifAb{}} and \emph{LTL \problemUVerifAb{}}.

\begin{theorem}
\label{thm:ltl_universal_verif_pspace}
The LTL \problemUVerifAb{} is \twoExptime{}-complete.
\end{theorem}

\begin{proof}
We first prove that the LTL \problemUVerifAb{} is in \twoExptime{}. Given an LTL \gameAb{} $\mathcal{G}$ and a nondeterministic Moore machine $\mathcal{M}$, we proceed as follows. We first perform the Cartesian product $G' = G_\lambda \times \mathcal{A}_0 \times \mathcal{A}_1 \times \dots \times \mathcal{A}_\nbrObjectives$ of the arena $G_\lambda$ with a Deterministic Parity Automaton (DPA) $\mathcal{A}_i$ for each LTL formula $\phi_i$, $i \in \{0, \ldots, \nbrObjectives\}$. The size of each automaton is at most double exponential in the size of its corresponding LTL formula, and the number of priorities it uses is exponential~\cite{SistlaVW87,Piterman07,EsparzaKRS17}. We thus have a parity \gameAb{} $\mathcal{G}'$ with arena $G'$ of double exponential size. We then use the \FPT{} algorithm of \autoref{thm:UPRV_fpt} on this \gameAb{} $\mathcal{G}'$, which is polynomial in $|G'|$ and exponential in the parameters $\nbrObjectives$ and $\max d'_i$ (the maximum priority used in the parity objectives). Therefore this algorithm is double exponential in $|G|$, single exponential in $\nbrObjectives$, and double exponential in the size of LTL formulas $\phi_i$, $i \in \{0, \ldots, \nbrObjectives\}$. This shows the \twoExptime{}-easyness.

Let us now prove the \twoExptime{}-hardness result by adapting the reduction of \autoref{prop:hardness_generic} for the case of the LTL \problemUVerifAb{}.
\begin{itemize}
    \item We consider the problem of deciding whether Player~$0$ has a winning strategy from $v_0$ in a two-player zero-sum game $(G_\lambda,\phi)$ where the $\phi$ is the LTL objective of Player~$0$. This problem is \twoExptime{}-complete~\cite{PnueliR89}.
    \item Given such a zero-sum game $(G_\lambda,\phi)$ and a vertex $v_0$, we construct an instance of the \problemUVerifAb{} on the same game arena $G'$ depicted in \autoref{fig:np-arena} which we used for the reduction of \autoref{prop:hardness_generic}. In this arena, $G$ is replaced by $G_\lambda$ and both $v'_0$ and $g_1$ are labelled with the set $\{x\}$ containing the single atomic proposition $x$ which does not appear in $\phi$. The nondeterministic machine $\mathcal{M}$ considered in the reduction is again the one with a single memory state that embeds every possible strategy of Player~$0$. The objective $\Omega_0$ of Player~$0$ is defined by LTL formula $\phi_0$ and the single objective $\Omega_1$ of Player~$1$ is defined by LTL formula $\phi_1$ as follows:
\begin{itemize}
    \item $\phi_0 = \neg \bigcirc x$,
    \item $\phi_1 = (\neg \bigcirc x) \land (\neg \bigcirc \phi)$
\end{itemize}
where $\bigcirc$ is the next operator in LTL. It is direct to see that objective $\Omega_0$ is not satisfied by the play $v'_0 g_1^\omega$ and is satisfied by all plays reaching $G_\lambda$. The objective $\Omega_1$ is not satisfied by the play $v'_0 g_1^\omega$ and is satisfied by plays reaching $G_\lambda$ if and only if the formula $\phi$ is not satisfied in those plays. 
    \item Using similar arguments as used in the proof of \autoref{prop:hardness_generic} and its adaptation to Boolean B\"uchi objectives, the following properties apply. A strategy $\sigma_0 \in \llbracket M \rrbracket$ makes the instance of the LTL \problemUVerifAb{} negative if every play $v'_0 \rho$ reaching $G_\lambda$ and consistent with this strategy falsifies objective $\Omega_1$ of Player~$1$ (as no payoff is then strictly larger than that of play $v'_0 g_1^\omega$, lost by Player~$0$). If this is the case, it follows that strategy $\sigma_0$ is a winning strategy for Player~$0$ from $v_0$ in the zero-sum game $(G_\lambda,\phi$) as every play $\rho$ consistent with this strategy satisfies formula $\phi$. The converse is also true. Player~$0$ therefore has a winning strategy from $v_0$ in $(G_\lambda,\phi)$ if and only if the corresponding instance of the LTL \problemUVerifAb{} is negative. It follows that the LTL \problemUVerifAb{} is \twoExptime{}-hard for LTL \gamesAb{} (as co-\twoExptime{} $=$ \twoExptime).
\end{itemize}
\end{proof}

\begin{theorem}
\label{thm:ltl_verif_pspace}
The LTL \problemVerifAb{} is \pspace{}-complete.
\end{theorem}

The proof of this theorem relies on two variants of the LTL model-checking problem that are both \pspaceComplete{} \cite{DBLP:journals/jacm/SistlaC85}. 

\subparagraph{LTL Model-Checking Problem.} Given a finite transition system $T$, an initial state, and an LTL formula $\psi$, the \emph{LTL existential} (resp.\! \emph{universal}) \emph{model-checking problem} is to decide whether $\psi$ is satisfied in at least one infinite path (resp.\! all infinite paths) of $T$ starting from the initial state. Notice that a finite transition system is the same model as a single-player labeled game arena and that an infinite path in $T$ corresponds to a play in this arena.

\begin{proof}[Proof of \autoref{thm:ltl_verif_pspace}]
We first prove that the LTL \problemVerifAb{} is in \pspace{} by adapting the \FPT{} algorithm provided for proving \autoref{thm:PRV_fpt}. Given an LTL \gameAb{} $\mathcal{G}$, this algorithm works as follows. For each payoff $p \in \{0,1\}^\nbrObjectives$, we check \emph{(i)} whether it is realizable and Pareto-optimal, if yes \emph{(ii)} whether there exists a play $\rho$ such that $\payoff{\rho} = p$ and $\won{\rho} = 0$. If for some payoff $p$, both tests succeed, then the given instance $\mathcal{G}$ is negative, otherwise it is positive. Checking that a payoff $p$ is realizable reduces to solving the LTL existential model-checking problem for the formula $\psi = (\bigwedge_{p_i = 1} \phi_i) \wedge (\bigwedge_{p_i = 0} \neg\phi_i$). This test can thus be performed in polynomial space. The second check in \emph{(i)} and the last check in \emph{(ii)} are similarly executed in polynomial space. The LTL \problemVerifAb{} is hence in \pspace{}.

We now prove that the LTL \problemVerifAb{} is \pspace-hard by showing that we can transform any instance of the LTL universal model-checking problem into an instance of the LTL \problemVerifAb{} such that the instance of the former is positive if and only if the corresponding instance of the latter is positive as well. Let $T$ be transition system and $\psi$ be an LTL formula. Given our previous remark, $T$ can be seen as a single-player labeled arena $G_\lambda$ for some labeling function $\lambda$. We create the following LTL \gameAb{} $\mathcal{G} = (G_\lambda, \psi, \phi_1)$ played on $G_\lambda = T$ where the objective of Player~$0$ is to satisfy the formula $\psi$ and the sole objective of Player~$1$ is to satisfy the formula $\phi_1 = true$. It is direct to see that any play in $G_\lambda$ satisfies the objective of Player~$1$ and therefore that every play in $G_\lambda$ is Pareto-optimal. It follows that the given instance of the LTL \problemVerifAb{} is positive if and only if every play in $G_\lambda$ satisfies the formula $\psi$. This corresponds exactly to the LTL universal model-checking problem.
\end{proof}

\section{Implementation and Evaluation}
\label{sec:implementation}

In this section, we aim to demonstrate the practical applicability of our verification framework for the \emph{\problemVerifAb{} and parity objectives}. We implemented Algorithms \ref{algo:fpt} and~\ref{algo:fpt-CE} and evaluated them on a parametric toy example generalizing \autoref{ex:intersection_general}, as well as on a family of randomly generated instances.

\input{intersection-figure-arena}

\subparagraph{Parametric Toy Example.}
We now develop \autoref{ex:intersection_general} into a proper instance of the \problemVerifAb{} for parity objectives, in such a way to later make it parametric when evaluating our algorithms. 

We assume that the behavior of Player~$0$, the system which controls car $c_1$, is to cross ahead when light $l_1$ is green and the intersection is clear of accidents. This behavior, corresponding to the single strategy $\sigma_0 \in \llbracket M \rrbracket$, is fixed and committed in advance. The resulting arena $G$ depicted in \autoref{fig:intersection_example_arena} describes the possible behaviors of the environment given this behavior of the system. When a play reaches a vertex in the arena, its content highlights whether the lights are red ($0$) or green ($1$) and whether the cars are waiting ($0$) or have tried to cross ($1$) at that point in the play. For simplicity, we assume that car $c_2$ (resp. $c_3$) tries to cross ahead when light $l_2$ (resp. $l_3$) turns green, that two lights can only be turned green at the same time if none of the lights are already green, and that once a light is green it stays so.

In the initial vertex, all lights are red and the cars are waiting. Player $1$ can decide to turn zero, one or two lights green. Notice that if lights $l_1$ and $l_2$ or $l_2$ and $l_3$ are turned green at the same time, an accident occurs as the corresponding cars are on crossing paths (we highlight vertices where this occurs in gray).

The objective $\Omega_0$ of Player~$0$ is to eventually cross the intersection without accident. The first objective of Player~$1$ is to ensure that no car waits infinitely often. 
The second (resp. third) objective of Player~$1$ is satisfied if car $c_2$ (resp. $c_3$) crosses the intersection before $c_3$ (resp. $c_2$) without accident. Finally, the fourth objective of Player~$1$ is satisfied if more than one car crosses at the same time.

It is easy to see that the single-player \gameAb{} $(G, \Omega_0, \Omega_1,  \dots, \Omega_4)$ is a positive instance of the \problemAb{}. For clarity, instead of specifying the parity objectives directly, we instead display the extended payoff of a play close to the edge on which it eventually loops. There are two Pareto-optimal payoffs in $G$ (highlighted in bold). First, if Player~$1$ turns both lights $l_1$ and $l_3$ green first, and then light $l_2$ then the payoff he obtains is $(1,0,1,1)$ as no car waits infinitely often, as car $l_3$ crosses first and as two cars have crossed at the same time. The second Pareto-optimal payoff is $(1,1,0,0)$ and occurs when Player~$1$ turns all lights green one at a time and turns light $l_2$ green before $l_3$.

\subparagraph{Implementation Details.} Both algorithms were implemented\footnote{Source code available at \url{https://github.com/skar0/pareto-rational-verification}.} in Python 3 using SPOT \cite{DBLP:conf/atva/Duret-LutzLFMRX16} (compiled to allow 64 acceptance sets) as a library to manipulate automata. We performed our experiments on a computer with an Intel Core i7-10875H CPU and 16GB of memory running Ubuntu 20.04 LTS. The game arena $G$ of a single-player parity \gameAb{} $\mathcal{G}$ is encoded as an automaton $\mathcal{A}$ in which $\nbrObjectives + 1$ priorities are assigned to each vertex, one for each priority function in $\mathcal{G}$.

\subparagraph{Payoff Realizability.} Our algorithms rely on different kinds of checks for the realizability of specific payoffs and other related properties, as discussed in \autoref{prop:parity_payoff_existence}. Such checks are performed in lines 6 and 8 of \autoref{algo:fpt} and in lines 3 and 4 of \autoref{algo:fpt-CE}. They are achieved by using SPOT's emptiness checking algorithm~\cite{DBLP:conf/atva/BaierBD00S19} to decide the existence of an accepting run in $\mathcal{A}$, given an acceptance condition expressed as a Boolean combination of priorities to be visited (in)finitely often. This acceptance condition encodes the desired check and such a run corresponds to a play satisfying it in $G$. In particular, a parity objective is easily translated into a Streett condition in that formalism. Therefore, checking for the existence of a play in $G$ with a payoff larger than or equal to $p$ amounts to deciding the emptiness of the automaton $\mathcal{A}$ representing $G$ for the language corresponding to the relevant intersection of parity objectives in $p$. We recall that the checks made in our algorithms are all performed in polynomial time except for the existence check of a play $\rho$ such that $\payoff{\rho} \not\in \; \downarrow^< \! A$ whose running time depends on the size of the antichain $A$.

\subparagraph{Theoretical Comparison of the Algorithms.} \autoref{thm:PRV_fpt} states that the \problemVerifAb{} is \FPT{} in the number $\nbrObjectives$ of objectives of Player~$1$ with a naive algorithm that constructs the antichain $\paretoSet{\sigma_0}$ and then checks for the existence of a Pareto-optimal play losing for Player~$0$. \autoref{algo:fpt} constructs the antichain $\paretoSet{\sigma_0}$ by descending in the lattice of payoffs level-by-level from payoff $(1,\ldots,1)$ while trying to find a Pareto-optimal play losing for Player~$0$ (and stops early upon finding such a play). \autoref{algo:fpt-CE} starts from the bottom of this lattice and climbs through it trying to find a counterexample (witness of a negative instance by \autoref{lem:certificate}) or to construct a certificate $A \sqsubseteq \paretoSet{\sigma_0}$ (witness of a positive instance). The latter algorithm is thus focused on a certificate instead of on $\paretoSet{\sigma_0}$ and avoids going through the lattice level-by-level. Nevertheless it contains a costly instruction in line 3 with the existence check of a play $\rho$ such that $\payoff{\rho} \not\in \; \downarrow^< \! A$ which depends on the current antichain $A$. 

\input{benchmarks-rerun}

\subparagraph{Comparison on Our Running Example.} 
As the \problemVerifAb{} is \FPT{} with an exponential dependence in the number $\nbrObjectives$ of objectives of Player~$1$, we evaluate the influence of parameters $\nbrObjectives$ and $|G|$ on the running time of \autoref{algo:fpt} and~\autoref{algo:fpt-CE} separately. We consider two families of arenas $G_k$ for some parameter $k$, corresponding to several copies of the arena of \autoref{fig:intersection_example_arena} all linked to a new initial vertex. \autoref{fig:benchmarkss}(a) reports the running time of both algorithms as a function of the parameter $k$ for the first family, in which the number of objectives remains $\nbrObjectives = 4$ and the number of copies (and therefore $|G|$) increases. Both positive ($+$) and negative ($-$) instances are considered (in the latter case by modifying the example so that a Pareto-optimal play is lost by Player~$0$). As expected, the running time increases polynomially with $|G|$ and remains overall low (even for thousands of copies). \autoref{algo:fpt-CE} performs better for both positive and negative instances, and we notice the early stopping of \autoref{algo:fpt} in case of negative instances. \autoref{fig:benchmarkss}(b) reports the running time as a function of $t$ for the second family, in which each copy of the arena has its own set of the two objectives for cars $c_2$ and $c_3$. \autoref{algo:fpt} exhibits an exponential running time in $\nbrObjectives$, whereas \autoref{algo:fpt-CE} executes much faster (for both positive and negative instances). Such a behavior can be explained by the fact that the overall number of realizable payoffs is low and grows slowly in this example (advantaging \autoref{algo:fpt-CE} which performs checks for lost plays with such payoffs), and that \autoref{algo:fpt} makes a systematic descent in the lattice from the top and performs an exponential number of checks as the realizable payoffs are low in the lattice.

\subparagraph{Comparison on Randomly Generated Instances.} We also evaluate the behavior of both algorithms on positive and negative randomly generated instances with $|G| = 500$, $d_i = 4$ for $i \in \{0, \dots, \nbrObjectives\}$, and for an increasing value of $\nbrObjectives$ (we fix the size of the arena and the number of priorities as they are not important factors for the complexity). \autoref{fig:benchmarkss}(c) and \autoref{fig:benchmarkss}(d) report the running time (on a logarithmic scale) of the 50 generated instances for each value of $\nbrObjectives$ as well as the average running time for both algorithms. This (average) running time appears to stay small even for large values of $t$. While \autoref{algo:fpt-CE} fared clearly better on the intersection example, this is not always the case for randomly generated instances. Even if \autoref{algo:fpt-CE} is slower on some instances, many other are solved faster with this algorithm than with \autoref{algo:fpt} (especially for larger values of $t$). We also notice that the distribution of the 50 running times of \autoref{algo:fpt} given some value of $t$ is denser than that of \autoref{algo:fpt-CE} which has larger variations. This may again be due to the fact that \autoref{algo:fpt} performs a systematic descent in the lattice. Assuming that the Pareto-optimal payoffs are in the middle of the lattice, numerous checks are first performed for non-realizable payoffs, constituting an overhead for the algorithm (which increases with $t$). With \autoref{algo:fpt-CE} exhibiting no such overhead, its running time is more dependent on individual examples.

To better understand the behavior of the algorithms, \autoref{table:statistics} reports the mean value of several parameters for each value of $\nbrObjectives$ and for positive and negative instances separately. The considered parameters are the average size of the antichain $\paretoSet{\sigma_0}$ and of the antichains constructed by both algorithms, as well as the ratio of payoffs realized by a play losing for Player~$0$ over the total number of realizable payoffs. The size of the antichain constructed by \autoref{algo:fpt} for positive instances is not indicated as it is equal to $|\paretoSet{\sigma_0}|$. In particular, we observe that the size of the antichain constructed in \autoref{algo:fpt-CE} is smaller than the one of \autoref{algo:fpt}, itself smaller or equal to (in the case of a positive instance) the size $|\paretoSet{\sigma_0}|$. We also observe that when $\nbrObjectives$ increases, the aforementioned ratio decreases. This could explain why \autoref{algo:fpt-CE} is more efficient than \autoref{algo:fpt} for larger values of $\nbrObjectives$. 

\begin{table}[t]
\caption{Mean value of specific parameters for randomly generated instances.}
\resizebox{1\textwidth}{!}{%
\begin{tabular}{|l|l|l|l|l|l|l|l|l|l|l|}
\hline
$t$  & 6     & 7     & 8    & 9     & 10     & 11     & 12 & 13 & 14 & 15   \\ \hline
$|\paretoSet{\sigma_0}|$ ($+$) &   6 &   7 &   9.82 &   18.28 &   49.50 &   100.96 &   210.88 &   418.44 &   794.24 &   1406.84 \\  
Ratio of lost payoffs ($+$) &  0.45 &  0.36 &  0.26 &  0.20 &  0.15 &  0.10 &  0.08 &  0.05 &  0.04 &  0.03 \\
$|A|$ in \autoref{algo:fpt-CE} ($+$) &   4.42 &   5.98 &   7.44 &   10.06 &   15.30 &   18.72 &   22.74 &   28.28 &   30.08 &   42.62 \\  \hline
$|\paretoSet{\sigma_0}|$ ($-$) &  6.02 &   11.02 &   21.74 &   45.84 &   88.08 &   171.22 &   304.76 &   527.30 &   847.28 &   1327.24 \\ 
Ratio of lost payoffs ($-$) &  0.90 &  0.80 &  0.71 &  0.64 &  0.55 &  0.48 &  0.42 &  0.38 &  0.34 &  0.31 \\
$|A|$ in \autoref{algo:fpt} ($-$) &    3.24 &   5.98 &   10.82 &   20.16 &   39.56 &   63.08 &   85.50 &   125.24 &   137.72 &   178.50 \\  
$|A|$ in \autoref{algo:fpt-CE} ($-$) &   3.68 &   5.48 &   8.12 &   12.32 &   16.90 &   24.26 &   33.04 &   36 &   35.94 &   35.92\\\hline
\end{tabular}%
}
\label{table:statistics}
\end{table}

In order to assess the cost of the test performed in line 3 of \autoref{algo:fpt-CE}, we have selected the positive and the negative randomly generated instance for which the algorithm exhibits the longest running time and retrieved the following data. \autoref{fig:benchmarks-CE}(a) reports the size of the antichain $A$ computed in the algorithm in each iteration. \autoref{fig:benchmarks-CE}(b) reports the running time (in seconds) for the call of line 3 in each iteration. We observe that on this example, the size of $A$ grows linearly and the time required to perform the check grows polynomially with it.

\input{benchmarks-CE}

\section{Conclusion}
In this paper, we have introduced the \problemVerifAb{} and its universal variant. The \problemVerifAb{} is \conpComplete{} for parity \gamesAb{} and \piComplete{} for Boolean B\"uchi \gamesAb{}. The \problemUVerifAb{} is in \pspace{} and \np{}- and \conp{}-hard for the former class of games, and \pspaceComplete{} for the latter class. The complexity of both problems was studied in the context where the objectives are defined using LTL formulas, and it was shown that the former was \pspaceComplete{} and the latter \twoExptime -complete. Both problems were also shown to be in \FPT{} for parity and Boolean B\"uchi objectives. Two variations of the FPT algorithm for the \problemVerifAb{} were introduced, and were also implemented and evaluated for the case of parity objectives on a parametric toy example as well as on randomly generated instances.

\bibliography{main}

\end{document}

%% file: macros.tex
\mathchardef\mhyphen="2D

\newcommand{\N}{\mathbb{N}}

\newcommand{\Plays}{\mathsf{Plays}}
\newcommand{\Hist}{\mathsf{Hist}}
\newcommand{\occ}[1]{\mathsf{Occ}({#1})}
\newcommand{\infOcc}[1]{\mathsf{Inf}({#1})}

\newcommand{\payoff}[1]{\mathsf{pay}({#1})}
\newcommand{\won}[1]{\mathsf{won}({#1})}
\newcommand{\payoffObj}[2]{\mathsf{pay}_{#1}({#2})}

\newcommand{\Playsigma}[1]{\mathsf{Plays}_{#1}}
\newcommand{\Histsigma}[1]{\mathsf{Hist}_{#1}}
\newcommand{\Playsigmazero}{\Playsigma{\sigma_0}}

\newcommand{\nbrObjectives}{t}
\newcommand{\Obj}{\Omega}
\newcommand{\ObjPlayer}[1]{\Omega_{#1}}
\newcommand{\target}{T} 

\newcommand{\parity}[1]{\mathsf{Parity}(#1)}

\newcommand{\BooleanBuchi}[1]{\mathsf{BooleanB\ddot{u}chi}(#1)}

\newcommand{\p}{$\mathsf{P}$}
\newcommand{\np}{$\mathsf{NP}$}
\newcommand{\npHard}{$\mathsf{NP}$-hard}
\newcommand{\npComplete}{$\mathsf{NP}$-complete}

\newcommand{\npspace}{$\mathsf{NPSPACE}$}

\newcommand{\copspace}{$\mathsf{co\mhyphen PSPACE}$}
\newcommand{\pspace}{$\mathsf{PSPACE}$}
\newcommand{\pspaceHard}{$\mathsf{PSPACE}$-hard}
\newcommand{\pspaceComplete}{$\mathsf{PSPACE}$-complete}
\newcommand{\nexptime}{$\mathsf{NEXPTIME}$}

\newcommand{\twoExptime}{$\mathsf{2EXPTIME}$}

\newcommand{\FPT}{$\mathsf{FPT}$}

\newcommand{\conp}{$\mathsf{co\mhyphen NP}$}
\newcommand{\conpHard}{$\mathsf{co\mhyphen NP}$-hard}
\newcommand{\conpComplete}{$\mathsf{co\mhyphen NP}$-complete}

\newcommand{\DP}{$\mathsf{BH_2}$}

\newcommand{\DPComplete}{$\mathsf{BH_2}$-complete}

\newcommand{\sigmaClass}{$\Sigma_2\mathsf{P}$}

\newcommand{\sigmaComplete}{$\Sigma_2\mathsf{P}$-complete}

\newcommand{\piClass}{$\Pi_2\mathsf{P}$}

\newcommand{\piComplete}{$\Pi_2\mathsf{P}$-complete}

\newcommand{\npNp}{$\mathsf{NP}^\mathsf{NP}$}
\newcommand{\npConp}{$\mathsf{NP}^\mathsf{co\mhyphen NP}$}

\newcommand{\problemSynt}{Stackelberg-Pareto Synthesis problem}
\newcommand{\problemAb}{SPS problem}

\newcommand{\SAT}{3SAT}
\newcommand{\coSAT}{co-3SAT}
\newcommand{\problemVerif}{Pareto-rational verification problem}
\newcommand{\problemVerifAb}{PRV problem}

\newcommand{\problemUVerif}{universal Pareto-rational verification problem}
\newcommand{\problemUVerifAb}{UPRV problem}
\newcommand{\problemPayoffVerifAb}{$p$-PRV problem}

\newcommand{\SATcoSAT}{SAT-unSAT}

\newcommand{\sigmaQBF}{$\Sigma_2$QBF}

\newcommand{\game}{Stackelberg-Pareto game}

\newcommand{\gameAb}{SP game}
\newcommand{\gamesAb}{SP games}
\newcommand{\paretoSet}[1]{P_{#1}}



%% file: intersection-figure.tex
\begin{figure}[t]
    \centering
    \resizebox{0.65\textwidth}{!}{%
	\begin{tikzpicture}
	
	\isogrid{10}{10}
	\begin{scope}[font=\huge]

	\draw[line width=2pt, rounded corners] (7-1.lower vertex) -- (5-3.right vertex) -- (3-1.upper vertex);
	
	\draw[line width=2pt, rounded corners] (9-3.lower vertex)--(7-5.right vertex) -- (9-8.lower vertex);
	
	\draw[line width=1pt] (6-4.right vertex)--(7-5.right vertex);
	
	\draw[line width=1pt, dashed] (9-2.upper vertex)--(6-5.left vertex);
	
	 \draw[line width=5pt, -{latex[slant=-0.6]}] ([yshift=-0.1cm, xshift=-0.3cm]3-7.upper vertex) -- ([xshift=0.1cm, yshift=0.1cm]4-6.left vertex);
	 \draw[line width=5pt, -{latex[slant=-0.6]}] ([yshift=-0.1cm, xshift=-0.1cm]7-5.left vertex) -- ([xshift=-0.1cm, yshift=-0.1cm]6-5.right vertex);
	 \draw[line width=5pt, -{latex[slant=0.6]}] ([yshift=-0.1cm, xshift=0.3cm]6-7.right vertex) -- ([yshift=-0.1cm, xshift=0.1cm]5-7.left vertex);
	 
	\draw[line width=1pt, fill=white, fill opacity=1] ([yshift=-0.5cm]8-4.left vertex) -- 
	([yshift=-0.5cm]7-3.lower vertex) --
	([yshift=-0.5cm]7-3.upper vertex) --
	([yshift=-0.5cm]6-4.upper vertex) --
	([yshift=-0.5cm]6-4.right vertex) --
	([yshift=-0.5cm]7-4.right vertex) -- 
	([yshift=-0.5cm]8-3.right vertex) -- 
	([yshift=-0.5cm]7-3.right vertex) -- 
	([yshift=-0.5cm]6-4.right vertex)
	([yshift=-0.5cm]7-4.left vertex) -- 
	([yshift=-0.5cm]7-3.upper vertex);
	
	\draw[line width=1pt, fill=gray, fill opacity=0.5] ([yshift=-0.5cm]8-4.left vertex) -- 
	([yshift=-0.5cm]7-3.lower vertex) --
	([yshift=-0.5cm]7-3.upper vertex) --
	([yshift=-0.5cm]6-4.upper vertex) --
	([yshift=-0.5cm]6-4.right vertex) --
	([yshift=-0.5cm]7-4.right vertex) -- 
	([yshift=-0.5cm]8-3.right vertex) -- 
	([yshift=-0.5cm]7-3.right vertex) -- 
	([yshift=-0.5cm]6-4.right vertex)
	([yshift=-0.5cm]7-4.left vertex) -- 
	([yshift=-0.5cm]7-3.upper vertex);
	
	 \node[] (c1) at ([yshift=-0.05cm]8-4.upper vertex){$c_1$};
	 
	\draw[line width=1pt, fill=white] ([yshift=0.07cm, xshift=0.55cm]7-5.upper vertex)--([yshift=0.25cm,xshift=0.25cm]7-5.right vertex) -- ([yshift=0.25cm, xshift=0.25cm]8-5.right vertex)--([yshift=0.07cm, xshift=0.55cm]8-5.upper vertex) -- cycle;
	
	\draw[rotate=30, fill=gray] (4.18,-9.26) ellipse (0.14cm and 0.19cm);
	\draw[rotate=30] (4,-9.6) ellipse (0.14cm and 0.19cm);

	 \node[] (l1) at ([yshift=-0.05cm, xshift=0.25cm]6-5.right vertex){$l_1$};

	\draw[line width=1pt, dashed] (6-7.left vertex)--(8-9.lower vertex);
	\draw[line width=1pt] (5-8.left vertex)--(6-7.left vertex);
	\draw[line width=2pt, rounded corners] (7-10.lower vertex)-- (5-7.right vertex)--(3-10.upper vertex);
	
	\draw[line width=1pt, fill=white] ([yshift=0.25cm, xshift=0.1cm]6-8.upper vertex)--([yshift=0.07cm, xshift=-0.25cm]5-8.right vertex) -- ([yshift=0.07cm, xshift=-0.25cm]4-8.right vertex)--([yshift=0.25cm, xshift=0.1cm]5-8.upper vertex) -- cycle;
	
	\draw[rotate=-30, fill=gray] (12.52,2.89) ellipse (0.14cm and 0.19cm);
	\draw[rotate=-30] (12.71,2.55) ellipse (0.14cm and 0.19cm);
	
	\node[] (l2) at ([yshift=0.0cm,xshift=0.25cm]4-8.upper vertex){$l_2$};

	\draw[line width=1pt, fill=white, fill opacity=1] ([yshift=0.5cm]7-8.left vertex) -- 
	([yshift=0.5cm]6-8.left vertex) --
	([yshift=0.5cm]6-8.upper vertex) --
	([yshift=0.5cm]7-9.upper vertex) --
	([yshift=0.5cm]7-9.lower vertex) --
	([yshift=0.5cm]8-9.left vertex) -- 
	([yshift=0.5cm]7-8.left vertex) 
	([yshift=0.5cm]7-9.left vertex) -- 
	([yshift=0.5cm]6-8.left vertex) 
	([yshift=0.5cm]7-9.upper vertex) -- 
	([yshift=0.5cm]7-9.lower vertex) --
	([yshift=0.5cm]8-9.left vertex)
	([yshift=0.5cm]8-9.left vertex) --
	([yshift=0.5cm]7-9.left vertex) --
	([yshift=0.5cm]7-9.upper vertex);
	
	\draw[line width=1pt, fill=gray, fill opacity=0.5] ([yshift=0.5cm]7-8.left vertex) -- 
	([yshift=0.5cm]6-8.left vertex) --
	([yshift=0.5cm]6-8.upper vertex) --
	([yshift=0.5cm]7-9.upper vertex) --
	([yshift=0.5cm]7-9.lower vertex) --
	([yshift=0.5cm]8-9.left vertex) -- 
	([yshift=0.5cm]7-8.left vertex) 
	([yshift=0.5cm]7-9.left vertex) -- 
	([yshift=0.5cm]6-8.left vertex) 
	([yshift=0.5cm]7-9.upper vertex) -- 
	([yshift=0.5cm]7-9.lower vertex) --
	([yshift=0.5cm]8-9.left vertex)
	([yshift=0.5cm]8-9.left vertex) --
	([yshift=0.5cm]7-9.left vertex) --
	([yshift=0.5cm]7-9.upper vertex);
	
	\node[] (c2) at ([yshift=0.05cm]7-8.upper vertex){$c_2$};

	\draw[line width=1pt, dashed] (4-7.left vertex)--(2-8.right vertex);
	
	\draw[line width=2pt, rounded corners] (1-8.upper vertex)-- (3-6.left vertex)--(1-3.upper vertex);

	\draw[line width=1pt] (3-5.right vertex)--(4-7.left vertex);

	\draw[line width=1pt, fill=white, fill opacity=1] ([yshift=0.5cm]3-7.lower vertex) -- 
	([yshift=0.5cm]3-7.left vertex) --
	([yshift=0.5cm]2-7.left vertex) --
	([yshift=0.5cm]1-7.right vertex) --
	([yshift=0.5cm]2-8.upper vertex) --
	([yshift=0.5cm]2-8.lower vertex) -- 
	([yshift=0.5cm]3-7.lower vertex) -- 
	([yshift=0.5cm]3-7.upper vertex) -- 
	([yshift=0.5cm]2-7.left vertex) --
	([yshift=0.5cm]3-7.upper vertex) -- 
	([yshift=0.5cm]2-8.upper vertex);
	
	\draw[line width=1pt, fill=gray, fill opacity=0.5] ([yshift=0.5cm]3-7.lower vertex) -- 
	([yshift=0.5cm]3-7.left vertex) --
	([yshift=0.5cm]2-7.left vertex) --
	([yshift=0.5cm]1-7.right vertex) --
	([yshift=0.5cm]2-8.upper vertex) --
	([yshift=0.5cm]2-8.lower vertex) -- 
	([yshift=0.5cm]3-7.lower vertex) -- 
	([yshift=0.5cm]3-7.upper vertex) -- 
	([yshift=0.5cm]2-7.left vertex) --
	([yshift=0.5cm]3-7.upper vertex) -- 
	([yshift=0.5cm]2-8.upper vertex);
	
    \node[] (c3) at ([yshift=0.05cm]2-8.left vertex){$c_3$};
        
	\draw[line width=1pt, fill=white] ([yshift=0.07cm, xshift=0.37cm]2-5.upper vertex)--([yshift=0.25cm,xshift=0.07cm]2-6.left vertex) -- ([yshift=0.25cm, xshift=0.07cm]3-6.left vertex)--([yshift=0.07cm, xshift=0.37cm]3-5.upper vertex) -- cycle;
	
	\draw[rotate=30, fill=gray] (6.19,-5.42) ellipse (0.14cm and 0.19cm);
	\draw[rotate=30] (6,-5.76) ellipse (0.14cm and 0.19cm);
	
	\node[] (l3) at ([xshift=-0.1cm]1-5.right vertex){$l_3$};

	\draw[line width=1pt, dashed] (2-2.upper vertex)--(4-4.right vertex);
	\draw[line width=1pt] (4-4.right vertex)--(5-4.left vertex);
	
	\end{scope}
	\end{tikzpicture}}%
    \caption{Traffic management at an intersection.}
    \label{fig:intersection_example}
\end{figure}

%% file: intersection-figure-arena.tex
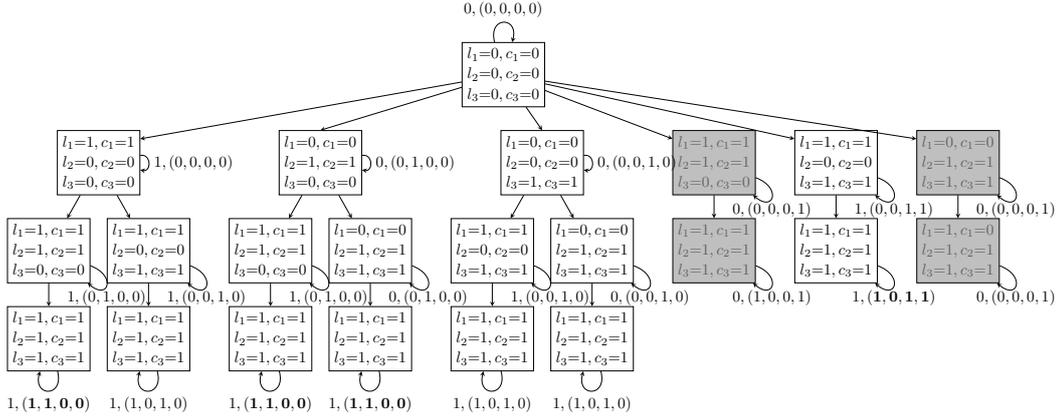
\begin{figure}[t]
	\centering
	\resizebox{\textwidth}{!}{%
		
		\begin{tikzpicture}

		\node[draw, rectangle,align=left] (v0) at (10.25,10){$l_1{=}0, c_1{=}0$\\$l_2{=}0, c_2{=}0$\\$l_3{=}0, c_3{=}0$};

    	\draw[-stealth] (v0) to [loop above, looseness=4] node []{$0,(0,0,0,0)$} (v0);

		\node[draw, rectangle,align=left, fill=gray, fill opacity=0.5] (v1) at (15,8){$l_1{=}1, c_1{=}1$\\$l_2{=}1, c_2{=}1$\\$l_3{=}0, c_3{=}0$};
		\node[draw, rectangle,align=left, fill=gray, fill opacity=0.5] (v2) at (15,6){$l_1{=}1, c_1{=}1$\\$l_2{=}1, c_2{=}1$\\$l_3{=}1, c_3{=}1$};

        \draw[-stealth] (v0) to [] node []{} (v1.150);        
        \draw[-stealth] (v1) to [] node []{} (v2);
    	\draw[-stealth] (v1) to [pos=0.5,out=340,in=320,looseness=4] node [below]{ $0,(0,0,0,1)$} (v1);
    	\draw[-stealth] (v2) to [pos=0.5,out=340,in=320,looseness=4] node [below]{$0,(1,0,0,1)$} (v2);

		\node[draw, rectangle,align=left] (v3) at (17.75,8){$l_1{=}1, c_1{=}1$\\$l_2{=}0, c_2{=}0$\\$l_3{=}1, c_3{=}1$};
		\node[draw, rectangle,align=left] (v4) at (17.75,6){$l_1{=}1, c_1{=}1$\\$l_2{=}1, c_2{=}1$\\$l_3{=}1, c_3{=}1$};

        \draw[-stealth] (v0) to [] node []{} (v3.150);        
        \draw[-stealth] (v3) to [] node []{} (v4);
        \draw[-stealth] (v3) to [pos=0.5,out=340,in=320,looseness=4] node [below]{$1,(0,0,1,1)$} (v3);
    	\draw[-stealth] (v4) to [pos=0.5,out=340,in=320,looseness=4] node [below]{$1,\mathbf{(1,0,1,1)}$} (v4);
    	
    	\node[draw, rectangle,align=left, fill=gray, fill opacity=0.5] (v5) at (20.5,8){$l_1{=}0, c_1{=}0$\\$l_2{=}1, c_2{=}1$\\$l_3{=}1, c_3{=}1$};
		\node[draw, rectangle,align=left, fill=gray, fill opacity=0.5] (v6) at (20.5,6){$l_1{=}1, c_1{=}0$\\$l_2{=}1, c_2{=}1$\\$l_3{=}1, c_3{=}1$};

        \draw[-stealth] (v0) to [] node []{} (v5.150);        
        \draw[-stealth] (v5) to [] node []{} (v6);
    	\draw[-stealth] (v5) to [pos=0.5,out=340,in=320,looseness=4] node [below]{$0,(0,0,0,1)$} (v5);
    	\draw[-stealth] (v6) to [pos=0.5,out=340,in=320,looseness=4] node [below]{$0,(0,0,0,1)$} (v6);
		
		\node[draw, rectangle,align=left] (v7) at (11.125,8){$l_1{=}0, c_1{=}0$\\$l_2{=}0, c_2{=}0$\\$l_3{=}1, c_3{=}1$};
		\node[draw, rectangle,align=left] (v8) at (12.25,6){$l_1{=}0, c_1{=}0$\\$l_2{=}1, c_2{=}1$\\$l_3{=}1, c_3{=}1$};
		\node[draw, rectangle,align=left] (v9) at (12.25,4){$l_1{=}1, c_1{=}1$\\$l_2{=}1, c_2{=}1$\\$l_3{=}1, c_3{=}1$};
        \node[draw, rectangle,align=left] (v10) at (10,6){$l_1{=}1, c_1{=}1$\\$l_2{=}0, c_2{=}0$\\$l_3{=}1, c_3{=}1$};
		\node[draw, rectangle,align=left] (v11) at (10,4){$l_1{=}1, c_1{=}1$\\$l_2{=}1, c_2{=}1$\\$l_3{=}1, c_3{=}1$};

        \draw[-stealth] (v0) to [] node []{} (v7.90);        
        \draw[-stealth] (v7) to [] node []{} (v8);
        \draw[-stealth] (v8) to [] node []{} (v9);
        \draw[-stealth] (v7) to [] node []{} (v10);
        \draw[-stealth] (v10) to [] node []{} (v11);
    	\draw[-stealth] (v7) to [pos=0.5,out=370,in=350,looseness=2] node [right]{$0,(0,0,1,0)$} (v7);
    	\draw[-stealth] (v8) to [pos=0.5,out=340,in=320,looseness=4] node [below]{$0,(0,0,1,0)$} (v8);
    	\draw[-stealth] (v9) to [loop below,looseness=4] node [below]{$1,(1,0,1,0)$} (v9);
    	\draw[-stealth] (v10) to [pos=0.5,out=340,in=320,looseness=4] node [below]{$1,(0,0,1,0)$} (v10);
    	\draw[-stealth] (v11) to [loop below,looseness=4] node [below]{$1,(1,0,1,0)$} (v11);
    	
		\node[draw, rectangle,align=left] (v12) at (6.125,8){$l_1{=}0, c_1{=}0$\\$l_2{=}1, c_2{=}1$\\$l_3{=}0, c_3{=}0$};
		\node[draw, rectangle,align=left] (v13) at (7.25,6){$l_1{=}0, c_1{=}0$\\$l_2{=}1, c_2{=}1$\\$l_3{=}1, c_3{=}1$};
		\node[draw, rectangle,align=left] (v14) at (7.25,4){$l_1{=}1, c_1{=}1$\\$l_2{=}1, c_2{=}1$\\$l_3{=}1, c_3{=}1$};
        \node[draw, rectangle,align=left] (v15) at (5,6){$l_1{=}1, c_1{=}1$\\$l_2{=}1, c_2{=}1$\\$l_3{=}0, c_3{=}0$};
		\node[draw, rectangle,align=left] (v16) at (5,4){$l_1{=}1, c_1{=}1$\\$l_2{=}1, c_2{=}1$\\$l_3{=}1, c_3{=}1$};

        \draw[-stealth] (v0) to [] node []{} (v12.90);        
        \draw[-stealth] (v12) to [] node []{} (v13);
        \draw[-stealth] (v13) to [] node []{} (v14);
        \draw[-stealth] (v12) to [] node []{} (v15);
        \draw[-stealth] (v15) to [] node []{} (v16);
    	\draw[-stealth] (v12) to [pos=0.5,out=370,in=350,looseness=2] node [right]{$0,(0,1,0,0)$} (v12);
    	\draw[-stealth] (v13) to [pos=0.5,out=340,in=320,looseness=4] node [below]{$0,(0,1,0,0)$} (v13);
    	\draw[-stealth] (v14) to [loop below,looseness=4] node [below]{$1,\mathbf{(1,1,0,0)}$} (v14);
    	\draw[-stealth] (v15) to [pos=0.5,out=340,in=320,looseness=4] node [below]{$1,(0,1,0,0)$} (v15);
    	\draw[-stealth] (v16) to [loop below,looseness=4] node [below]{$1,\mathbf{(1,1,0,0)}$} (v16);
    	
		\node[draw, rectangle,align=left] (v17) at (1.125,8){$l_1{=}1, c_1{=}1$\\$l_2{=}0, c_2{=}0$\\$l_3{=}0, c_3{=}0$};
		\node[draw, rectangle,align=left] (v18) at (2.25,6){$l_1{=}1, c_1{=}1$\\$l_2{=}0, c_2{=}0$\\$l_3{=}1, c_3{=}1$};
		\node[draw, rectangle,align=left] (v19) at (2.25,4){$l_1{=}1, c_1{=}1$\\$l_2{=}1, c_2{=}1$\\$l_3{=}1, c_3{=}1$};
        \node[draw, rectangle,align=left] (v20) at (0,6){$l_1{=}1, c_1{=}1$\\$l_2{=}1, c_2{=}1$\\$l_3{=}0, c_3{=}0$};
		\node[draw, rectangle,align=left] (v21) at (0,4){$l_1{=}1, c_1{=}1$\\$l_2{=}1, c_2{=}1$\\$l_3{=}1, c_3{=}1$};

        \draw[-stealth] (v0) to [] node []{} (v17.30);        
        \draw[-stealth] (v17) to [] node []{} (v18);
        \draw[-stealth] (v18) to [] node []{} (v19);
        \draw[-stealth] (v17) to [] node []{} (v20);
        \draw[-stealth] (v20) to [] node []{} (v21);
    	\draw[-stealth] (v17) to [pos=0.5,out=370,in=350,looseness=2] node [right]{$1,(0,0,0,0)$} (v17);
    	\draw[-stealth] (v18) to [pos=0.5,out=340,in=320,looseness=4] node [below]{$1,(0,0,1,0)$} (v18);
    	\draw[-stealth] (v19) to [loop below,looseness=4] node [below]{$1,(1,0,1,0)$} (v19);
    	\draw[-stealth] (v20) to [pos=0.5,out=340,in=320,looseness=4] node [below]{$1,(0,1,0,0)$} (v20);
    	\draw[-stealth] (v21) to [loop below,looseness=4] node [below]{$1,\mathbf{(1,1,0,0)}$} (v21);
		\end{tikzpicture}
		}%
	
	\caption{The arena $G$ of the intersection example.}
	\label{fig:intersection_example_arena}
\end{figure}

%% file: benchmarks-rerun.tex
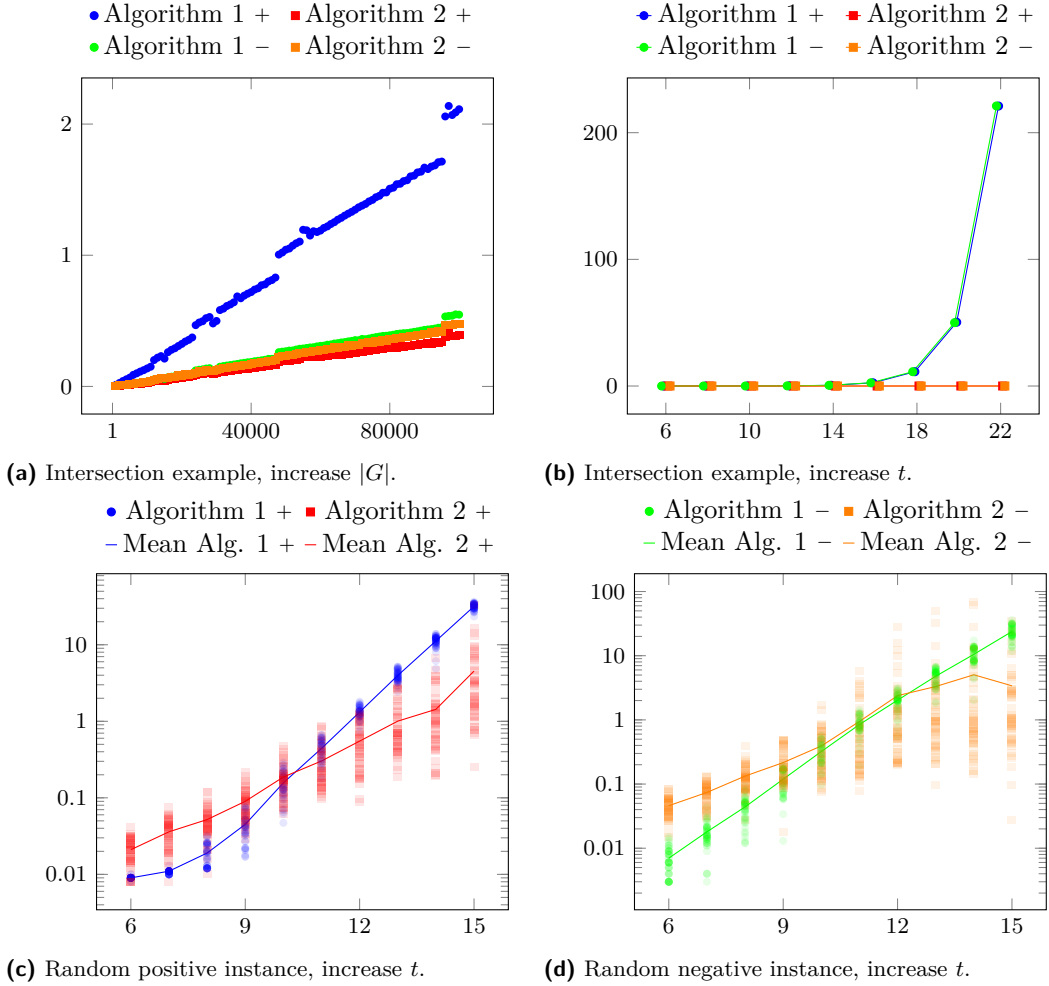
\begin{figure}[t]
\begin{subfigure}[t]{0.5\textwidth}
\centering
\begin{tikzpicture}
\begin{axis}[
    x tick label style={/pgf/number format/.cd, set thousands separator={},
    fixed},   
    width=\textwidth,
    xtick={1, 40000, 80000},    scaled x ticks = false,
    tick label style={font=\footnotesize},
    legend style={
    draw={none}, 
    at={(0,1.25)},
    anchor=north west,
    legend columns=2,
    text height=1.5ex,
    /tikz/every even column/.append style={column sep=0.2cm},
    legend image post style={
    opacity=1,
    mark size=4.5pt, 
    xscale=0.33, 
    yscale=0.33}}
    ]
	\addplot[
	only marks, 
    color=blue,
    mark=*, 
    fill opacity=1,
    mark size=1.5pt,
    draw opacity = 0]
    table [y=AO_time, x=nbr_vertices]{benchmarks-rerun/intersection-vertices-positive.dat};
    \addlegendentry{Algorithm 1 \raisebox{0.2ex}{\text{\scriptsize $+$}}}
	\addplot[
	only marks, 
    color=red,
    mark=square*, 
    fill opacity=1,
    mark size=1.5pt,
    draw opacity = 0]
	table [y=CE_time, x=nbr_vertices]{benchmarks-rerun/intersection-vertices-positive.dat};
    \addlegendentry{Algorithm 2 \raisebox{0.2ex}{\text{\scriptsize $+$}}}
	\addplot[
	only marks, 
    color=green,
    mark=*, 
    fill opacity=1,
    mark size=1.5pt,
    draw opacity = 0]
    table [y=AO_time, x=nbr_vertices]{benchmarks-rerun/intersection-vertices-negative.dat};
    \addlegendentry{Algorithm 1 \raisebox{0.2ex}{\text{\scriptsize $-$}}}
	\addplot[
	only marks, 
    color=orange,
    mark=square*, 
    fill opacity=1,
    mark size=1.5pt,
    draw opacity = 0]
	table [y=CE_time, x=nbr_vertices]{benchmarks-rerun/intersection-vertices-negative.dat};
    \addlegendentry{Algorithm 2 \raisebox{0.2ex}{\text{\scriptsize $-$}}}
\end{axis}%
\end{tikzpicture}%
\caption{Intersection example, increase $|G|$.}
\end{subfigure}
\hfill 
\begin{subfigure}[t]{0.5\textwidth}
\centering
\begin{tikzpicture}
\begin{axis}[
    x filter/.expression={(x < 4)? nan : x},
    width=\textwidth,
    xtick={6, 10, 14, 18, 22},
    scaled x ticks = false,
    tick label style={font=\footnotesize},
    legend style={
    draw={none}, 
    at={(0,1.25)},
    anchor=north west,
    legend columns=2,
    text height=1.5ex,
    /tikz/every even column/.append style={column sep=0.2cm},
    legend image post style={
    opacity=1,
    mark size=4.5pt, 
    xscale=0.33, 
    yscale=0.33}}
    ]
	\addplot[
	color=blue,
    mark=*,
    mark size=1.5pt,
    x filter/.code={\pgfmathadd{\pgfmathresult}{-0.1}}]
	table [y=AO_time, x=nbr_objectives, ]{benchmarks-rerun/intersection-objectives-positive.dat};
    \addlegendentry{Algorithm 1 \raisebox{0.2ex}{\text{\scriptsize $+$}}}
	\addplot[
	color=red,
    mark=square*,
    mark size=1.5pt,
    x filter/.code={\pgfmathadd{\pgfmathresult}{0.1}}]
	table [y=CE_time, x=nbr_objectives]{benchmarks-rerun/intersection-objectives-positive.dat};
    \addlegendentry{Algorithm 2 \raisebox{0.2ex}{\text{\scriptsize $+$}}}
    \addplot[
    color=green,
    mark=*,
    mark size=1.5pt,
    x filter/.code={\pgfmathadd{\pgfmathresult}{-0.2}}]
	table [y=AO_time, x=nbr_objectives]{benchmarks-rerun/intersection-objectives-negative.dat};
    \addlegendentry{Algorithm 1 \raisebox{0.2ex}{\text{\scriptsize $-$}}}
	\addplot[
	color=orange,
    mark=square*,
    mark size=1.5pt,
    x filter/.code={\pgfmathadd{\pgfmathresult}{0.2}}]
	table [y=CE_time, x=nbr_objectives]{benchmarks-rerun/intersection-objectives-negative.dat};
    \addlegendentry{Algorithm 2 \raisebox{0.2ex}{\text{\scriptsize $-$}}}
\end{axis}%
\end{tikzpicture}%
\caption{Intersection example, increase $\nbrObjectives$.}
\end{subfigure}\\
\begin{subfigure}[t]{0.5\textwidth}
\centering
\begin{tikzpicture}
\begin{axis}[
    ymode=log,
    width=\textwidth,
    xtick={6, 9, 12, 15},
    log ticks with fixed point,
    tick label style={font=\footnotesize},
    legend style={
    draw={none}, 
    at={(0,1.25)},
    anchor=north west,
    legend columns=2,
    /tikz/every even column/.append style={column sep=0.05cm},
    legend image post style={
    opacity=1,
    mark size=6pt, 
    xscale=0.25, 
    yscale=0.25}}
    ]
    \addplot[
    only marks, 
    color=blue,
    mark=*, 
    fill opacity=0.1, 
    draw opacity = 0,
    mark size=1.5pt]
	table [y=AO_time, x=nbr_objectives]{benchmarks-rerun/random-positive.dat};   \addlegendentry{Algorithm 1 \raisebox{0.2ex}{\text{\scriptsize $+$}}}
    \addplot[
    only marks, 
    color=red,
    mark=square*, 
    fill opacity=0.1, 
    draw opacity = 0,
    mark size=1.5pt]
	table [y=CE_time, x=nbr_objectives]{benchmarks-rerun/random-positive.dat};   \addlegendentry{Algorithm 2 \raisebox{0.2ex}{\text{\scriptsize $+$}}}
    \addplot[
    color=blue]
	table [y=mean_AO_time, x=nbr_objectives]{benchmarks-rerun/random-positive.dat};  
	\addlegendentry{Mean Alg. 1 \raisebox{0.2ex}{\text{\scriptsize $+$}}}
    \addplot[
    color=red]
	table [y=mean_CE_time, x=nbr_objectives]{benchmarks-rerun/random-positive.dat};  
	\addlegendentry{Mean Alg. 2 \raisebox{0.2ex}{\text{\scriptsize $+$}}}
\end{axis}%
\end{tikzpicture}%
\caption{Random positive instance, increase $\nbrObjectives$.}
\end{subfigure}
\hfill 
\begin{subfigure}[t]{0.5\textwidth}
\centering
\begin{tikzpicture}
\begin{axis}[
    ymode=log,
    width=\textwidth,
    xtick={6, 9, 12, 15},
    log ticks with fixed point,
    scaled x ticks = false,
    tick label style={font=\footnotesize},
    legend style={
    draw={none}, 
    at={(0,1.25)},
    anchor=north west,
    legend columns=2,
    text height=1.5ex,
    /tikz/every even column/.append style={column sep=0.05cm},
    legend image post style={
    opacity=1,
    mark size=6pt, 
    xscale=0.25, 
    yscale=0.25}}
    ]
	\addplot[
	only marks, 
	color=green,
    mark=*, 
    fill opacity=0.1, 
    draw opacity = 0,
    mark size=1.5pt]
	table [y=AO_time, x=nbr_objectives]{benchmarks-rerun/random-negative.dat};  
    \addlegendentry{Algorithm 1 \raisebox{0.2ex}{\text{\scriptsize $-$}}}
	\addplot[
	only marks, 
	color=orange,
    mark=square*, 
    fill opacity=0.1, 
    draw opacity = 0,
    mark size=1.5pt]
	table [y=CE_time, x=nbr_objectives]{benchmarks-rerun/random-negative.dat};  
    \addlegendentry{Algorithm 2 \raisebox{0.2ex}{\text{\scriptsize $-$}}}
    \addplot[
    color=green]
	table [y=mean_AO_time, x=nbr_objectives]{benchmarks-rerun/random-negative.dat};  
	\addlegendentry{Mean Alg. 1 \raisebox{0.2ex}{\text{\scriptsize $-$}}}
    \addplot[
    color=orange]
	table [y=mean_CE_time, x=nbr_objectives]{benchmarks-rerun/random-negative.dat}; 
    \addlegendentry{Mean Alg. 2 \raisebox{0.2ex}{\text{\scriptsize $-$}}}
\end{axis}%
\end{tikzpicture}%
\caption{Random negative instance, increase $\nbrObjectives$.}
\end{subfigure}
\caption{Running time of both algorithms (in seconds) as a function of $\nbrObjectives$ or $|G|$ ($x$-axis).}
\label{fig:benchmarkss}
\end{figure}

%% file: benchmarks-CE.tex
\begin{figure}[t] 
    \begin{subfigure}[t]{0.5\textwidth}
        \centering
        \begin{tikzpicture}
        \begin{axis}[
            x tick label style={/pgf/number format/.cd, set thousands separator={},
            fixed},   
            width=\textwidth,
            xtick={0, 100, 200},
            scaled x ticks = false,
            tick label style={font=\footnotesize},
            legend style={
            draw={none}, 
            at={(0,1.25)},
            anchor=north west,
            legend columns=1,
            text height=1.5ex,
            /tikz/every even column/.append style={column sep=0.2cm},
            legend image post style={
            opacity=1,
            mark size=4.5pt, 
            xscale=0.33, 
            yscale=0.33}},
            xlabel={iteration},
            ylabel={$|A|$},
            ylabel near ticks,
            xlabel near ticks
            ]
        	\addplot[
        	only marks, 
            color=red,
            mark=square*, 
            fill opacity=1,
            mark size=1.5pt,
            draw opacity = 0]
        	table[y=A_size, x=iteration]{benchmarks-rerun/CE_difficult_positive.dat};
            \addlegendentry{Positive instance}
        	\addplot[
        	only marks, 
            color=orange,
            mark=square*, 
            fill opacity=1,
            mark size=1.5pt,
            draw opacity = 0]
        	table[y=A_size, x=iteration]{benchmarks-rerun/CE_difficult_negative.dat};
            \addlegendentry{Negative instance}
        \end{axis}%
        \end{tikzpicture}%
        \caption{Evolution of $|A|$.}
    \end{subfigure}\hfill \begin{subfigure}[t]{0.5\textwidth}
        \centering
        \begin{tikzpicture}
        \begin{axis}[
            x tick label style={/pgf/number format/.cd, set thousands separator={},
            fixed},   
            width=\textwidth,
            xtick={0, 100, 200},
            scaled x ticks = false,
            tick label style={font=\footnotesize},
            legend style={
            draw={none}, 
            at={(0,1.25)},
            anchor=north west,
            legend columns=1,
            text height=1.5ex,
            /tikz/every even column/.append style={column sep=0.2cm},
            legend image post style={
            opacity=1,
            mark size=4.5pt, 
            xscale=0.33, 
            yscale=0.33}},
            xlabel={iteration},
            ylabel={running time (seconds)},
            ylabel near ticks,
            xlabel near ticks
            ]
        	\addplot[
        	only marks, 
            color=red,
            mark=square*, 
            fill opacity=1,
            mark size=1.5pt,
            draw opacity = 0]
        	table[y=call_time, x=iteration]{benchmarks-rerun/CE_difficult_positive.dat};
            \addlegendentry{Positive instance}
        	\addplot[
        	only marks, 
            color=orange,
            mark=square*, 
            fill opacity=1,
            mark size=1.5pt,
            draw opacity = 0]
        	table[y=call_time, x=iteration]{benchmarks-rerun/CE_difficult_negative.dat};
            \addlegendentry{Negative instance}
        \end{axis}%
        \end{tikzpicture}%
        \caption{Evolution of the running time of line 3.}
    \end{subfigure}
    \caption{Statistics for each iteration of Algorithm 2 ($x$-axis) on the most difficult positive and negative randomly generated instance from the benchmarks.}
    \label{fig:benchmarks-CE}
\end{figure}
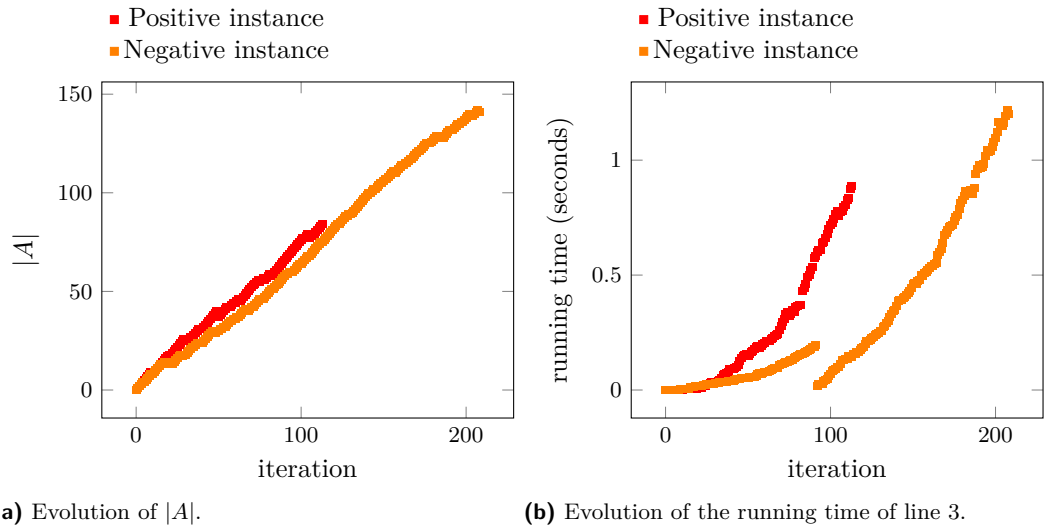